%% file: Manuscript.tex
\documentclass[preprint,12pt, 3p, sort&compress]{elsarticle}
\usepackage{fullpage}
\usepackage{color}
\usepackage{caption}
\usepackage{subcaption}
\usepackage{multirow}
\usepackage{amsmath}
\usepackage{amssymb}
\usepackage{amsthm}
\usepackage[colorlinks   = true, urlcolor     = cyan, linkcolor    = blue, citecolor   = red]{hyperref}
\usepackage{bm}
\usepackage{pict2e}
\usepackage{epstopdf}
\usepackage{comment}
\usepackage{epsfig}
\usepackage{url}
\usepackage[ruled,vlined]{algorithm2e}

\newcommand{\expchar}[2]{\textrm{e}^{\frac{2\pi \textrm{i} #1}{#2}}}
\newcommand{\expcharconj}[2]{\textrm{e}^{-\frac{2\pi \textrm{i} #1}{#2}}}

\journal{Journal of the Mechanics and Physics of Solids}

\input{preamble}

\begin{document}


\begin{frontmatter}

\title{Cyclic Density Functional Theory :  A route to the first principles simulation of bending in nanostructures}

\author[lbl]{Amartya S.\ Banerjee}
\ead{asb@lbl.gov}
\author[phanish_address]{Phanish Suryanarayana\corref{cor1}}
\ead{phanish.suryanarayana@ce.gatech.edu}

\cortext[cor1]{Corresponding author}

\address[lbl]{Computational Research Division,
Lawrence Berkeley National Laboratory, Berkeley, CA 94720, U.S.A}
\address[phanish_address]{College of Engineering, Georgia Institute of Technology, Atlanta, GA 30332, U.S.A}

\begin{abstract}
We formulate and implement Cyclic Density Functional Theory (Cyclic DFT) --- a self-consistent first principles  simulation method for nanostructures with cyclic symmetries. Using arguments based on Group Representation Theory, we rigorously demonstrate that the Kohn-Sham eigenvalue problem for such systems can be reduced to a fundamental domain (or cyclic unit cell) augmented with cyclic-Bloch boundary conditions. Analogously, the equations of electrostatics appearing in Kohn-Sham theory can be reduced to the fundamental domain augmented with cyclic boundary conditions. By making use of this symmetry cell reduction, we show that the electronic ground-state energy and the Hellmann-Feynman forces on the atoms can be calculated using quantities defined over the fundamental domain. We develop a symmetry-adapted finite-difference discretization scheme to obtain a fully functional numerical realization of the proposed approach. We verify that our  formulation and implementation of Cyclic DFT is both accurate and efficient through selected examples. 

The connection of cyclic symmetries with uniform bending deformations provides an elegant route to the ab-initio study of bending in nanostructures using Cyclic DFT. As a demonstration of this capability, we simulate the uniform bending of a silicene nanoribbon and obtain its energy-curvature relationship from first principles. A self-consistent ab-initio simulation of this nature is unprecedented and well outside the scope of any other systematic first principles method in existence. Our simulations reveal that the bending stiffness of the silicene nanoribbon is intermediate between that of graphene and molybdenum disulphide --- a trend which can be ascribed to the variation in effective thickness of these materials. We describe several future avenues and applications of Cyclic DFT, including its extension to the study of non-uniform bending deformations and its possible use in the study of the nanoscale flexoelectric effect.
\end{abstract}


\end{frontmatter}

\section{Introduction}
\label{sec:introduction}
In recent decades, quantum mechanical calculations using Kohn-Sham Density Functional Theory (Kohn-Sham DFT) \citep{HK_DFT, KohnSham_DFT} have become the de facto workhorse of computational materials science. The pseudopotential plane-wave method (\emph{Plane-wave DFT}) --- perhaps the most widely used implementation of the Kohn-Sham theory --- is currently available in a number of mature software packages \citep{Kresse_abinitio_iterative, CASTEP_1, Quantum_Espresso_1, Gonze_ABINIT_1}. The Plane-wave DFT approach involves expanding the unknowns (e.g., the Kohn-Sham orbitals and the electron density) into a linear combination of plane-waves\footnote{Plane-waves are functions of the form $\mathrm{e}^{\mathrm{i}\bfk\cdot\bfx}$. They form eigenfunctions of translational symmetry operators.}, and subsequently carrying out various computations through the use of Fast Fourier Transforms (FFTs) to seamlessly switch between quantities expressed on real space grids and their plane-wave expansion coefficients. Due to the periodic nature of plane-waves, Plane-wave DFT is well suited for the study of systems with translational symmetry. However, the study of structures which are non-periodic may require the use of large computational supercells \citep{Martin_ES}, thereby rendering the method inefficient for such problems.\footnote{Alternatives to Plane-wave DFT include real-space methods based on finite-differences \citep{Octopus_1, PARSEC, Phanish_SPARC_1, ghosh2016sparc2} and finite-elements \cite{Gavini_Kohn_Sham, Pask_FEM_review_1, Pask_FEM_review_2,Gavini_higher_order}. Although these methods allow non-periodic boundary conditions to be imposed more readily, their use in general symmetry-adapted self-consistent first principles calculations has not been considered prior to this work.} 

A vast number of materials systems of interest today are non-periodic, but are associated with other physical symmetry groups. This includes for example, nanoclusters and molecules associated with point group symmetries, and helical nanostructures associated with screw transformation symmetries. The importance of such nanostructures associated with alternative (non-periodic) symmetries cannot be overstated. For instance, they are anticipated to exhibit unprecedented  materials properties --- particularly, collective properties such as ferromagnetism and ferroelectricity --- in manners that are otherwise unavailable in the bulk phase \citep{James_OS}. Consequently, a substantial body of work has been devoted in recent years to the theoretical framework and mathematical classification of such structures\footnote{In the mechanics of materials literature, such materials systems have been referred to as  \emph{Objective Structures} \citep{James_OS}.}  \citep{James_OS, DEJ_ObjForm}. Further, judicious use of the symmetries associated with these structures has been made in designing novel computational methods -- both at the level of atomistic simulations \citep{Dumitrica_James_OMD, dayal2010nonequilibrium, aghaei2012symmetry} as well as electronic structure calculations \citep{My_PhD_Thesis, banerjee2015spectral}.

While the study of electronic properties of nanostructures associated with various non-periodic physical symmetries is of much scientific and technological interest in itself, an additional outcome of having access to computational methods specifically designed to exploit the underlying physical symmetries is that these methods allow one to study the associated deformation modes. In the setting of conventional Plane-wave DFT for example, the application of a homogeneous deformation to a periodic system still results in a structure with translational symmetry, and therefore the deformed system can be easily accommodated through a unit cell that has also been deformed in the same manner. On the other hand, it is far more difficult to account for non-homogeneous deformations such as bending and torsion in the periodic setting, whereby their study \citep{wei2012bending, naumov2011gap} is likely to involve various complications, inaccuracies and inefficiencies. 

To the best of our knowledge, as far as atomistic systems are concerned, the connections between various physical symmetry groups and specific non-homogeneous deformation modes appears first in \citep{James_OS}. This idea has been subsequently exploited for simulating the effects of bending deformations by means of cyclic symmetry groups \citep{Dumitrica_James_OMD, Dumitrica_Bending_Graphene, ma2015thermal, Pekka_Efficient_Approach, Pekka_CNT_Bending, Pekka_GNR_Bending, Pekka_Revised_Periodic}, and torsional deformations by means of helical symmetry groups \citep{Dumitrica_James_OMD, zhang2008stability, zhang2009electromechanical, Pekka_Efficient_Approach, kit2012twisting, Pekka_Twisted_GNR, Pekka_Revised_Periodic}. A pervasive issue with these aforementioned works however, is that the simulations in question have been carried out within the framework of interatomic potentials or tight binding methods (classical semi-empirical or Density Functional Tight Binding (DFTB)).  The failure of these more approximate models in simulating various physical systems is well known \citep{ismail2000ab,hauch1999dynamic, cocco2010gap, koskinen2009density, naumov2011gap} and it has been understood for some time that access to a systematic self-consistent first principles simulation methodology would be highly desirable \citep{Pekka_Revised_Periodic, James_OS, My_PhD_Thesis}. However, such a computational methodology appears to have been out of reach prior to this work.\footnote{Indeed, as remarked in \citep{Pekka_Revised_Periodic}, a central difficulty has been to formulate an analog of plane-waves for non-periodic symmetries. This issue has been subsequently addressed and resolved in \citep{My_PhD_Thesis}.}

\begin{figure}[ht]
\centering
\begin{subfigure}{\textwidth}
\centering
\includegraphics[width=0.5\textwidth]{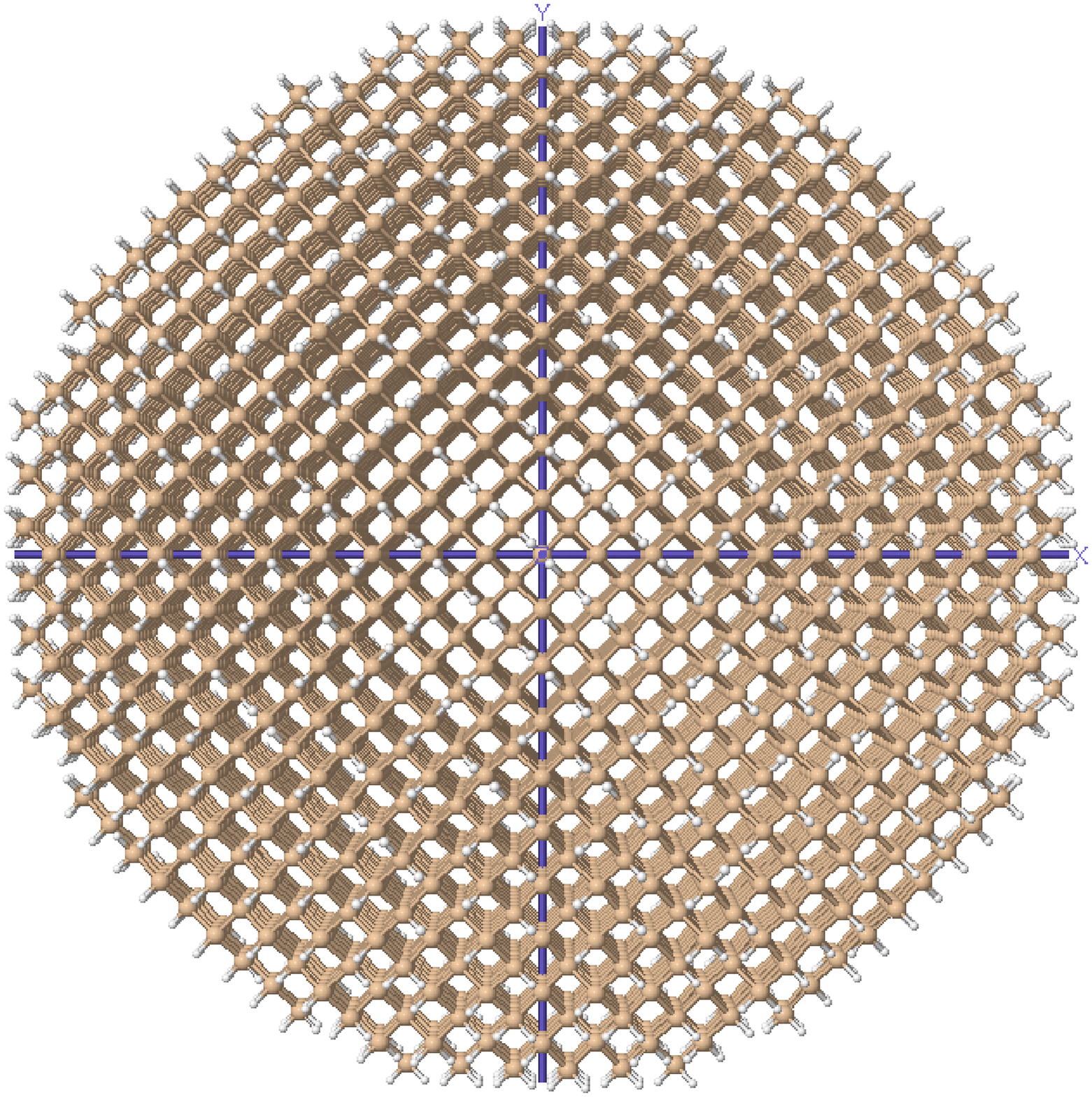}
\caption{Hydrogen passivated silicon quantum nanodot with 4-fold cyclic symmetry. We thank Yunkai Zhou (Southern Methodist University) for providing us with the nanodot structure.}
\label{fig_ex:subfig_a}
\end{subfigure}
\begin{subfigure}{.48\textwidth}
\centering
\includegraphics[width=0.9\textwidth]{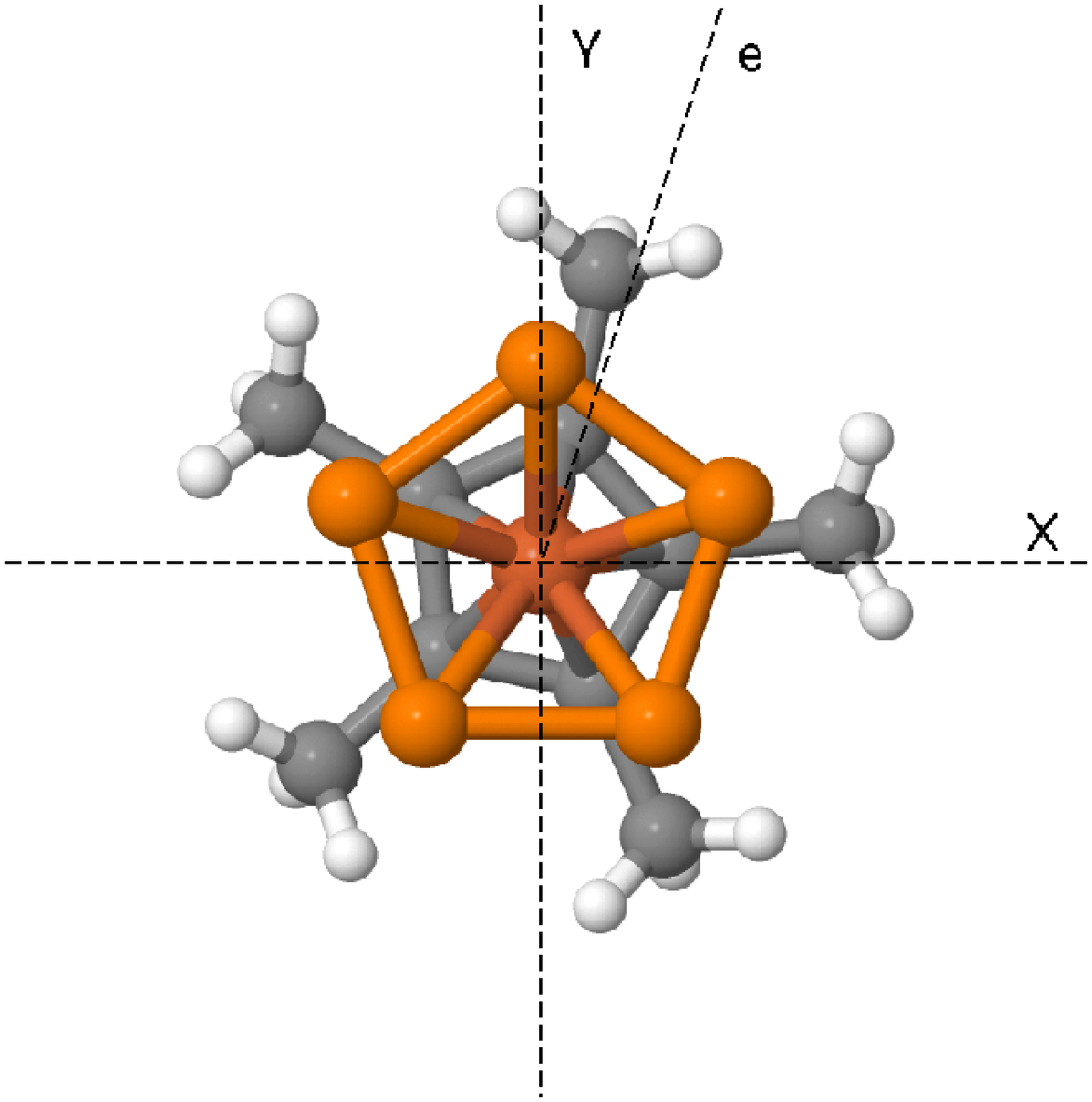}
\caption{Pentaphosphaferrocene molecule with a 5-fold symmetry: the atoms between the X axis and the line denoted as e can be rotated in steps of $72^{\circ}$ and replicated to produce the entire structure.}
\label{fig_ex:subfig_b}
\end{subfigure}%
$\quad$
\begin{subfigure}{.48\textwidth}
\centering
\includegraphics[width=0.9\textwidth]{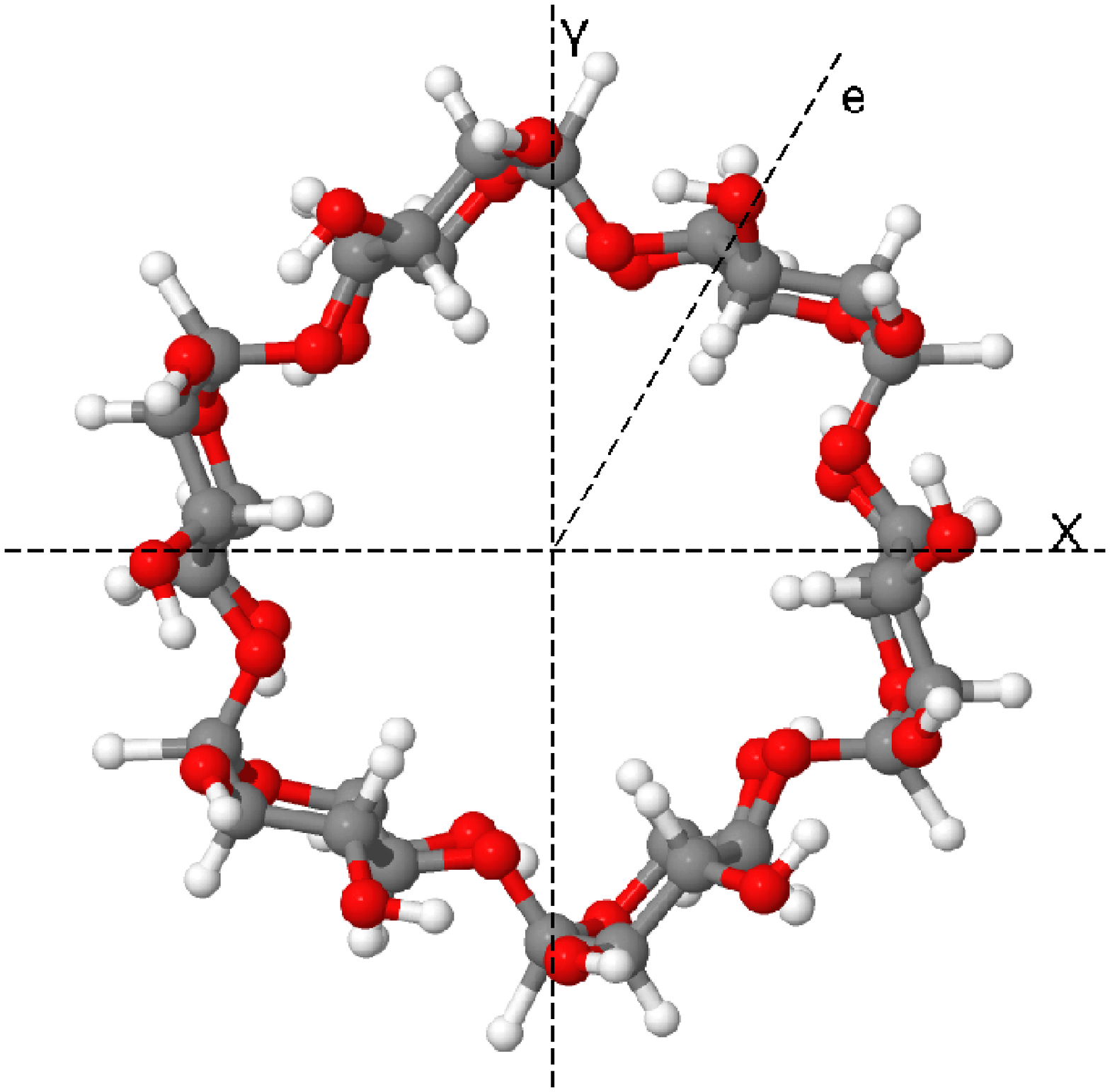}
\caption{Alpha-cyclodextrin molecule with a 6-fold symmetry: the atoms between the X axis and the line denoted as e can be rotated in steps of $60^{\circ}$ and replicated to produce the entire structure.}
\label{fig_ex:subfig_c}
\end{subfigure}
\caption{Examples of cyclic structures from nanotechnology and chemistry.}
\label{fig:cyclic_structures}
\end{figure}

In view of the above discussion, we formulate and implement Cyclic Density Functional Theory  (Cyclic DFT) --- a self-consistent first principles simulation method for cyclic nanostructures.\footnote{The crystallographic restriction theorem \citep{senechal1996quasicrystals} prevents any periodic code --- including all Plane-wave DFT codes --- from making full use of the symmetry of a general cyclic structure.} Cyclic symmetries are ubiquitous in various clusters and molecular systems \citep{wiki_cyclic_compound, hargittai2009symmetry, willock2009, go1973ring}  (see Fig.~\ref{fig:cyclic_structures} for some examples), and therefore the present methodology can be used to study a large variety of nanostructures from first principles in a systematic and efficient manner. Notably, even general point group symmetries associated with complex nanosystems usually contain cyclic groups as proper subgroups \citep{altmann1994point}, which extends the scope of the various materials systems that can be studied with Cyclic DFT.\footnote{For example, the icosahedral group that is associated with many fullerenes contains a cyclic subgroup originating from a 5-fold rotational symmetry.} Furthermore, due to the connections between cyclic symmetries with bending deformations in nanostructures (i.e., the idea that cyclic boundary conditions locally simulate the behavior of a system subjected to uniform bending), Cyclic DFT makes it possible to carry out systematic ab-initio simulations of nanostructures subjected to bending deformations. In particular, this opens up the possibility that electro-mechanical or other multi-physics coupling effects in nanostructures can be faithfully simulated from first principles by means of this novel framework.

Broadly, the present contribution can be viewed as a particular flavor of so called Objective DFT \citep{My_MS_Thesis, My_PhD_Thesis}, i.e., a self-consistent first principles method for an atomistic or molecular system with a non-periodic symmetry (i.e., an \textit{Objective Structure}). While Objective DFT is far more general and even allows systems associated with non-Abelian or non-compact symmetry groups to be simulated, it reduces to a particularly simple and efficient form for the case of cyclic structures, as we show in this work. Furthermore, the aforementioned connection of cyclic symmetries with the efficient simulation of bending deformations in nanostructures provides a compelling case as to why a thorough description of the theoretical and practical aspects of Cyclic DFT is warranted. Finally, Cyclic DFT can easily be extended to study systems of infinite extent (e.g., nanotubes, bending of nanosheets), which is beyond the scope of the spectral scheme originally presented in Objective DFT. This provides the motivation for the current work.

One of the key theoretical steps involved in the formulation of Cyclic DFT is the symmetry cell reduction of the Kohn-Sham equations posed on the entire cyclic structure to the fundamental domain (or cyclic unit cell) associated with the cyclic symmetry group. In order to accomplish this task, we use tools from Group Representation Theory to rigorously formulate and prove a Bloch theorem for cyclic structures (Theorem \ref{thm:Bloch_Theorem}). In addition, we perform a symmetry cell reduction of the electrostatics problem, the system's electronic ground-state free energy, and the Hellmann-Feynman forces on the atoms. Consequently, on every iteration in the Self-Consistent Field (SCF) method \citep{slater1974self}, the reduction due to symmetry is expected to result in computational savings that varies in direct proportion to the cyclic group order of the system under study.\footnote{Asymptotically, the reduction in computational cost is actually quadratic with respect to group order (Footnote \ref{Footnote:QuadraticReduction}).} An additional outcome of the Bloch theorem for cyclic structures is that it allows the Kohn-Sham eigenvalues associated with the cyclic structure to be presented in a manner similar to that of traditional band diagrams for periodic structures. As a result, the onset of instabilities and transitions in such systems can be more easily detected.

We develop a symmetry-adapted finite-difference method to discretize the equations of Cyclic DFT resulting from the symmetry cell reduction. In addition to the systematic convergence properties and relative simplicity of this discretization strategy, it allows bending in nanostructures to be simulated more efficiently by allowing us to place finite-difference nodes only in regions of interest. Subsequently, we develop a completely functional numerical realization of the method using state of the art algorithms. Numerical experiments and comparison with benchmark calculations allow us to establish the accuracy and efficiency of our Cyclic DFT implementation. Further, these experiments reveal that the symmetry cell reduction not only results in the expected computational speed-up on every SCF step, but it also allows overall faster convergence of the SCF fixed-point iteration. As a consequence of these computational savings, highly accurate first principles simulations of cyclic systems containing many hundreds of atoms can be done quite routinely using even a serial implementation of Cyclic DFT.

Finally, as a demonstration of how Cyclic DFT enables the study of materials properties from first principles, we simulate uniform bending deformations in a silicene nanoribbon and obtain its energy-curvature relationship. In particular, our simulations allow us to obtain the bending stiffness of this material in the linear Euler--Bernoulli regime and to compare our results with the bending stiffness of other two-dimensional materials that have been studied in the literature. The usage of first principles approaches in the mechanics of materials brings with it the hope that these techniques will enable engineers to understand the true atomistic dependence of constitutive laws. Our application of Cyclic DFT to directly evaluate the bending stiffness can be broadly viewed in this light and it motivates us to carry out more sophisticated simulations of this nature on other important nanomaterials systems in the near future.\footnote{Exploiting point group symmetries in ab-initio calculations has also been considered in the chemistry literature in the context of Linear Combination of Atomic Orbitals (LCAO) methods (see \citep{LCAO_famous,Roothan_SCF_Symmetry,LCAO_1,LCAO_2,atkins2011molecular}, for example). However, the focus of these works has been towards using symmetry-adapted basis functions for reducing the effort associated with the computation of the multi-center integrals and the entries of the Hamiltonian matrix that appear in LCAO calculations. Thus, they differ in perspective from our own approach which focuses on formulation and solution of symmetry-adapted cell problems instead. Additionally, in contrast to finite-difference methods, it is usually non-trivial to systematically improve the quality of solution obtained via LCAO  methods due to basis completeness issues. Finally, the application of cyclic symmetries to the study of bending deformations in nanostructures does not appear to have been considered by chemists before.} 

The rest of the paper is organized as follows.\footnote{The atomic unit system with $m_{\text{e}}=1,e=1,\hbar=1,\frac{1}{4\pi\epsilon_0}=1$, is chosen for the rest of the work, unless otherwise mentioned.} In Section \ref{Sec:Formulation}, we present the mathematical underpinnings and symmetry-adapted finite-difference discretization scheme that constitute Cyclic DFT. In Section \ref{Sec:Results}, we verify the accuracy and efficiency of Cyclic DFT, and then use it to study the uniform bending of a silicene nanoribbon. Finally, we provide concluding remarks in Section \ref{Conclusions}. 


\section{Formulation} \label{Sec:Formulation}
In this section, we describe the key aspects of Cyclic DFT. We begin by a formal discussion of cyclic groups and cyclic structures in Section \ref{Subsec:CyclicGroups}, and then discuss the Kohn-Sham theory for such structures in Section \ref{Subsec:KohnShamProblem}. Next, we show how cyclic symmetry leads to an appropriate symmetry cell reduction of the Kohn-Sham problem in Section \ref{Subsec:CyclicBlochTheorem}, and pose the resulting equations on the fundamental domain in Section \ref{Subsec:CyclicDFTEq}.  Finally, we formulate a symmetry-adapted finite-difference discretization scheme for solving the cell problem in Section \ref{Subsec:FD}.

\subsection{Cyclic groups and cyclic structures} \label{Subsec:CyclicGroups}
Let $\calG$ denote a cyclic group of order $\mathfrak{N}$ generated by the single element $g$, i.e., 
\begin{align}
\label{abstract_cyclic_group}
\displaystyle\calG=\{g^0, g^1,g^2,\ldots,g^{\mathfrak{N}-1}\}\,.
\end{align}
Identifying $g^{\mathfrak{N}} = g^0 = e$ as the group identity element, we see that the inverse element of $g^{\gamma}\in \calG$ is the element $g^{\mathfrak{N}-\gamma} \in \calG$. A physical realization of this abstract group\footnote{Also identifiable as an additive group of integers modulo $\mathfrak{N}$.} can be obtained by considering a discrete group of rotations about a common axis. Let $(\bfe_1,\bfe_2,\bfe_3)$ represent the standard basis of $\rz^3$. Here, we form a faithful linear representation of $\calG$ on $\rz^3$ by using the following rotation matrices about $\bfe_3$:
\begin{align}
\label{rotation_cyclic_group}
\displaystyle\{\mathfrak{R}^{\gamma}:\gamma=0,\ldots,\mathfrak{N}-1\} \quad\text{with}\quad
\mathfrak{R}^{\gamma}=\begin{pmatrix}
              \cos\frac{2\pi\gamma}{\mathfrak{N}} & -\sin\frac{2\pi\gamma}{\mathfrak{N}} & 0\\ 
               \sin\frac{2\pi\gamma}{\mathfrak{N}} & \;\cos\frac{2\pi\gamma}{\mathfrak{N}} & 0 \\ 
               0 & 0 & 1
             \end{pmatrix}\,.
\end{align}
With this representation, the multiplication operation of the group simply becomes multiplication of the matrices $\mathfrak{R}^{\gamma}$. Henceforth, we will identify a cyclic group of order $\mathfrak{N}$ by the collection of matrices shown in Eq.~\ref{rotation_cyclic_group}; we will use $\Gamma_{\mathfrak{N}}$ to denote the collection of integers $\{0,1,\ldots,\mathfrak{N}-1\}$; and for any matrix $\mathfrak{R}^{\gamma} \in \calG$ and set $S \subset \rz^3$, we will use $\mathfrak{R}^{\gamma}(S)$ to denote the set $\{\mathfrak{R}^{\gamma}\bfx : \bfx \in S \}$. 

Let $\calC \subseteq \rz^3$ be an open set with a regular boundary which is invariant under the action of the group $\calG$, i.e., for every $\gamma \in \Gamma_{\mathfrak{N}}$, we have:
\begin{align}
\label{domain_invariance}
\mathfrak{R}^{\gamma}(\calC) = \calC\,.
\end{align}
Then, the \emph{symmetry cell} or \emph{fundamental domain} or \emph{cyclic unit cell} of $\calG$ in $\calC$ is a set $\calD_{\calG} \subset \calC$ such that: 
\begin{align}
\bigcup_{\gamma}\mathfrak{R}^{\gamma}(\calD_{\calG}) &= \calC\,,
\label{fundamental_domain_1}
\end{align}
and for every $\gamma_1, \gamma_2 \in \Gamma_{\mathfrak{N}}$
\begin{align}
\mathfrak{R}^{\gamma_1}(\calD_{\calG}) \bigcap \mathfrak{R}^{\gamma_2}(\calD_{\calG}) &= \textrm{a set of Lebesgue measure }0, \textrm{ for } \gamma_1 \neq \gamma_2\,. 
\label{fundamental_domain_2}
\end{align}
We will denote the boundary surfaces common to the original domain and the fundamental domain by $\partial\calD_{\calG}^{0} = \partial\calD_{\calG} \cap \partial\calC$, and those that are unique to the fundamental domain by $\partial\calD_{\calG}^{C} = \partial\calD_{\calG} \setminus \partial\calD_{\calG}^{0} $. For example, if $\calC$ is a finite cylinder with axis $\bfe_3$, the  fundamental domain $\calD_{\calG}$ will be a sector of the cylinder with angle  $2\pi / \mathfrak{N}$.

Consider a finite collection\footnote{In a forthcoming contribution, we will consider the case in which $\calP$ consists of an infinite number of points. This can be used for example, to represent a structure such as a bent silicene sheet that is periodic in the $\bfe_3$ direction while having cyclic symmetry around the $\bfe_3$ axis.} of distinct points\footnote{For the sake of simplicity, we will assume that none of the points are at the origin, although this assumption is not really required as far as theory and methods developed in this work are concerned.} $\calP \subset \calD_{\calG}$, which we will denote by $\big\{ \bfx_{0,k} \big\}_{k=1}^{M_{\calP}}$. These points are representative of atomic positions within the fundamental domain and will be referred to as the \emph{simulated points} or \emph{simulated atoms}. A \emph{cyclic structure}\footnote{Alternately, an \emph{objective structure generated by a cyclic group}, using the terminology employed in the work of \citep{My_PhD_Thesis, DEJ_ObjForm}.} $\calS_{\cal_G, \calP}$ is the group orbit of a given cyclic group $\calG$ acting on the simulated atoms in $\calP$:
\begin{align}
\label{cyclic_structure}
\displaystyle\calS_{\cal_G, \calP}=\bigcup_{\gamma}\mathfrak{R}^{\gamma}(\calP)= \bigcup_{\gamma , k}\mathfrak{R}^{\gamma}\bfx_{0,k}\,.
\end{align}
We will represent a general atom of such a structure using the notation $\bfx_{\gamma,k} = \mathfrak{R}^{\gamma}\bfx_{0,k}$, $\gamma \in \Gamma_{\mathfrak{N}}$. Further, we will denote the nuclear charge\footnote{Following usual convention adopted in ab-initio theories, we will consider nuclear charge to be negative and therefore the electronic charge to be positive. Additionally, we will neglect the core charges and consider only the valence charges, consistent with the use of the pseudopotential approximation.} of the atom at position $\bfx_{0,k}  \in \calP$ by $Z_k$. It follows from the cyclic symmetry of the structure that the atom at the position $\bfx_{\gamma,k} = \mathfrak{R}^{\gamma}\bfx_{0,k}$ (for any $\gamma \in \Gamma_{\mathfrak{N}}$) also has nuclear charge $Z_k$.


\subsection{Kohn-Sham problem for cyclic structures} \label{Subsec:KohnShamProblem}
A detailed formulation of the Kohn-Sham problem \citep{KohnSham_DFT} as it applies to the computation of the electronic and structural properties of a finite collection of atoms can be found in numerous references (e.g., see \citep{banerjee2015spectral, Gavini_Kohn_Sham, Phanish_SPARC_1}). In the present work, we are interested specifically in the first principles simulations of cyclic structures $\calS_{\cal_G, \calP}$, as have been defined in the previous subsection. For the sake of simplicity, we assume that the system as a whole is charge neutral, i.e., there are a total of $\mathfrak{N}N_e$ electrons, where $N_{\text{e}} = - \displaystyle \sum_{k=1}^{M_\calP}Z_k$ represents the number of electrons in the fundamental domain.\footnote{The electrons are not physically restricted to be confined to the fundamental domain. This situation is analogous to the band theory of crystalline solids \citep{Reed_Simon4, LeBris_Defranceschi_ReviewBook}, where a fixed number of electrons are prescribed to be in the periodic unit cell, although the electrons can be spatially delocalized over the infinite crystal.} Consequently, we pose the Kohn-Sham equations in terms of $\mathfrak{N} N_{\text{s}}$ orbitals with $N_{\text{s}} > N_{\text{e}} / 2$ and $N_{\text{s}} \approx N_{\text{e}} / 2$.\footnote{In practical calculations, $N_{\text{s}} = N_{\text{e}} / 2 + N_{\text{extra}}$, and $N_{\text{extra}} = 10 - 20$ for typically used electronic temperatures \citep{zhou2014chebyshev}.} Additionally, we employ the local density approximation (LDA) \citep{KohnSham_DFT}, thermalization via the Fermi-Dirac smearing \citep{Mermin_Finite_Temp}, the local pseudopotential approximation \citep{Kohanoff}, and the local reformulation of the electrostatics \citep{suryanarayana2014augmented,Phanish_SPARC_1}. 

In the aforementioned setting, the Kohn-Sham equations in $\calC$ can be written as:\footnote{Henceforth, the space of square integrable functions on the domain $\calC$ will be denoted by $\mathsf{L}^2(\calC)$, while the Sobolev space of functions in $\mathsf{L}^2(\calC)$ whose $\mathsf{k}^{\textrm{th}}$ order weak derivatives also lie in $\mathsf{L}^2(\calC)$ will be denoted by $\mathsf{H^k}(\calC)$. The subspace of $\mathsf{H}^1(\calC)$ functions that vanish on the boundary $\partial \calC$ in the trace sense will be denoted by $\mathsf{H}^1_0(\calC)$. The space of $k-$times continuously differentiable functions that are compactly supported on $\calC$ will be denoted by $\textrm{C}^\mathsf{k}_c(\calC)$. In particular, continuous functions with compact support on $\calC$ will be denoted as $\textrm{C}_c^0(\calC)$. }
\begin{subequations} \label{Eqns:KS}
\begin{align}
\label{KS_equations}
K(\rho)\,\phi_i&=\lambda_i\,\phi_i\,,\;\innprod{\phi_i}{\phi_j}{\mathsf{L}^2(\calC)}=\delta_{ij}\;,\\
\label{KS_explain1}
\text{where}\quad
K(\rho)&=-\half \Delta + V_{\text{es}}(\rho, \calS_{\cal_G, \calP})  + V_{\text{xc}}(\rho)\;,\\
\label{KS_explain2}
\text{with} \quad
\rho(\bfx)&=2 \sum_{i=1}^{\mathfrak{N} N_{\text{s}}} f_i\abs{\phi_i(\bfx)}^2\,,\\
\label{poisson_problem}
\text{and} \quad
-\frac{1}{4\pi}\Delta V_{\text{es}} &= b + \rho\,. \\
\label{KS_explain3}
\text{Here,} \quad f_i &= \left( 1 + \exp\left( \frac{\lambda_i - \lambda_f}{\sigma} \right) \right)^{-1}, \\
\label{KS_explain4}
\text{with $\lambda_f$ s.t.} \quad &2\sum_{i=1}^{\mathfrak{N} N_{\text{s}}} f_i = \mathfrak{N} N_{\text{e}}\,.
\end{align}
\end{subequations}
The corresponding boundary conditions are:
\begin{subequations} \label{Eqns:KS:BC}
\begin{align} 
\text{for} \,\,\, \bfx \in \partial\calC \,, \quad  \phi_i(\bfx) &= 0 \,,  \\
\rho(\bfx) &= 0 \,, \\
\text{and} \quad V_{\text{es}}(\rho(\bfx),\calS_{\cal_G, \calP}) &=  \int_{\calC}\!\frac{\rho(\bfy) + b(\bfy, \calS_{\cal_G, \calP})}{\abs{\bfx-\bfy}}\; \mathrm{d\bfy} \,. \label{Eqn:KS:BC:Ves} 
\end{align}
\end{subequations}
Above, $\rho$ is the electron density; $K(\rho)$ is the Kohn-Sham operator with eigenvalues $\lambda_i$ and eigenfunctions $\phi_i \in \mathsf{H}^1(\calC)$, commonly referred to as the Kohn-Sham orbitals; $V_{\text{es}}$ is the electrostatic potential \citep{Pask2005,Gavini_Kohn_Sham, suryanarayana2014augmented}, written as the solution to the Poisson problem in Eqs.~\ref{poisson_problem} and \ref{Eqn:KS:BC:Ves}; $V_{\text{xc}}$ is the exchange-correlation potential; $\mathfrak{N} N_{\text{s}}$ is the total number of states\footnote{This can be thought of as choosing $N_{\text{s}}$ states corresponding to the fundamental domain.}; $0 \leq f_i \leq 1$ are the thermalized orbital occupations arising from the Fermi-Dirac smearing (Eq.~\ref{KS_explain3}) with parameter $\sigma$; $\lambda_f$ is the Fermi level of the system, determined by the solution of Eq.~\ref{KS_explain4} \footnote{As a result, the electron density satisfies the constraint on the total number of electrons:
\begin{align}
\label{density_normalized}
\int_{\calC}\!\rho(\bfx)\, \mathrm{d\bfx}= \mathfrak{N} N_{\text{e}}\,.
\end{align}}; 
and $b(\bfy, \calS_{\cal_G, \calP}) = \displaystyle \sum_{\gamma = 0}^{\mathfrak{N}-1} \sum_{k = 1}^{M_{\calP}} b_k(\bfy, \bfx_{\gamma,k})$ is the total nuclear pseudocharge density, with $b_k$ being the pseudocharge of the $k^{th}$ nucleus.\footnote{Each $b_k(\cdot, \bfx_{\gamma,k})$ is a spherically symmetric smooth function that is compactly supported in a small ball centered at $\bfx_{\gamma,k}$, with net enclosed charge $Z_k$. This implies that $b_k$ can be expressed using the ansatz $b_k(\bfx, \bfx_{\gamma,k}) = \eta_k(\abs{\bfx - \bfx_{\gamma,k}})$,  where $\eta_k(\cdot)$  is a smooth one-dimensional function that satisfies $\eta_k(r)=0 \, \forall \, r>r_k$ and $\displaystyle \int_{0}^{r_k}\!\eta_k(r) 4\pi r^2 \mathrm{dr} = Z_k$\,. As a result, the total nuclear pseudocharge density satisfies the relation
\begin{align}
\displaystyle \int_{\calC}\! b(\bfx, \calS_{\cal_G, \calP})\, \mathrm{d\bfx} = \mathfrak{N}\sum_{k=1}^{M_\calP}Z_k\,.
\end{align}} 
Due to the finite extent of the system under study and its overall charge neutrality, zero-Dirichlet boundary conditions are prescribed on the Kohn-Sham orbitals $\phi_i$ \citep{wavefunc_decay1, wavefunc_decay2, banerjee2015spectral, Gavini_Kohn_Sham}. It therefore follows from Eq.~\ref{KS_explain2} that the electron density $\rho$ also obeys this boundary condition. The Dirichlet boundary conditions prescribed on $V_{\text{es}}$ are obtained by expressing the solution of Eq.~\ref{poisson_problem} in terms of the Green's function of the Laplacian \citep{Phanish_SPARC_1}.\footnote{As explained later, the boundary conditions for $V_{\text{es}}$ can be numerically evaluated by direct quadrature or through the use of cylindrical multipole moments.} 

The self-consistent solution of the Kohn-Sham problem described by Eqs. \ref{Eqns:KS} and \ref{Eqns:KS:BC} leads to the electronic ground-state\footnote{Following standard terminology \citep{LeBris_ReviewBook}, we will refer to the electronic ground state to mean the self-consistent solution of the Kohn-Sham equations for a given set of nuclear positions.}  for the cyclic structure $\calS_{\cal_G, \calP}$. The nonlinear eigenvalue problem is typically solved using a fixed-point iteration with respect to either the electron density $\rho$ or the total effective potential
\begin{align}
\label{effective_potential}
V_{\textrm{eff}}(\bfx) = V_{\textrm{es}}(\rho(\bfx), \calS_{\cal_G, \calP}) + V_{{\textrm{xc}}}(\rho(\bfx)) \,,
\end{align}
and this is commonly referred to as the SCF method \citep{slater1974self}. In each  iteration of this method, the Poisson problem in Eqs.~\ref{poisson_problem} and \ref{Eqn:KS:BC:Ves} needs to be solved to calculate $V_{\textrm{eff}}$, and the lowest $\mathfrak{N}N_{\text{s}}$ eigenstates of the linearized Kohn-Sham operator: 
\begin{align}
\label{linear_operator}
\hamil = -\half \Delta + V_{\textrm{eff}}(\bfx)
\end{align}
need to be determined to calculate $\rho$.\footnote{It is possible to calculate the electron density directly from $\hamil$, without actually calculating the orbitals (e.g., see \cite{suryanarayana2013optimized,suryanarayana2013spectral,pratapa2015spectral}). } The eigenfunctions of $\hamil$ are subjected to zero--Dirichlet boundary conditions, consistent with the Kohn-Sham orbitals. 

At the electronic ground-state, the system's free energy can be evaluated by means of a Harris-Foulkes \citep{harris1985simplified,foulkes1989tight} type functional \citep{Phanish_SPARC_1}:\footnote{In some cases, it might be necessary to add a term $E_{\text{overlap}}({\calS_{\cal_G, \calP}})$ to the free energy expression in order to account for the possible overlaps of the pseudocharges \citep{suryanarayana2014augmented, Phanish_SPARC_1}.} 
\begin{align}
\nonumber
\mathcal{F}({\calS_{\cal_G, \calP}})  =& \;2\sum_{i = 1}^{\mathfrak{N}N_{\text{s}}} f_i \lambda_i + \int_{\calC}\!\varepsilon_{\text{xc}} (\rho(\bfx)) \rho(\bfx) \, \mathrm{d \bfx} - \int_{\calC}\!V_{\text{xc}}(\rho(\bfx))\rho(\bfx)\,\mathrm{d\bfx} \\\nonumber 
&+ \frac{1}{2} \int_{\calC}\!\big (b(\bfx, \calS_{\cal_G, \calP}) - \rho(\bfx)\big) V_{\text{es}}(\bfx, \calS_{\cal_G, \calP}) \, \mathrm{d\bfx} \\\nonumber
&-\half \sum_{\gamma = 0}^{\mathfrak{N}-1} \sum_{k = 1}^{M_{\calP}} \int_{\calC}\!b_k(\bfx, \bfx_{\gamma,k})\, V_k(\bfx, \bfx_{\gamma,k})\,\mathrm{d\bfx} \nonumber \\
&+ {2\,\sigma} \sum_{i=1}^{\mathfrak{N} N_{s}} \big( f_i \log f_i + (1-f_i) \log (1 - f_i) \big)\,, 
 \label{Harris_Foulkes_Energy} 
\end{align}
where the first term is the so called Kohn-Sham band energy; the second and third terms arise from the exchange-correlation effects; the fourth and fifth terms arise due to electrostatic interactions and pseudocharge self energies, respectively; and the last term represents the free energy associated with the electronic entropy of the system. The corresponding Hellmann-Feynman force \citep{Finnis_book, Parr_Yang, Martin_ES} on the atom $\bfx_{\gamma,k}$\footnote{More specifically, its nucleus.} can be written as \citep{suryanarayana2014augmented, Phanish_SPARC_1}:\footnote{If $E_{\text{overlap}}({\calS_{\cal_G, \calP}})$ is included in the free energy expression in Eq. \ref{Harris_Foulkes_Energy}, it becomes necessary to add the term 
\begin{align}
\bff_{\text{overlap}}(\bfx_{\gamma,k}) = - \pd{E_{\text{overlap}}({\calS_{\cal_G, \calP}})}{\bfx_{\gamma,k}}
\end{align}
to the atomic force in order to account for possible overlaps of pseudocharges \citep{suryanarayana2014augmented, Phanish_SPARC_1}. \label{footnote:fn_1}}
\begin{align}
\label{force_formula_1}
\bff_{\bfx_{\gamma,k}} =- \pd{\mathcal{F}({\calS_{\cal_G, \calP}})}{\bfx_{\gamma,k}} = \int_{\calC}\!\nabla b_k(\bfx, \bfx_{\gamma,k})\big(V_{\text{es}}(\bfx, \calS_{\cal_G, \calP}) - V_k(\bfx, \bfx_{\gamma,k}) \big)\,\mathrm{d\bfx}\,.
\end{align}

The problem of determining the equilibrium geometry\footnote{In principle, an equilibrium geometry (i.e., a configuration in which the internal forces on the atoms are all zero) can be associated with a local minimum, maximum or saddle point of the energy landscape. However, most applications in mechanics are usually concerned with energy minimizing structures and we will also restrict our attention to such cases.} of the cyclic structure can be stated as solving the minimization problem:\footnote{In practice, several local structural minima (and not just a single global minimum) may be involved in the energy landscape of a structure.}
\begin{align}
\label{geometry_optimization_original}
\infm{\{\bfx_{\gamma,k} \in \calS_{\cal_G, \calP}\}}{\mathcal{F}({\calS_{\cal_G, \calP}})}\,,\,\text{subject to}\;\bfx_{\gamma,k} = \mathfrak{R}^{\gamma}\bfx_{0,k} \,,\, \text{for} \,\, \gamma \in \Gamma_{\mathfrak{N}}\,. 
\end{align}
For every atomic configuration that arises during this geometry optimization procedure, the electronic ground-state needs to be determined by solving the Kohn-Sham equations in Eqs.~\ref{Eqns:KS} and \ref{Eqns:KS:BC}, after which the corresponding free energy and atomic forces can be calculated using Eqs.~\ref{Harris_Foulkes_Energy} and \ref{force_formula_1}, respectively.


\subsection{Symmetry cell reduction of the Kohn-Sham problem for cyclic structures} \label{Subsec:CyclicBlochTheorem}
We now describe how the underlying symmetry of cyclic structures can be utilized to perform a symmetry cell reduction of the Kohn-Sham DFT problem described in the previous subsection to the fundamental domain. 


\subsubsection{Reduction of the electron density} \label{Subsubsec:Symm:rho}
A natural starting point is to make the assumption that the electron density $\rho$ inherits the symmetry of the cyclic structure. Thus, for (almost) every $\bfx \in \calC$ and for $\gamma \in \Gamma_{\mathfrak{N}}$, we may write:\footnote{As a consequence of Proposition \ref{prop:invariance_to_BC}, Eq.~\ref{rho_symmetry} can be interpreted as cyclic boundary conditions (on the surfaces $\partial\calD_{\calG}^{C}$) applied to the electron density when restricted to the fundamental domain.} 
\begin{align}
\label{rho_symmetry}
\rho(\bfx) = \rho(\mathfrak{R}^{\gamma} \bfx)\,.
\end{align}
This assumption is very similar in nature to that commonly used in crystal lattice calculations, wherein it is assumed that the electron density inherits the translational symmetry of the lattice \citep{Defranceschi_LeBris, rhodes2010crystallography, rohrer2001structure, giustino2014materials}. Analogous to that case, it is conceivable that period doubling (i.e., cyclic group order halving in our case) charge density waves or Pierels instabilities \citep{peierls1955quantum,gruner2000density, kennedy2004proof} can occur in the system, leading to a structural transition and a breakdown of the assumption in Eq.~\ref{rho_symmetry}. In such a case, a larger cyclic unit cell that corresponds to a lower group order needs to be employed. However, the study of such instabilities, including the detection of their onset as well as their physical consequences is beyond the scope of the present contribution and is a compelling topic for future research.\footnote{It is likely however, that the cyclic band structure and cyclic phonon diagrams obtained from Cyclic DFT computations (Section \ref{subsec:silicene_results}) will be instrumental in the prediction of such instabilities.}


\subsubsection{Reduction of the effective potential} \label{Subsubsec:Symm:Veff}
Let us now consider the consequences of the cyclic group invariance of the electron density (i.e., $\rho(\bfx) = \rho(\mathfrak{R}^{\gamma} \bfx)$ for $\gamma \in \Gamma_{\mathfrak{N}}$) on the effective potential. First we observe: 
\begin{proposition} \label{prop:PseudochargeDensity}The following conditions hold for the pseudocharges and the corresponding potentials for a cyclic structure:
\begin{enumerate}
\item For any $\gamma \in \Gamma_{\mathfrak{N}}$, the nuclear pseudocharges $b_k(\cdot)$ and the corresponding pseudopotentials $V_k(\cdot)$ obey the conditions: 
\begin{align}
\label{bk_condition}
b_k(\mathfrak{R}^{\gamma}\bfx, \bfx_{0,k}) & = b_k(\bfx, \bfx_{\mathfrak{N}-\gamma,k})\,,\\
\label{vk_condition}
 V_k(\mathfrak{R}^{\gamma}\bfx, \bfx_{0,k}) &= V_k(\bfx, \bfx_{\mathfrak{N}-\gamma,k})\,.
\end{align}
\item The total nuclear pseudocharge \begin{align} \label{Prop:Invariance_b}
b(\bfx, \calS_{\cal_G, \calP}) = \sum_{\gamma = 0}^{\mathfrak{N}-1} \sum_{k = 1}^{M_{\calP}} b_k(\bfx, \bfx_{\gamma,k}) = \sum_{\gamma = 0}^{\mathfrak{N}-1}\sum_{k = 1}^{M_{\calP}} \eta_k(\abs{\bfx - \bfx_{\gamma,k}})
\end{align}
is invariant under $\calG$.
\end{enumerate}
\end{proposition}
\begin{myproof}\hfill
\begin{enumerate}
\item The radial symmetry of $b_k(\cdot)$ implies that $b_k(\mathfrak{R}^{\gamma}\bfx, \bfx_{0,k}) = b_k(\bfx, \mathfrak{R}^{\mathfrak{N}-\gamma}\bfx_{0,k})$, from which Eq.~\ref{bk_condition} follows. Next, using the change of variables $\mathfrak{R}^{\mathfrak{N}-\gamma}\bfy = \bfz$, we get:
\begin{align}
\nonumber
V_k(\mathfrak{R}^{\gamma}\bfx, \bfx_{0,k}) &= \int_{\calC}\!\frac{b_k(\bfy, \bfx_{0,k})}{\abs{\mathfrak{R}^{\gamma}\bfx-\bfy}}\; \mathrm{d\bfy} = \int_{\calC}\!\frac{b_k(\bfy, \bfx_{0,k})}{\abs{\mathfrak{R}^{\gamma}\bfx-\mathfrak{R}^{\gamma}\mathfrak{R}^{\mathfrak{N}-\gamma}\bfy}}\; \mathrm{d\bfy} \\
&=\int_{\calC}\!\frac{b_k(\mathfrak{R}^{\gamma}\bfz, \bfx_{0,k})}{\abs{\bfx-\bfz}}\;\mathrm{d\bfz}\, = \int_{\calC}\!\frac{b_k(\bfz, \mathfrak{R}^{\mathfrak{N}-\gamma}\bfx_{0,k})}{\abs{\bfx-\bfz}}\; \mathrm{d\bfz}
= V_k(\bfx, \mathfrak{R}^{\mathfrak{N}-\gamma}\bfx_{0,k})\,,\label{Vk_calculation}
\end{align}
from which Eq.~\ref{vk_condition} follows.
\item We note that for any $\gamma,\gamma_1 \in \Gamma_{\mathfrak{N}}$, the rotation matrix $\mathfrak{R}^{\mathfrak{N}-\gamma}$ acting on the point $\bfx_{\gamma_1,k}= \mathfrak{R}^{\gamma_1}\bfx_{0,k}$ produces the point $\bfx_{\gamma_2,k} = \mathfrak{R}^{\gamma_2}\bfx_{0,k}$, with $\gamma_2 = (\mathfrak{N}-\gamma+\gamma_1)\,\text{mod}\,\mathfrak{N}$. In this situation, if $\gamma$ is held fixed while $\gamma_1$ is allowed to cycle through the elements of $\Gamma_{\mathfrak{N}}$, $\gamma_2$ also cycles through each element of $\Gamma_{\mathfrak{N}}$, thus yielding the group orbit of $\bfx_{0,k}$ under $\calG$. Thus, we may write for any $\gamma \in \Gamma_{\mathfrak{N}}$:
\begin{align}
\nonumber
b(\mathfrak{R}^{\gamma}\bfx, \calS_{\cal_G, \calP}) &= \sum_{\gamma_1 = 0}^{\mathfrak{N}-1}\sum_{k = 1}^{M_{\calP}} \eta_k(\abs{\mathfrak{R}^{\gamma}\bfx - \mathfrak{R}^{\gamma_1}\bfx_{0,k}}) \\\nonumber
&= \sum_{\gamma_1 = 0}^{\mathfrak{N}-1}\sum_{k = 1}^{M_{\calP}} \eta_k(\abs{\mathfrak{R}^{\gamma}\bfx - \mathfrak{R}^{\gamma}\,\mathfrak{R}^{\mathfrak{N} - \gamma}\,\mathfrak{R}^{\gamma_1}\bfx_{0,k}})\\\nonumber
&= \sum_{\gamma_1 = 0}^{\mathfrak{N}-1}\sum_{k = 1}^{M_{\calP}} \eta_k(\abs{\bfx - \mathfrak{R}^{\mathfrak{N} - \gamma}\,\mathfrak{R}^{\gamma_1}\bfx_{0,k}})\\\nonumber
&= \sum_{\gamma_2 = 0}^{\mathfrak{N}-1}\sum_{k = 1}^{M_{\calP}} \eta_k(\abs{\bfx - \mathfrak{R}^{\gamma_2}\bfx_{0,k}})\;, \textrm{with}\; \gamma_2 = (\mathfrak{N}-\gamma+\gamma_1)\,\text{mod}\,\mathfrak{N}\\\label{b_invariant}
&= b(\bfx, \calS_{\cal_G, \calP})\,,
\end{align}
where we have used the spherical symmetry of $\eta_k$ to arrive at the third equality. 
\end{enumerate}
\end{myproof}
Next, we use the above result to show:
\begin{proposition}
\label{prop:invariant_potential} The total effective potential 
\begin{align}
\nonumber
\displaystyle V_{\emph{\textrm{eff}}}(\bfx) = V_{\emph{\textrm{es}}}(\rho(\bfx), \calS_{\cal_G, \calP}) + V_{\emph{\textrm{xc}}}(\rho(\bfx))
\end{align}
is invariant under $\calG$.
\end{proposition}
\begin{myproof}
It suffices to show that both the terms that constitute $V_{\textrm{eff}}(\bfx)$ are individually group invariant. The exchange-correlation potential $V_{{\textrm{xc}}}(\rho(\bfx))$ is invariant by virtue of the invariance of $\rho(\bfx)$. To show the invariance of the electrostatic potential, we recall that it can be written as
\begin{align}
V_{\text{es}}(\rho(\bfx),\calS_{\cal_G, \calP}) &= \int_{\calC}\!\frac{\rho(\bfy) + b(\bfy, \calS_{\cal_G, \calP})}{\abs{\bfx-\bfy}}\; \mathrm{d\bfy} \,.
\end{align}
Therefore, for any $\gamma \in \Gamma_{\mathfrak{N}}$, we may write
\begin{align}
V_{\text{es}}(\rho(\mathfrak{R}^{\gamma} \bfx),\calS_{\cal_G, \calP}) &= \int_{\calC}\!\frac{\rho(\bfy)+b(\bfy, \calS_{\cal_G, \calP})}{\abs{\mathfrak{R}^{\gamma}\bfx-\bfy}}\; \mathrm{d\bfy}  \nonumber \\
&= \int_{\calC}\!\frac{\rho(\mathfrak{R}^{\gamma} \bfz) + b(\mathfrak{R}^{\gamma} \bfz, \calS_{\cal_G, \calP})}{\abs{\mathfrak{R}^{\gamma}\bfx-\mathfrak{R}^{\gamma} \bfz}}\; \mathrm{d\bfz}  \nonumber \\
&= \int_{\calC}\!\frac{\rho(\bfz)+b(\bfz, \calS_{\cal_G, \calP})}{\abs{\bfx-\bfz}}\; \mathrm{d\bfz} \nonumber \\
&= V_{\text{es}}(\rho(\bfx),\calS_{\cal_G, \calP}) \,,
\end{align}
where we have used the change of variables $\bfy = \mathfrak{R}^{\gamma}\bfz$ and the invariance of the Lebesgue measure under rotations to arrive at the second equality. Further, we have used the invariance of $\rho$ (Eq.~\ref{rho_symmetry}) and $b$ (Eq.~\ref{Prop:Invariance_b}) to arrive at the third equality. 
\end{myproof}

As a consequence of the above results --- specifically, the cyclic group invariance of $b(\bfx, \calS_{\cal_G, \calP})$ and $V_{\text{es}}(\rho(\bfx),\calS_{\cal_G, \calP})$ --- the Poisson problem in Eqs.~\ref{poisson_problem} and \ref{Eqn:KS:BC:Ves} can be reduced to the fundamental domain, with cyclic boundary conditions on the surfaces $\partial\calD_{\calG}^{C}$.


\subsubsection{Reduction of the linear eigenvalue problem} \label{Subsubsec:EigProblem}
We recall that in the every iteration of the SCF method, it is necessary to compute part of the spectrum associated with the linearized Kohn-Sham operator: 
\begin{align}
\hamil = -\half \Delta + V_\textrm{eff}(\bfx) \,,
\end{align}
where the effective potential $V_\textrm{eff}(\bfx)$ has been evaluated from the electron density calculated during the previous SCF iteration. Assuming that $V_\textrm{eff}(\bfx)$ is a continuous function,\footnote{This is indeed true if $\rho$ is continuous \citep{My_PhD_Thesis}, as is evident from Eq. \ref{effective_potential}.} it can be shown that the operator $\hamil$ on $\mathsf{L}^2(\calC)$, with domain $\textrm{Dom}(\hamil) = \mathsf{H}^2(\calC) \cap \mathsf{H}^1_0(\calC)$, is an elliptic self-adjoint operator with a compact resolvent \citep{My_PhD_Thesis, Evans_PDE, Renardy_Rogers, Kato}. Therefore, $\hamil$ has an increasing sequence of real eigenvalues, with the associated eigenfunctions\footnote{Although a finite cylinder has edges on its boundary, the eigenfunctions are expected to be at least $\mathsf{H}^2(\calC)$ regular since the domain is convex \citep{grisvard2011elliptic}.} forming an orthonormal basis of $\mathsf{L}^2(\calC)$.

Let us assume that the electron density calculated during the previous SCF iteration is group invariant.\footnote{This is indeed the case in Cyclic DFT, as shown in Section \ref{Subsubsec:ConsistencySCF}.} It follows from Proposition \ref{prop:invariant_potential} that the effective potential $V_\textrm{eff}(\bfx)$ in the current SCF iteration is also group invariant. This leads us to the following observation:

\begin{proposition}
\label{prop:invariant_operator}
With \emph{$V_\textrm{eff}(\bfx)$} continuous over $\calC$ and group invariant, the operator $\hamil$ commutes with the action of the group $\calG$ on functions in $\mathsf{L}^2(\calC)$, i.e., for any $\gamma \in \Gamma_{\mathfrak{N}}$, if \emph{$T_\gamma : \mathsf{L}^2(\calC) \to \mathsf{L}^2(\calC)$} is an operator such that \emph{$T_\gamma f(\bfx) \mapsto f\big((\mathfrak{R}^{\gamma})^{-1}\bfx\big)$}, then for any function \emph{$f \in \textrm{Dom}(\hamil)$}, we have \emph{$\hamil T_\gamma f = T_\gamma \hamil f$}.
\end{proposition}
\begin{myproof}
Since the function space $\textrm{C}^2_c(\calC)$ is dense in $\textrm{Dom}(\hamil)$, it suffices to verify this result for $f \in \textrm{C}^2_c(\calC)$. Using a change of variables calculation, it can be shown that the Laplacian operator commutes with rotations, and therefore with the operators $T_\gamma$. This combined with the group invariance of $V_\textrm{eff}(\bfx)$ establishes the required result. 
\end{myproof}

It can be shown  that the collection of operators $\{ T_\gamma: \gamma \in \Gamma_{\mathfrak{N}}\}$ form a faithful unitary representation\footnote{Each operator $T_\gamma$ is unitary and together the collection $\{ T_\gamma: \gamma \in \Gamma_{\mathfrak{N}}\}$ forms an identity preserving homomorphism of $\calG$ on the carrier space $\mathsf{L}^2(\calC)$. Since the representation is faithful, the map $\calG \ni \mathfrak{R}^{\gamma} \mapsto T_\gamma$ is in fact an isomorphism.} of the group $\calG$ on the space $\mathsf{L}^2(\calC)$ \citep{My_PhD_Thesis}. Proposition \ref{prop:invariant_operator} shows that the operator $\hamil$ is left invariant by these representations, i.e., we may write $T_\gamma \hamil (T_\gamma)^{-1} = \hamil$. In order to exploit this invariance property for further analysis, it becomes natural to use tools from the representation theory of finite groups \citep{My_PhD_Thesis, Folland_Harmonic, Barut_Reps}. We proceed to do so by first noting the following result (see Theorem 2.3.18 of \citep{My_PhD_Thesis}, as well as \citep{Bossavit_Old, Bossavit_New}) stated here without proof:
\begin{lemma}
For $\nu, \gamma \in \Gamma_{\mathfrak{N}}$, let \emph{$\chi_{\nu}(\gamma) = \expchar{\nu \gamma}{\mathfrak{N}}$} and let
\begin{align}
\label{peter_weyl_projectors}
P^{\nu} = \displaystyle \frac{1}{\mathfrak{N}}\sum_{\gamma = 0}^{\mathfrak{N} - 1}\overline{\chi_{\nu}(\gamma)}\,T_{\gamma} = \displaystyle \frac{1}{\mathfrak{N}}\sum_{\gamma = 0}^{\mathfrak{N} - 1}\expcharconj{\nu \gamma}{\mathfrak{N}}\,T_{\gamma}.
\end{align}
The operators $P^{\nu}$ are projection operators on $\mathsf{L}^2(\calC)$, and the ranges of these projectors $V^{\nu} = P^{\nu}\big(\mathsf{L}^2(\calC)\big)$ form closed, mutually orthogonal subspaces of  $\mathsf{L}^2(\calC)$ for different values of $\nu$. The following direct sum decomposition holds:
\begin{align}
\label{direct_sum_subspaces}
\mathsf{L}^2(\calC) = \bigoplus_{\nu = 0}^{\mathfrak{N}-1} V^{\nu}\,,
\end{align}
and further $f \in V^{\nu}$ if and only if it obeys the condition:
\begin{align}
\label{Bloch_cyclic}
T_{\gamma}f (\bfx) = f\big((\mathfrak{R}^{\gamma})^{-1}\bfx\big) = \chi_\nu(\gamma)f(\bfx)\,.
\end{align}
\label{lemma:rep_theory_result}
\end{lemma}

The complex scalars $\displaystyle \chi_{\nu}(\gamma) $ are the so called characters of the group $\calG$, and the projectors $P^{\nu}$ are the associated Peter-Weyl projectors \citep{Folland_Harmonic}. The range $V^{\nu}$ of each projector  forms an invariant subspace of $\mathsf{L}^2(\calC)$ . A useful corollary of the above result follows immediately by looking at the case $\nu = 0$:
\begin{corollary}
For any given scalar function $f$ on $\calC$, the function $\displaystyle\sum_{\gamma = 0}^{\mathfrak{N}-1} f(\mathfrak{R}^{\gamma} \bfx)$ is group invariant. 
\label{coro:group_invariance}
\end{corollary}
\noindent The condition $T_\gamma \hamil (T_\gamma)^{-1} = \hamil$ implies that the subspaces $V^{\nu}$ are invariant subspaces for the operator $\hamil$ as well. This allows us to \emph{block-diagonalize} $\hamil$ so that we may write:
\begin{align}
\hamil = \bigoplus_{\nu = 0}^{\mathfrak{N}-1} \hamil^{\nu}\,,
\label{block_diagonal_hamil}
\end{align}
with each $\hamil^{\nu} = \hamil P^{\nu}$ denoting the projection of $\hamil$ onto the invariant subspace $V^{\nu}$. As a result, the eigenstates  of $\hamil$ may be computed by solving each of the projected problems associated with $\hamil^{\nu}, \nu \in \Gamma_{\mathfrak{N}}$. Consequently, the eigenfunctions of $\hamil$ obey the cyclic-Bloch condition shown in Eq.~\ref{Bloch_cyclic}. We may summarize this discussion and rigorously establish the result as follows:
\begin{theorem}[Cyclic Bloch Theorem]
Given a potential \emph{$V_\textrm{eff}(\bfx)$} which is continuous on $\calC$ and invariant under the cyclic group $\calG$, there exists a basis of $\mathsf{L}^2(\calC)$ such that every eigenstate $v$ of the operator \emph{$\hamil = -\half \Delta + V_\textrm{eff}$} obeys the following condition (for any $\gamma \in \Gamma_{\mathfrak{N}}$):
\begin{align}
\label{Bloch_cyclic_in_theorem}
v\big((\mathfrak{R}^{\gamma})^{-1}\bfx\big) = \chi_\nu(\gamma)v(\bfx) = \expchar{\nu \gamma}{\mathfrak{N}} v(\bfx)\,,
\end{align}
for some $\nu \in \Gamma_{\mathfrak{N}}$. Conversely, for every $\nu \in \Gamma_{\mathfrak{N}}$, there exists at least one eigenvector of $\hamil$ obeying the above condition.
\label{thm:Bloch_Theorem}
\end{theorem}
\begin{myproof}
Since $\hamil$ commutes with each of the operators in $\{ T_\gamma: \gamma \in \Gamma_{\mathfrak{N}}\}$, and since the projectors $P^{\nu}$ introduced in Eq.~\ref{peter_weyl_projectors} are linear combinations of the operators in $\{ T_\gamma: \gamma \in \Gamma_{\mathfrak{N}}\}$, $\hamil$ also commutes with each of the projectors. This means that $\hamil$ leaves the range $V^{\nu}$ of each of the projectors invariant. This is because if $f \in V^{\nu}$, then $P^{\nu} f = f$ and further:
\begin{align}
P^{\nu} (\hamil f) =  \hamil (P^{\nu} f) = \hamil f\,,
\end{align}
which implies that $\hamil f \in V^{\nu}$. Note that this also implies that $\hamil$ leaves $(V^{\nu})^{\perp}$ --- the orthogonal complement of $V^{\nu}$ --- invariant as well. This can be easily checked by looking at the projection operator $\calI - P^{\nu}$ (with $\calI$ denoting the identity operator on $\mathsf{L}^2(\calC)$), which projects onto the orthogonal complement of $V^{\nu}$. 

The above conditions imply that $\hamil$ must have at least one eigenvector $v \in V^{\nu}$ for every $\nu \in \Gamma_{\mathfrak{N}}$. To establish this (for any $\nu \in \Gamma_{\mathfrak{N}}$), we first recall that the eigenvectors of $\hamil$ form a complete orthonormal basis set in $\mathsf{L}^2(\calC)$. Since $V^{\nu}$ is a closed subspace of $\mathsf{L}^2(\calC)$, there exists an eigenvector $v$ of $\hamil$ which is not completely orthogonal to $V^{\nu}$. We write this eigenvector as $v = v_1 + v_2$ with $v_1 \in V^{\nu}, v_1 \neq 0$ and $v_2 \in (V^{\nu})^{\perp}$ . Subsequently, we have:
\begin{align}
\label{eigenvector_calc}
\hamil v = \hamil (v_1 + v_2) = \lambda v = \lambda(v_1 + v_2)\,,
\end{align}
which implies that $\hamil v_1 -  \lambda v_1 = \hamil v_2 -  \lambda v_2$. Since $\hamil$ leaves both $V^{\nu}$ and its orthogonal complement invariant, the left hand side of this expression lies in $V^{\nu}$ while the right hand side lies in $(V^{\nu})^{\perp}$. Since $V^{\nu} \cap (V^{\nu})^{\perp} = \{0\}$, it must be that $\hamil v_1 =  \lambda v_1$ and $\hamil v_2 =  \lambda v_2$ and so, (possibly after an appropriate choice of basis of $\mathsf{L}^2(\calC)$) $v_1 \in V^{\nu}$ and $v_2 \in  (V^{\nu})^{\perp}$ are individual eigenvectors of $\hamil$, as required.\footnote{We may now repeat this argument for the orthogonal complement of $v_1$ within $V^{\nu}$ --- since this subspace is also left invariant by $\hamil$, an eigenvector of $\hamil$ can be found in it.  Proceeding in this fashion (i.e., considering the orthogonal complement within $V^{\nu}$ of the linear span of the eigenvectors found so far), we can in fact show the existence of infinitely many eigenvectors of $\hamil$ in $V^{\nu}$ as long as the subspace $V^{\nu}$ is infinite-dimensional.} 

Next, we observe that by suitable choice of an orthonormal basis of $\mathsf{L}^2(\calC)$, every eigenvector of $\hamil$ can be made to lie in one of the subspaces $V^{\nu}$. To see this, we note that since the decomposition of $\mathsf{L}^2(\calC)$ into the invariant subspaces $V^{\nu}$ is exhaustive (Eq.~\ref{direct_sum_subspaces}), every eigenvector of $\hamil$ either lies in one of the subspaces $V^{\nu}$ or it can be written as a linear combination of eigenvectors from these subspaces. If the eigenstates of $\hamil$ are non-degenerate (i.e., there is no repeated eigenvalue), the second option is ruled out as follows: writing the eigenvector as $ v = \displaystyle \sum_{\nu = 0}^{\mathfrak{N} - 1} c_{\nu} v_{\nu}$, with $v_{\nu} \in V^{\nu}, c_{\nu} \in \cz$, a calculation similar to the one in the previous paragraph shows that each individual component $v_{\nu}$ must be an eigenvector as well (since $V^{\nu} \cap V^{\nu'} = \{0\}$ for $\nu \neq \nu'$). However, this would violate the orthogonality of the eigenvectors of $\hamil$ since $\innprod{v}{v_{\nu}}{} \neq 0$ unless $c_{\nu} = 0$. In the presence of degeneracies,\footnote{While it might be tempting to simplify the statement and proof of Theorem \ref{thm:Bloch_Theorem} by assuming lack of any degeneracies, examples appear to indicate that degeneracies in the spectrum of $\hamil$ might generically appear (e.g., see \citep{My_Elliott_Symmetry_paper}). Thus, we find it useful to add the caveat that the cyclic-Bloch boundary conditions  (Eq.~\ref{Bloch_cyclic_in_theorem}) can be expected to hold for any eigenstate of $\hamil$ provided an appropriate choice of basis vectors of $\mathsf{L}^2(\calC)$ has been made. We are grateful to the anonymous reviewer for pointing out this subtle issue to us.\\} any linear combination of the degenerate eigenstates from different invariant subspaces is still an eigenstate for $\hamil$. In this case, we may choose the basis vectors of $\mathsf{L}^2(\calC)$ appropriately so that the individual eigenstates from the different invariant subspaces (and not their linear combinations) are identified as the eigenvectors of $\hamil$.

Finally, since every eigenvector $v$ of $\hamil$ lies in one of the invariant subspaces $V^{\nu}$, the condition in Eq.~\ref{Bloch_cyclic_in_theorem} can be simply identified as the  characterization of the subspaces presented in Eq.~\ref{Bloch_cyclic}, thus establishing the theorem.\footnote{The proof presented here is slightly different in technical details from that presented in \citep{My_MS_Thesis, My_PhD_Thesis}. In the latter case, a more direct use of Schur's Lemma \citep{Folland_Harmonic}, as applied to the resolvent operator of the Hamiltonian was made.}
\end{myproof}

We may use the Cyclic Bloch theorem to reduce the physical domain of the eigenvalue problem associated with $\hamil$ from $\calC$ to the symmetry cell $\calD_{\calG}$. Roughly speaking, since every point in $\calC$ can be obtained as the action of a rotation matrix $\mathfrak{R}^{\gamma}$ (for some $\gamma \in \Gamma_{\mathfrak{N}}$) acting on a point in the fundamental domain $\calD_{\calG}$, it suffices to specify a function $f \in V^{\nu}$ (in particular, an eigenvector $v\in  V^{\nu}$ of $\hamil$) using only the points within the fundamental domain, as long as boundary conditions consistent with Eq.~\ref{Bloch_cyclic} are prescribed on $\partial\calD_{\calG}^{C}$. We summarize this result as follows:\footnote{It becomes necessary to work with functions in $\mathsf{H}^1_0(\calC)$ so that boundary conditions on $\partial\calD_{\calG}$ can be interpreted in the trace sense. The eigenvectors of $\hamil$ are actually more regular than functions in $\mathsf{H}^1_0(\calC)$.} 
\begin{proposition}
\label{prop:invariance_to_BC}
Let $\bfx \in \partial\calD_{\calG}^{C}$ such that $\mathfrak{R}^{\gamma}\bfx \in \partial\calD_{\calG}^{C}$ for some $\gamma \in \Gamma_{\mathfrak{N}}$ and \footnote{For the cylindrical geometries that we are considering, $\gamma \in \{0,1,\mathfrak{N}-1\}$. } let \emph{$\text{cl.}(\calD_{\calG})$} denote the (topological) closure of the fundamental domain. A function $f \in \big(V^{\nu} \cap \mathsf{H}^1_0(\calC) \big)$ if and only if the restriction of the function to \emph{$\text{cl.}(\calD_{\calG})$}, denoted by \emph{$\tilde{f} = f|_{\text{cl.}(\calD_{\calG})}$}, satisfies
\begin{align}
\label{Bloch_cyclic_restatement}
\tilde{f}(\mathfrak{R}^{\gamma}\bfx) = \overline{\chi_{\nu}(\gamma)} \, \tilde{f}(\bfx) = \expcharconj{\nu \gamma}{\mathfrak{N}}\,\tilde{f}(\bfx)\,\,\,\, \text{for} \,\,\,\,\bfx \in \partial\calD_{\calG}^{C}\,.
\end{align}
\end{proposition}
\begin{myproof}
Since the function space $\textrm{C}_c^0(\calC)$ is dense in $\mathsf{H}^1_0(\calC)$, it suffices to work with such functions. If $f \in V^{\nu} \cap \textrm{C}_c^0(\calC)$, then it obeys Eq.~\ref{Bloch_cyclic} for all $\bfx\in \calC$ and in particular on the surfaces $\partial\calD_{\calG}^{C} \subset \calC$. On the other hand, if $\tilde{f}$ is a continuous function defined over the fundamental domain such that it obeys Dirichlet boundary conditions on $\partial\calD_{\calG}^{0}$ and the condition in Eq.~\ref{Bloch_cyclic_restatement} on $\partial\calD_{\calG}^{C}$ for some $\gamma, \nu \in \Gamma_{\mathfrak{N}}$, we may define an extension $f$ of $\tilde{f}$ to all of $\calC$ as follows. For any $\bfy \in \calC$ (such that $\bfy$ does not lie on the axis $\bfe_3$) we may identify a point $\bfx \in \calD_{\calG}$ and a rotation matrix $\mathfrak{R}^{\gamma_2}\in \calG$ such that $\bfy = \mathfrak{R}^{\gamma_2} \bfx$. Then, we set
\begin{align}
\label{how_to_define_f}
f(\bfy) = f(\mathfrak{R}^{\gamma_2} \bfx) = \expcharconj{\nu \gamma_2}{\mathfrak{N}}\,\tilde{f}(\bfx)\,,
\end{align}
and $f = 0$ on $\partial \calC$. The function $f$ defined this way is compactly supported on $\calC$ and is continuous across each of the surfaces $\mathfrak{R}^{\gamma_1}(\partial\calD_{\calG})$ for any $\gamma_1 \in \Gamma_{\mathfrak{N}}$, by virtue of Eq.~\ref{Bloch_cyclic_restatement}. By the continuity of $\tilde{f}$ in the interior of $\calD_{\calG}$, $f$ is also continuous in the interior of $\mathfrak{R}^{\gamma_1}(\calD_{\calG})$. Thus, $f \in \textrm{C}_c^0(\calC)$. Further, Eq.~\ref{how_to_define_f} implies that $f$ obeys Eq.~\ref{Bloch_cyclic} and therefore by Lemma \ref{lemma:rep_theory_result}, $f \in V^{\nu}$. This establishes the sought result.
\end{myproof}

As a consequence of the above result, the Cyclic Bloch Theorem may be reinterpreted to mean that the eigenstates of $\hamil$ can be obtained by solving $\mathfrak{N}$ independent eigenvalue problems associated with the operators $\hamil^{\nu}$ ($\nu \in \Gamma_{\mathfrak{N}}$) over the fundamental domain. Each problem corresponds to a projection of the original problem to one of the subspaces $V^{\nu}$, and therefore it obeys zero-Dirichlet and cyclic-Bloch (Eq.~\ref{Bloch_cyclic_restatement}) boundary conditions on the surfaces $\partial\calD_{\calG}^{0}$ and $\partial \calD_{\calG}^{C}$, respectively. Generically, each of these problems results in an infinite sequence of orthonormal eigenstates.

It is worth noting that Theorem \ref{thm:Bloch_Theorem} represents an extension of the Bloch theorem used in classical solid state physics \citep{Ashcroft_Mermin, Reed_Simon4, Odeh_Keller} to cyclic structures. A further generalization of this result to the case of structures with more complex symmetries (or Objective Structures) can be found in the first author's thesis work \citep{My_PhD_Thesis, My_MS_Thesis}.  Theorem \ref{thm:Bloch_Theorem} appears in a somewhat more rudimentary form with certain key details omitted in \cite{Sattlegger_Thesis}. The gist of the result also appears in the physics literature \citep{Pekka_1,Pekka_2} --- however, only in the context of tight binding and not Kohn-Sham DFT --- wherein the derivation follows a line of heuristic reasoning (aimed at guessing the correct extension of the classical Bloch Theorem), and therefore lacks the mathematical rigor adopted here.


\subsubsection{Consistency of the SCF method after symmetry cell reduction} \label{Subsubsec:ConsistencySCF}
While performing the symmetry cell reduction of the linear eigenvalue problem, we have assumed that the electron density calculated during the previous SCF iteration is group invariant. It is therefore important to ensure that the electron density remains group invariant throughout the complete fixed-point iteration. To do so, we note that as a consequence of the reduction achieved by the Cyclic Bloch Theorem, the electron density in any SCF iteration can be calculated using the expression:\footnote{Henceforth, a tilde will be used to denote that the quantity has been defined over the fundamental domain $\calD_{\calG}$. }
\begin{align} \label{Eqn:rho:Bloch-cyclic}
\tilde{\rho}(\bfx)&=2 \sum_{\nu = 0}^{\mathfrak{N} - 1}\sum_{i=1}^{N_{\text{s}}} {f}_i^{\nu}\abs{\tilde{\phi}_i^{\nu}(\bfx)}^2 \,,
\end{align}
where $\tilde{\phi}_i^{\nu}$ are the eigenfunctions of $\hamil^{\nu}$, and ${f}_i^{\nu}$ are the corresponding occupations obtained using the Fermi-Dirac smearing.  

An inspection of Eq.~\ref{Eqn:rho:Bloch-cyclic} reveals that the electron density calculated in any SCF iteration satisfies (for any $\gamma \in \Gamma_{\mathfrak{N}}$):
\begin{align}
\tilde{\rho}(\mathfrak{R}^{\gamma}\bfx) =2 \sum_{\nu = 0}^{\mathfrak{N} - 1}\sum_{i=1}^{N_{\text{s}}} {f}_i^{\nu}\abs{\tilde{\phi}_i^{\nu}(\mathfrak{R}^{\gamma}\bfx)}^2
=2 \sum_{i=1}^{N_{\text{s}}} {f}_i^{\nu}\abs{\expcharconj{\nu \gamma}{\mathfrak{N}}\,\tilde{\phi}_i^{\nu}(\bfx)}^2 = 2 \sum_{i=1}^{N_{\text{s}}} {f}_i^{\nu}\abs{\tilde{\phi}_i^{\nu}(\bfx)}^2 = \tilde{\rho}(\bfx)\,.
\label{density_invariant_calc}
\end{align}
Thus, the extension of this density to the entire domain $\calC$ is continuous and is group invariant. Consequently, it follows from Proposition \ref{prop:invariant_potential} and Theorem \ref{thm:Bloch_Theorem} that the Kohn-Sham Hamiltonian resulting from this density can once again be reduced to the fundamental domain with the aid of the cyclic-Bloch boundary conditions. This ensures that SCF iterations remain consistent provided one starts performing these iterations with an electron density that is group invariant. In principle, such an electron density may be generated by considering any $\rho' \in \textrm{C}_c^0(\calC)$ obeying $\displaystyle \int_{\calC}\rho'(\bfx) \, \mathrm{d\bfx}= \mathfrak{N}N_{\text{e}}$ and then constructing 
\begin{align}
\rho_{0}(\bfx)= \frac{1}{\mathfrak{N}}\sum_{\gamma = 0}^{\mathfrak{N}-1}\rho'(\mathfrak{R}^{\gamma}\bfx)\,.
\end{align}
It follows from Corollary \ref{coro:group_invariance} that the guess electron density $\rho_{0}$ is group invariant, and therefore its restriction to the fundamental domain serves as a suitable starting point. 
\subsubsection{Reduction of the free energy and the atomic forces}
Having performed a symmetry cell reduction of the self-consistent Kohn-Sham equations, we now express the cyclic structure's free energy at the electronic ground-state, as well as the Hellmann Feynman forces on the atoms, in terms of quantities defined over the fundamental domain. 

To simplify the free-energy, we use the cyclic-Bloch reduction to rewrite the first (band energy) and last (electronic entropy energy) terms of Eq. \ref{Harris_Foulkes_Energy} as follows:
\begin{align}
{E}_{\text{band}} &= 2\sum_{\nu =0}^{\mathfrak{N}-1}\sum_{i = 1}^{N_{\text{s}}} f_i^{\nu} \lambda_i^{\nu} \,, \\
{E}_{\text{entropy}} &= 2\,\sigma \sum_{\nu=0}^{\mathfrak{N}-1}\sum_{i=1}^{N_{s}} \big( f_i^{\nu} \log f_i^{\nu} + (1-f_i^{\nu}) \log (1 - f_i^{\nu}) \big) \,,
\end{align}
where $\lambda_i^{\nu}$ are the eigenvalues of $\hamil^{\nu}$ at self-consistency. In addition, ${f}_i^{\nu}$ are the corresponding occupations obtained using the Fermi-Dirac smearing. The second, third, and fourth terms of Eq.~\ref{Harris_Foulkes_Energy} are integrals over the complete domain $\calC$ with integrands that are group invariant. Consequently, they can be reduced to the fundamental domain as follows:
\begin{align}
\int_{\calC}\!\varepsilon_{\text{xc}} (\rho(\bfx)) \rho(\bfx) \, \mathrm{d \bfx} &= \mathfrak{N} \int_{\calD_{\calG}}\!\varepsilon_{\text{xc}} (\tilde{\rho}(\bfx)) \tilde{\rho}(\bfx) \, \mathrm{d \bfx} \,, \\
\int_{\calC}\!V_{\text{xc}}(\rho(\bfx))\rho(\bfx)\,\mathrm{d\bfx} &= \mathfrak{N} \int_{\calD_{\calG}}\!V_{\text{xc}}(\tilde{\rho}(\bfx))\tilde{\rho}(\bfx)\,\mathrm{d\bfx} \,, \\
\frac{1}{2} \int_{\calC}\!\big (b(\bfx, \calS_{\cal_G, \calP})-\rho(\bfx)\big) V_{\text{es}}(\bfx, \calS_{\cal_G, \calP}) \, \mathrm{d\bfx} &= \frac{\mathfrak{N}}{2} \int_{\calD_{\calG}}\!\big (\tilde{b}(\bfx, \calP)-\tilde{\rho}(\bfx)\big) \tilde{V}_{\text{es}}(\bfx, \calP) \, \mathrm{d\bfx} \,.
\end{align}
To see how the fifth term in Eq.~\ref{Harris_Foulkes_Energy}, i.e., pseudocharge self energy term:
\begin{align}
{E}_{\text{self}}(\calS_{\cal_G, \calP}) = \half \sum_{\gamma = 0}^{\mathfrak{N}-1} \sum_{k = 1}^{M_{\calP}} \int_{\calC}\!b_k(\bfx, \bfx_{\gamma,k}) \, V_k(\bfx, \bfx_{\gamma,k}) \,\mathrm{d\bfx}\,,
\end{align}
can be reduced to the fundamental domain, we recall Proposition \ref{prop:PseudochargeDensity} (part 1), to write:
\begin{align}
\nonumber
{E}_{\text{self}}(\calS_{\cal_G, \calP}) &= \half \sum_{\gamma = 0}^{\mathfrak{N}-1} \sum_{k = 1}^{M_{\calP}} \int_{\calC}\!b_k(\bfx, \mathfrak{R}^{\gamma}\bfx_{0,k})\, V_k(\bfx, \mathfrak{R}^{\gamma}\bfx_{0,k})\, \mathrm{d\bfx} \\\nonumber
&= \half \sum_{\gamma = 0}^{\mathfrak{N}-1} \sum_{k = 1}^{M_{\calP}} \int_{\calC}\!b_k(\mathfrak{R}^{\mathfrak{N}-\gamma}\bfx, \bfx_{0,k})\,V_k(\mathfrak{R}^{\mathfrak{N}-\gamma}\bfx, \bfx_{0,k})\, \mathrm{d\bfx}  \\ \nonumber
&= \half \sum_{k = 1}^{M_{\calP}} \int_{\calC}\,\sum_{\gamma = 0}^{\mathfrak{N}-1} b_k(\mathfrak{R}^{\mathfrak{N}-\gamma}\bfx, \bfx_{0,k})\, V_k(\mathfrak{R}^{\mathfrak{N}-\gamma}\bfx, \bfx_{0,k})\, \mathrm{d\bfx} \\ \nonumber
&= \frac{\mathfrak{N}}{2}  \sum_{\gamma = 0}^{\mathfrak{N}-1} \sum_{k = 1}^{M_{\calP}} \int_{\calD_{\calG}}\!b_k(\mathfrak{R}^{\gamma}\bfx, \bfx_{0,k})\, V_k(\mathfrak{R}^{\gamma}\bfx, \bfx_{0,k})\, \mathrm{d\bfx} \\
&= \frac{\mathfrak{N}}{2}  \sum_{\gamma = 0}^{\mathfrak{N}-1} \sum_{k = 1}^{M_{\calP}} \int_{\calD_{\calG}}\!\tilde{b}_k(\bfx, \bfx_{\gamma,k})\,\widetilde{V}_k(\bfx, \bfx_{\gamma,k})\, \mathrm{d\bfx}\,,
\end{align}
where the fourth equality is obtained by observing that each of the integrands in the previous expression are group invariant via Corollary \ref{coro:group_invariance}. Note that, $\tilde{b}_k$ and $\widetilde{V}_k$ are the restriction of $b_k$ and $V_k$ to the fundamental domain.  

Next, we consider the reduction of the atomic forces. As a consequence of the frame indifference of the free energy $\mathcal{F}({\calS_{\cal_G, \calP}})$, the Hellmann-Feynman force on the atom located at $\bfx_{\gamma,k} = \mathfrak{R}^{\gamma} \bfx_{0,k}$ is related to the force experienced by the corresponding atom ${\bfx_{0,k}} \in \calP$ within the fundamental domain $\calD_{\calG}$ through the relation:\footnote{This result is not restricted only to the case of Kohn-Sham DFT however -- it holds for more general theories such as the many body Schr\"odinger equation in Born-Oppenheimer quantum mechanics \cite{Dumitrica_James_OMD}.}
\begin{align}
\label{force_equivariance}
\bff_{\bfx_{\gamma,k}} = (\mathfrak{R}^{\gamma})^T\,\bff_{\bfx_{0,k}} = (\mathfrak{R}^{\mathfrak{N}-\gamma})\,\bff_{\bfx_{0,k}} \,,
\end{align}
where
\begin{align}
\label{force_formula_revisited}
\bff_{\bfx_{0,k}} = \int_{\calC}\!\nabla b_k(\bfx, \bfx_{0,k})\big(V_{\text{es}}(\bfx, \calS_{\cal_G, \calP}) - V_k(\bfx, \bfx_{0,k}) \big)\,\mathrm{d\bfx}\,,
\end{align}
and $V_{\text{es}}$ is the solution of the Poisson problem in Eqs. \ref{poisson_problem} and \ref{Eqn:KS:BC:Ves} at the electron density associated with the (electronic) ground-state . Therefore, it suffices to calculate the forces for only those atoms that are located within the fundamental domain, the expression for which can be rewritten in terms of quantities expressed over the  fundamental domain as follows:
\begin{align}
\label{force_formula_first_term_rhs}
\bff_{\bfx_{0,k}} &= \sum_{\gamma = 0}^{\mathfrak{N}-1}\mathfrak{R}^{\mathfrak{N}-\gamma}\int_{\calD_{\calG}}\!\nabla b_k(\mathfrak{R}^{\gamma}\bfx, \bfx_{0,k})\big(V_{\text{es}}(\mathfrak{R}^{\gamma}\bfx,\calS_{\cal_G, \calP}) - V_k(\mathfrak{R}^{\gamma}\bfx, \bfx_{0,k}) \big)\, \mathrm{d\bfx} \nonumber \\
&= \sum_{\gamma = 0}^{\mathfrak{N}-1}\mathfrak{R}^{\mathfrak{N}-\gamma}\int_{\calD_{\calG}}\!\nabla \tilde{b}_k(\bfx, \mathfrak{R}^{\gamma}\bfx_{0,k})\big(\widetilde{V}_{\text{es}}(\bfx,\calP) - \widetilde{V}_k(\bfx, \mathfrak{R}^{\gamma}\bfx_{0,k}) \big)\,\mathrm{d\bfx} \nonumber \\
&= \sum_{\gamma = 0}^{\mathfrak{N}-1}\mathfrak{R}^{\mathfrak{N}-\gamma}\int_{\calD_{\calG}}\!\nabla \tilde{b}_k(\bfx, \bfx_{\gamma,k})\big(\widetilde{V}_{\text{es}}(\bfx,\calP) - \tilde{V}_k(\bfx, \bfx_{\gamma,k}) \big)\,\mathrm{d\bfx}\,,
\end{align}
where the second equality is obtained by using cyclic boundary conditions: 
\begin{align}
\nonumber
V_{\text{es}}(\mathfrak{R}^{\gamma}\bfx,\calS_{\cal_G, \calP}) = V_{\text{es}}(\bfx,\calS_{\cal_G, \calP})=\widetilde{V}_{\text{es}}(\bfx,\calP)\,,
\end{align}
and Proposition \ref{prop:PseudochargeDensity} (part 1). 


\subsection{Cyclic DFT: Kohn-Sham problem on the fundamental domain for cyclic structures} \label{Subsec:CyclicDFTEq}
\begin{figure}[ht]
\begin{subfigure}{\linewidth}
\centering
\includegraphics[scale=0.4]{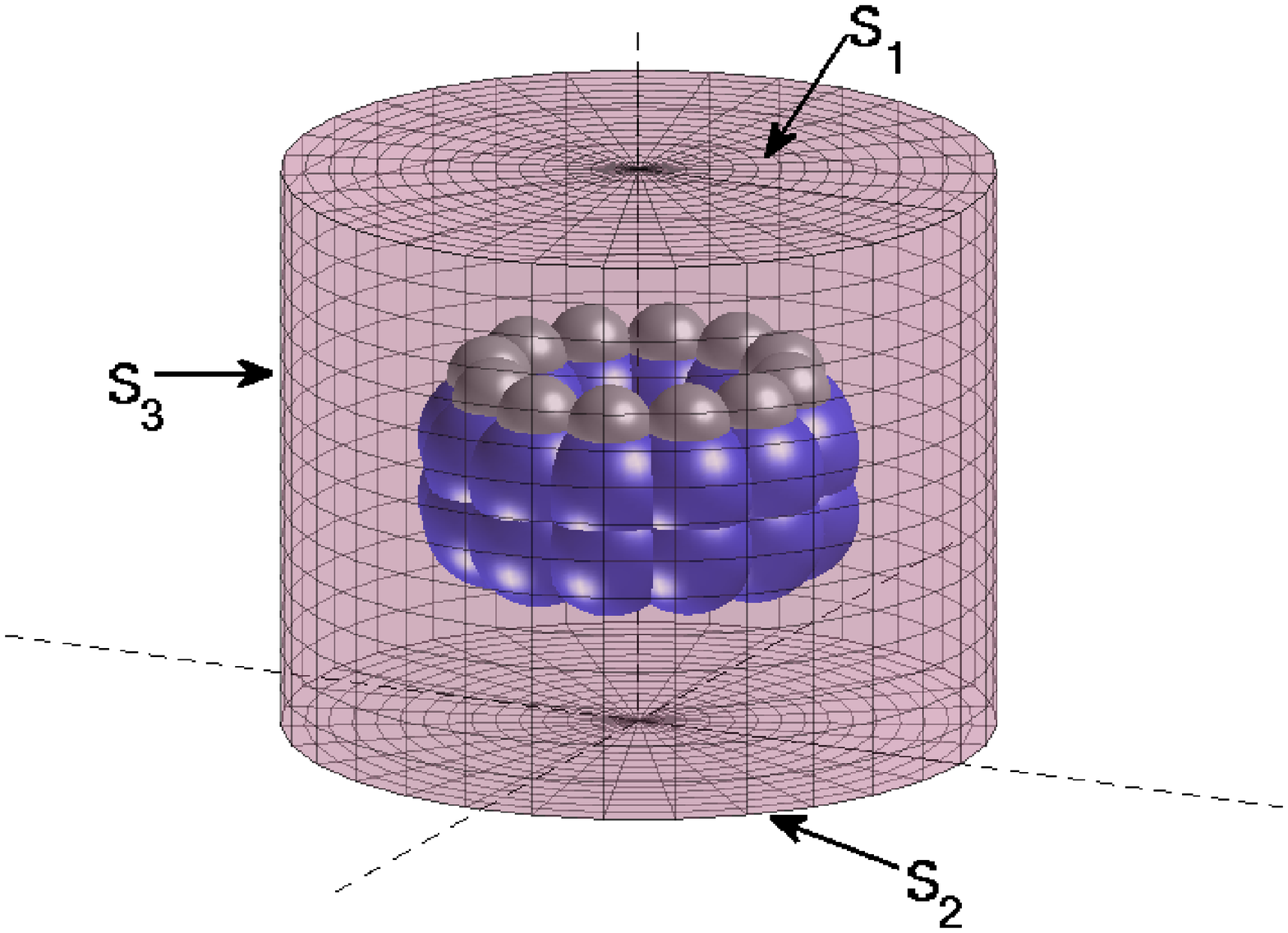}
\caption{The original problem for the cyclic structure posed in the cylinder $\calC$. The orbitals and the electron density are subjected to zero-Dirichlet boundary conditions on the surfaces $\text{S}_1,\text{S}_2,\text{S}_3$. On these surfaces, the electrostatic potential is given by Eq.~\ref{Eqn:KS:BC:Ves}. Note that $\partial \calC = \text{S}_1 \cup \text{S}_2 \cup \text{S}_3$ in the notation of the text.}
\label{fig1:subfig_a}
\end{subfigure}\\[1ex]
\begin{subfigure}{.45\linewidth}
\centering
\includegraphics[scale=0.4]{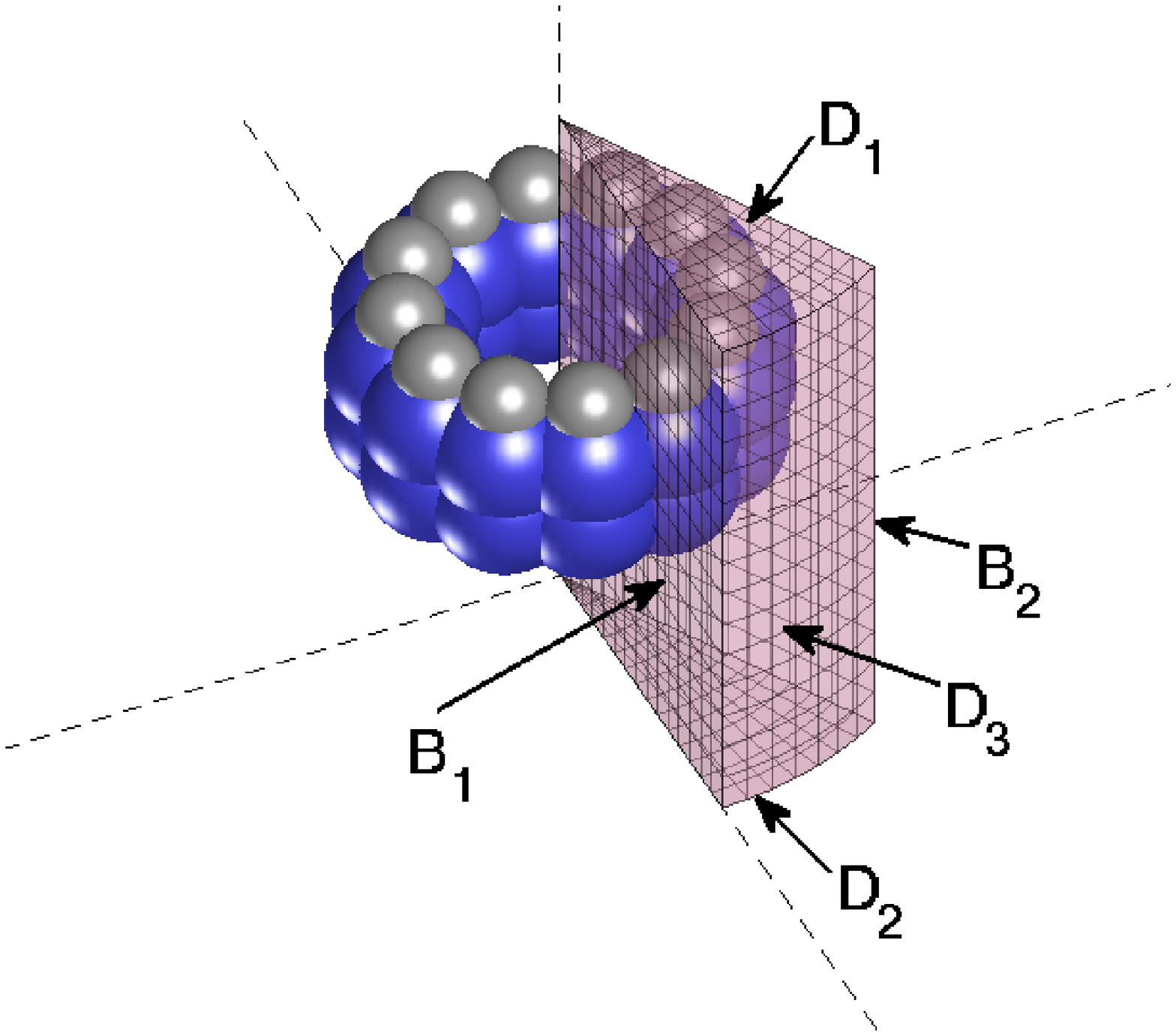}
\caption{The symmetry cell reduction results in cyclic-Bloch boundary conditions for the orbitals and cyclic boundary conditions for the density on the surfaces $\text{B}_1, \text{B}_2$. Both the electron density and the orbitals are subjected to zero-Dirichlet boundary conditions on the surfaces $\text{D}_1,\text{D}_2,\text{D}_3$.}
\label{fig1:subfig_b}
\end{subfigure}%
$\quad$
\begin{subfigure}{.45\linewidth}
\centering
\includegraphics[scale=0.4]{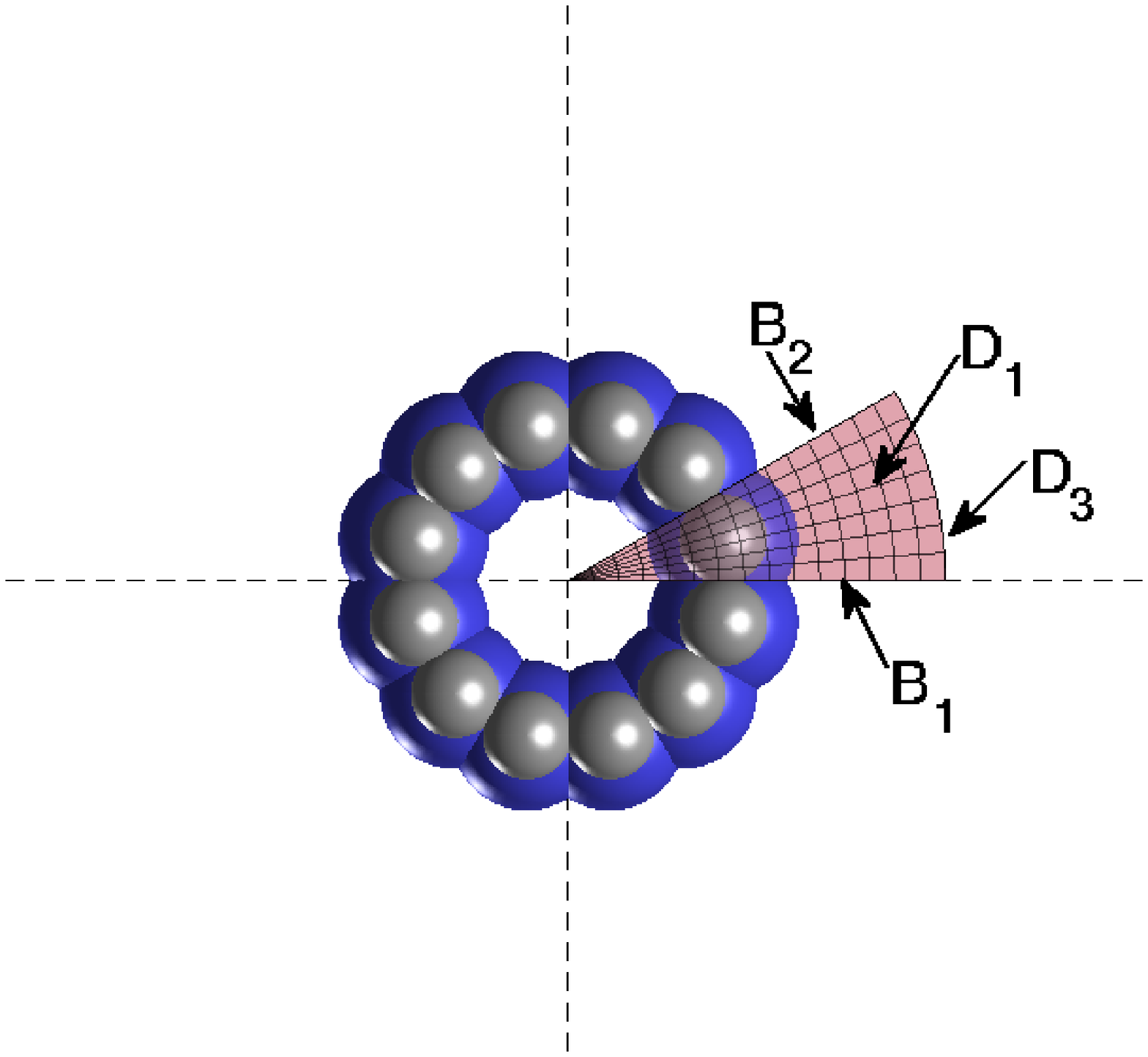}
\caption{Another view of the cyclic unit cell $\calD$ and the cyclic cell reduction. Bottom surface $\text{D}_2$ has not been shown. The electrostatic potential obeys cyclic boundary conditions on the surfaces $\text{B}_1, \text{B}_2$ and is given by Eq.~\ref{Eqn:SymmetryReducedBC_Ves} on $\text{D}_1,\text{D}_2,\text{D}_3$. Note that $\partial\calD_{\calG}^{0}= D_1 \cup D_2 \cup D_3$, while $\partial\calD_{\calG}^{C}= B_1 \cup B_2$ in the notation of the text. }
\label{fig1:subfig_c}
\end{subfigure}
\caption{Schematic of cyclic unit cell reduction.}
\label{fig:cyclic_cell_reduction}
\end{figure}

We now summarize the formulation of Cyclic DFT developed in the previous subsections. After the symmetry cell reduction depicted in Fig.~\ref{fig:cyclic_cell_reduction}, the Kohn-Sham equations on the fundamental domain $\calD_{\calG}$ can be written as (for $\nu \in \Gamma_{\mathfrak{N}}$) 
\begin{subequations} \label{Eqns:CyclicDFT:KS}
\begin{align}
\label{Cyclic_DFT_1}
\widetilde{K}(\tilde{\rho})\,\tilde{\phi}_i^{\nu}&={\lambda}_i^{\nu}\,\tilde{\phi}_i^{\nu}\;;\;\innprod{\tilde{\phi}_i^{\nu}}{\tilde{\phi}_j^{\nu}}{\mathsf{L}^2(\calD_{\calG})}=\frac{1}{\mathfrak{N}}\,\delta_{ij}\,,\\
\label{Cyclic_DFT_2}
\text{where}\quad
\widetilde{K}(\tilde{\rho})&=-\half \Delta + \widetilde{V}_{\text{es}} + {V}_{\text{xc}}(\tilde{\rho})\,,\\
\text{with}\quad
\label{Cyclic_DFT_4}
\tilde{\rho}(\bfx)&=2 \sum_{\nu = 0}^{\mathfrak{N} - 1}\sum_{i=1}^{N_{\text{s}}} {f}_i^{\nu}\abs{\tilde{\phi}_i^{\nu}(\bfx)}^2\,,\\
\label{Cyclic_DFT_3}
\text{and} \quad
-\frac{1}{4\pi}\Delta \widetilde{V}_{\text{es}} &= \,(\tilde{\rho} + \tilde{b})\,,\, \tilde{b} = b|_{\calD_{\calG}}\,. \\
\label{Cyclic_DFT_5}
\text{Here,} \quad
{f}_i^{\nu} &= \left( 1 + \exp\left( \frac{{\lambda}_i^{\nu} - \lambda_f}{\sigma} \right) \right)^{-1},\\
\label{Cyclic_DFT_6}
\text{with $\lambda_f$ s.t.} \quad &2 \sum_{\nu = 0}^{\mathfrak{N} - 1}\sum_{i=1}^{N_{\text{s}}} {f}_i^{\nu} = \mathfrak{N} N_{\text{e}}\,.
\end{align}
\end{subequations}
The corresponding boundary conditions on $ \partial\calD_{\calG} = \partial\calD_{\calG}^{C} \cup \partial\calD_{\calG}^{0}$ are 
\begin{subequations} \label{Eqns:CyclicDFT:KS:BC}
\begin{align}
\text{for} \,\,\, \bfx \in \partial\calD_{\calG}^{C} \,, \quad \tilde{\phi}_i^{\nu}(\mathfrak{R}^{\gamma}\bfx) &=  \expcharconj{\nu \gamma}{\mathfrak{N}}\,\tilde{\phi}_i^{\nu}(\bfx) \,, \label{Eq:CyclicDFT:phi_cyclicBlochBC} \\
\tilde{\rho}(\mathfrak{R}^{\gamma}\bfx) &= \tilde{\rho}(\bfx) \,, \\
\text{and} \quad \widetilde{V}_{\text{es}} (\mathfrak{R}^{\gamma}\bfx \calP) &= \widetilde{V}_{\text{es}}(\bfx,\calP) \,. \label{Eq:CyclicDFT:Ves:cyclic} \\
\text{For} \,\,\, \bfx \in \partial\calD_{\calG}^{0} \,, \quad \tilde{\phi}_i^{\nu}(\bfx) &= 0 \,, \label{Eq:CyclicDFT:phi_zeroBC} \\
\tilde{\rho}(\bfx) &= 0 \,, \\
\text{and} \quad \widetilde{V}_{\text{es}}(\bfx,\calP) &= \sum_{\gamma = 0}^{\mathfrak{N}-1} \int_{\calD_{\calG}}\!\frac{\tilde{b}(\bfy, \calP)+\tilde{\rho}(\bfy)}{\abs{\bfx-\mathfrak{R}^{\gamma}\bfy}}\; \mathrm{d\bfy} \,. \label{Eqn:SymmetryReducedBC_Ves}
\end{align}
\end{subequations}

In each iteration of the SCF method in Cyclic DFT, the following linear eigenproblems (there are $\mathfrak{N}$ of them, one for each $\nu$):
\begin{align} \label{Eq:CyclicDFT:linearEigenproblem}
\left( -\frac{1}{2} \Delta + \widetilde{V}_{\text{eff}} \right) \tilde{\phi}^{\nu}_{i} = \lambda_i^{\nu} \tilde{\phi}^{\nu}_{i} \,, \quad \nu=0, 1, \ldots, \mathfrak{N}-1 \,,
\end{align}
where
\begin{align}
\label{effective_potential_reduced}
\widetilde{V}_{\textrm{eff}}(\bfx) = \widetilde{V}_{\textrm{es}}(\tilde{\rho}(\bfx), \calP) + V_{{\textrm{xc}}}(\tilde{\rho}(\bfx)) \,,
\end{align}
need to be solved for the lowest $N_s$ eigenvalues and corresponding eigenfunctions. The eigenfunctions are subject to the cyclic-Bloch boundary conditions in Eq.~\ref{Eq:CyclicDFT:phi_cyclicBlochBC} and zero-Dirichlet boundary conditions in Eq.~\ref{Eq:CyclicDFT:phi_zeroBC}. The symmetry cell reduced electrostatic potential $\widetilde{V}_{\textrm{es}}$ is obtained by solving the Poisson equation in Eq.~\ref{Cyclic_DFT_3} subject to the cyclic boundary conditions in Eq.~\ref{Eq:CyclicDFT:Ves:cyclic} and Dirichlet boundary conditions in Eq.~\ref{Eqn:SymmetryReducedBC_Ves}. 

At the electronic ground-state, the free energy \emph{per unit cell} of the cyclic structure, i.e., the quantity $\widetilde{\mathcal{F}}(\calP) ={\mathcal{F}}({\calS_{\cal_G, \calP}})/\mathfrak{N}$, can be expressed in terms of quantities over the fundamental domain as:\footnote{If the term ${E}_{\text{overlap}}(\calS_{\cal_G, \calP})$ is included in the expression in Eq.~\ref{Harris_Foulkes_Energy}, it is also necessary to include a term of the form $\widetilde{E}_{\text{overlap}}(\calP) ={E}_{\text{overlap}}(\calS_{\cal_G, \calP}) / \mathfrak{N}$ in Eq.~\ref{Harris_Foulkes_Energy_per_cell}. This term can be expressed in terms of integrals over the fundamental domain, analogous to other the terms on the right hand side of Eq.~\ref{Harris_Foulkes_Energy_per_cell}.}
\begin{align}
\nonumber
\widetilde{\mathcal{F}}(\calP)  =& \;\frac{2}{\mathfrak{N}}\bigg(\sum_{\nu =0}^{\mathfrak{N}-1}\sum_{i = 1}^{N_{\text{s}}} f_i^{\nu} \lambda_i^{\nu}\bigg) + \int_{\calD_{\calG}}\!\varepsilon_{\text{xc}} (\tilde{\rho}(\bfx)) \tilde{\rho}(\bfx) \, \mathrm{d \bfx} - \int_{\calD_{\calG}}\!{V}_{\text{xc}}(\tilde{\rho}(\bfx))\tilde{\rho}(\bfx)\, \mathrm{d\bfx} \\\nonumber
&+\frac{1}{2} \int_{\calD_{\calG}}\!\big (\tilde{b}(\bfx, \calP)-\tilde{\rho}(\bfx)\big) \widetilde{V}_{\text{es}}(\bfx, \calP) \, \mathrm{d\bfx}-\frac{1}{2}  \sum_{\gamma = 0}^{\mathfrak{N}-1} \sum_{k = 1}^{M_{\calP}} \int_{\calD_{\calG}}\!\widetilde{V}_k(\bfx, \bfx_{\gamma,k})\,\tilde{b}_k(\bfx, \bfx_{\gamma,k})\,\mathrm{d\bfx}\\
 &+ \frac{2\,\sigma}{\mathfrak{N}} \bigg(\sum_{\nu=0}^{\mathfrak{N}-1}\sum_{i=1}^{N_{s}} \big( f_i^{\nu} \log f_i^{\nu} + (1-f_i^{\nu}) \log (1 - f_i^{\nu}) \big)\bigg)\,.
 \label{Harris_Foulkes_Energy_per_cell} 
\end{align}
The corresponding Hellmann-Feynman atomic force takes the form:\footnote{It might be necessary to include the term $\bff_{\text{overlap}}(\bfx_{0,k})$ --- described earlier in footnote \ref{footnote:fn_1} --- in Eq.~\ref{force_formula_revisited} to account for possible overlaps of pseudocharges. This quantity can be expressed in terms of integrals over the fundamental domain, analogous to the other terms in Eq.~\ref{force_formula_FD}.}
\begin{align}
\label{force_formula_FD}
\bff_{\bfx_{0,k}} = \sum_{\gamma = 0}^{\mathfrak{N}-1}\mathfrak{R}^{\mathfrak{N}-\gamma}\int_{\calD_{\calG}}\!\nabla \tilde{b}_k(\bfx, \bfx_{\gamma,k})\big(\widetilde{V}_{\text{es}}(\bfx,\calP) - \widetilde{V}_k(\bfx, \bfx_{\gamma,k}) \big)\,\mathrm{d\bfx}\,.
\end{align}

The problem of determining the equilibrium geometry\footnote{Once again, this should be interpreted in the sense of computing local structural minima (i.e. local minima in the ground state electronic free energy as the atomic positions are varied).} of the cyclic structure as described by Eq.~\ref{geometry_optimization_original} can be reformulated as:
\begin{align}
\label{geometry_optimization_reduced}
\infm{\{\bfx_{0,k} \in {\calP}\}}{\widetilde{\mathcal{F}}(\calP)}\,.
\end{align}
Indeed, it follows from Eq.~\ref{force_equivariance} that if the atoms in the fundamental domain are in equilibrium, so is every other atom in the cyclic structure. 

\subsection{Symmetry-adapted finite-difference discretization} \label{Subsec:FD}
Having formulated the governing equations of Cyclic DFT, we now describe a discretization strategy. We fix $\calC$ to be an annular cylinder\footnote{The use of an annular region enhances the efficiency of the discretization scheme not only due to reduced degrees of freedom, but also due to the improved conditioning of the various matrices resulting from the discretization. In addition, it avoids the coordinate system singularity along the line $r = 0$, which would otherwise require special treatment (see for e.g. \citep{mohseni2000numerical, lai2001note}).} with axis $\bfe_3$, height $H$, inner radius $R_{\text{in}}$, and outer radius $R_{\text{out}}$. Let the cyclic structure have a symmetry group of order $\mathfrak{N}$. It follows that the fundamental domain $\calD_{\calG}$ is a sector of the annular cylinder with angle  $2\pi / \mathfrak{N}$. To ensure that the resulting discretization scheme is compatible with the cyclic symmetry of the system, we work in cylindrical coordinates $(r,\vartheta,z)$. In this setting\footnote{In the notation introduced earlier, the points with cylindrical coordinates $\big(r \in [R_{\text{in}}, R_{\text{out}}],\vartheta = 0,z \in [0,H]\big)$ and $\big(r \in [R_{\text{in}}, R_{\text{out}}],\vartheta = \frac{2\pi}{\mathfrak{N}},z \in [0,H]\big)$ form the surfaces $\partial\calD_{\calG}^{C}$, while the points $\big(r \in [R_{\text{in}}, R_{\text{out}}],\vartheta \in [0,\frac{2\pi}{\mathfrak{N}}],z = 0\big)$, $\big(r \in [R_{\text{in}}, R_{\text{out}}],\vartheta \in [0,\frac{2\pi}{\mathfrak{N}}],z = H\big)$, $\big(r = R_{\text{in}},\vartheta \in [0,\frac{2\pi}{\mathfrak{N}}],z \in [0,H]\big)$ and  $\big(R_{\text{out}},\vartheta \in [0,\frac{2\pi}{\mathfrak{N}}],z \in [0,H]\big)$ form the surfaces $\partial\calD_{\calG}^{0}$.}, the linear eigenproblems in Eq.~\ref{Eq:CyclicDFT:linearEigenproblem} take the form:
\begin{align}
\nonumber
-\half \bigg(\hpd{\tilde{\phi}^{\nu}_i(r,\vartheta,z)}{r}{2} &+ \frac{1}{r} \pd{\tilde{\phi}^{\nu}_i(r,\vartheta,z)}{r} + \frac{1}{r^2}\hpd{{\tilde{\phi}}^{\nu}_i(r,\vartheta,z)}{\vartheta}{2}+ \hpd{\tilde{\phi}^{\nu}(r,\vartheta,z)}{z}{2}\bigg )\\ 
&+\widetilde{V}_{\text{eff}}(r,\vartheta,z)\,\tilde{\phi}^{\nu}_i(r,\vartheta,z) = \lambda_{i}^{\nu}\,\tilde{\phi}^{\nu}_i(r,\vartheta,z)\,,
\label{cylindrical_linear_eigen}
\end{align}
with boundary conditions (Eqs.~\ref{Eq:CyclicDFT:phi_cyclicBlochBC} and \ref{Eq:CyclicDFT:phi_zeroBC}):
\begin{align}
\label{Bloch_BC_in_cylindrical_coordinates}
\tilde{\phi}^{\nu}_i(r, \vartheta = \frac{2\pi}{\mathfrak{N}},z) &=  \expcharconj{\nu \gamma}{\mathfrak{N}}\,\tilde{\phi}^{\nu}_i(r, \vartheta = 0,z)\,,\\
\label{Dirichlet_BC_in_cylindrical_coordinates}
\tilde{\phi}^{\nu}_i(r, \vartheta ,z = 0) = \tilde{\phi}^{\nu}_i(r, \vartheta ,z = H) &= \tilde{\phi}^{\nu}_i(r = R_{\text{in}}, \vartheta ,z) =  \tilde{\phi}^{\nu}_i(r = R_{\text{out}}, \vartheta, z) = 0\,.
\end{align}
Similarly, the Poisson equation in Eq.~\ref{Cyclic_DFT_3} can be written as:
\begin{align}
\nonumber
-\frac{1}{4 \pi} \bigg( \hpd{\widetilde{V}_{\text{es}}(r,\vartheta,z)}{r}{2} &+ \frac{1}{r} \pd{\widetilde{V}_{\text{es}}(r,\vartheta,z)}{r} + \frac{1}{r^2}\hpd{\widetilde{V}_{\text{es}}(r,\vartheta,z)}{\vartheta}{2}+ \hpd{\widetilde{V}_{\text{es}}(r,\vartheta,z)}{z}{2} \bigg) \\
&=\tilde{\rho}(r,\vartheta,z)+\tilde{b}(r,\vartheta,z) \,,
\label{electrostatics_cylindrical_fd}
\end{align}
with boundary conditions (Eqs.~\ref{Eq:CyclicDFT:Ves:cyclic} and \ref{Eqn:SymmetryReducedBC_Ves}):
\begin{align}
\label{Electrostatic_BC_in_cylindrical_coordinates_1}
\widetilde{V}_{\text{es}}(r,\vartheta = 0 ,z) &= \widetilde{V}_{\text{es}}(r,\vartheta = \frac{2\pi}{\mathfrak{N}},z)\,,\\
\widetilde{V}_{\text{es}}(r = R_{\text{in}},\vartheta ,z) = \mathfrak{v}(r = R_{\text{in}},\vartheta ,z)\,&, \, \widetilde{V}_{\text{es}}(r = R_{\text{out}},\vartheta ,z) = \mathfrak{v}(r = R_{\text{out}},\vartheta ,z)\,,\\\label{Electrostatic_BC_in_cylindrical_coordinates_2}
\widetilde{V}_{\text{es}}(r ,\vartheta ,z = 0) = \mathfrak{v}(r ,\vartheta ,z = 0)\,&, \,\widetilde{V}_{\text{es}}(r,\vartheta ,z = H) = \mathfrak{v}(r ,\vartheta ,z=H)\,.
\end{align}
Above, $\mathfrak{v}(r,\vartheta,z)$ denotes the quantity obtained by evaluation of Eq.~\ref{Eqn:SymmetryReducedBC_Ves} at the point with cylindrical coordinates $(r,\vartheta,z)$.

We discretize the equations of Cyclic DFT using the finite-difference method with a uniform grid spacing of $\Delta r$, $\Delta \vartheta$ and $\Delta z$ in the radial, angular and $\bfe_3$ directions, respectively. Specifically, we employ a finite-difference mesh that consists of the points
\begin{align}
\calM = \calM_{R} \times \calM_{\vartheta} \times  \calM_{z}\,,
\end{align}
where $\calM_{R}$, $\calM_{\vartheta}$, and $\calM_{z}$ represent the nodes in the radial, angular and $\bfe_3$ directions, respectively, i.e., 
\begin{align}
\calM_{R} &= \{R_{\text{in}},R_{\text{in}}+\Delta r,\ldots, R_{\text{in}} + (N_{r}-1)\Delta r = R_{\text{out}}\}\,,\\
\calM_{\vartheta} &= \{0,\Delta \vartheta,\ldots,  (N_{\vartheta}-1)\Delta \theta = \frac{2\pi}{\mathfrak{N}}-\Delta \theta\}\,,\\
\calM_{z} &= \{0,\Delta z, (N_{z}-1)\Delta z = H\}\,,
\end{align}
and $N_{r}$, $N_{\vartheta}$ and $N_{z}$ denotes the number of grid points in the corresponding directions. We index the nodes by the triplet of natural numbers $(k_r,k_{\vartheta},k_z)$, with $k_r \in \{1,2,\ldots, N_r\}$, $k_{\vartheta} \in \{1,2,\ldots, N_{\vartheta} \}$ and $k_{z} \in \{1,2,\ldots,N_{z}\}$, so that the grid point with the indices $(k_r,k_{\vartheta},k_z)$ refers to the physical point  $(r_{k_r},\vartheta_{k_{\vartheta}},z_{k_z}) \in \calM$. While discretizing the governing equations, we represent the values of the various fields involved by their values at the finite-difference nodes. This casts the infinite dimensional problem to one posed on a linear space of dimension $N_rN_{\vartheta}N_{z}$.

We express the various derivatives in Eqs.~\ref{cylindrical_linear_eigen} and \ref{electrostatics_cylindrical_fd} using high-order finite-differences: 
\begin{align}
\label{higher_order_fd_1}
\pd{f}{s}\bigg\vert_{s=s_{k_s}} &\approx \sum_{q=1}^{n_{o}} w_{q,1} \big(f{(s_{k_s+q})} - f{(s_{k_s-q})}\big)\, := D_sf(s_{k_s}),\\
\label{higher_order_fd_2}
\hpd{f}{s}{2}\bigg\vert_{s=s_{k_s}}  &\approx \sum_{q=0}^{n_{o}} w_{q,2} \big(f{(s_{k_s+q})} + f{(s_{k_s-q})}\big)\, := D^{2}_sf(s_{k_s})\,,
\end{align}
where the $s$-coordinate represents any one of the $r$, $\vartheta$ or $z$-coordinates. The weights appearing in the above equations can be written as \citep{suryanarayana2014augmented, mazziotti1999spectral}:
\begin{align}
\label{fd_weights_1}
{w}_{q,1} &= \frac{(-1)^{q+1}}{ (\Delta s) q} \frac{(n_{o}!)^2}{(n_{o}-q)! (n_{o}+q)!} \,, \,\, q=1, 2, \ldots, n_{o}\,,\\
\label{fd_weights_2}
w_{0,2} &=  - \frac{1}{(\Delta s)^2} \sum_{q=1}^{n_o} \frac{1}{q^2} \,, \\\label{fd_weights_3}
w_{q,2} &= \frac{2 (-1)^{q+1}}{(\Delta s)^2 q^2} \frac{(n_{o}!)^2}{(n_{o}-q)! (n_{o}+q)!} \,, \,\, q=1, 2, \ldots, n_{o}\,,
\end{align} 
where $\Delta s$ denotes the mesh spacing along the $s$-coordinate. With these weights, the finite-difference expressions in Eqs.~\ref{higher_order_fd_1} and \ref{higher_order_fd_2} represent $2 n_{o}$-order accurate approximations, i.e., the error is $\mathcal{O}({\Delta s}^{2n_{o}})$. 

On approximating the derivatives using finite-differences, the eigenvalue equations in Eq.~\ref{cylindrical_linear_eigen} take the form:  
\begin{align}
\nonumber
-\half\bigg(&D^{2}_r\,\tilde{\phi}^{\nu}_{i}(r_{k_r},\vartheta_{k_{\vartheta}},z_{k_z})+\frac{1}{r_{k_r}}D_{r}\,\tilde{\phi}^{\nu}_i(r_{k_r},\vartheta_{k_{\vartheta}},z_{k_z})+ \frac{1}{r_{k_r}^2}D^{2}_{\vartheta}\,\tilde{\phi}^{\nu}_{i}(r_{k_r},\vartheta_{k_{\vartheta}},z_{k_z})\\
&+ D^{2}_{z}\,\tilde{\phi}^{\nu}_{i}(r_{k_r},\vartheta_{k_{\vartheta}},z_{k_z})\bigg )+\widetilde{V}_{\text{eff}}(r_{k_r},\vartheta_{k_{\vartheta}},z_{k_z})\,\tilde{\phi}^{\nu}_i(r_{k_r},\vartheta_{k_{\vartheta}},z_{k_z}) = \lambda_{i}^{\nu}\,\tilde{\phi}^{\nu}_i(r_{k_r},\vartheta_{k_{\vartheta}},z_{k_z})\,.
\label{discretized_KS_Hamiltonian}
\end{align}
During the application of the finite-difference stencils, any reference to grid points which do not lie in $\calM$ is resolved by using the discrete representation of the boundary conditions in Eqs.~\ref{Bloch_BC_in_cylindrical_coordinates} and \ref{Dirichlet_BC_in_cylindrical_coordinates}: 
\begin{align}
\tilde{\phi}^{\nu}_{i}(r_{k_r},\vartheta_{k_{\vartheta}},z_{k_z}) &=  \expcharconj{\nu \gamma}{\mathfrak{N}}\,\tilde{\phi}^{\nu}_{i}(r_{k_r},\vartheta_{k_{\vartheta}+N_{\vartheta}},z_{k_z}) \,\,\,\, \text{if} \,\,\,\, k_{\vartheta} < 1 \,, \\
\tilde{\phi}^{\nu}_{i}(r_{k_r},\vartheta_{k_{\vartheta}},z_{k_z}) &= \expchar{\nu \gamma}{\mathfrak{N}}\,\tilde{\phi}^{\nu}_{i}(r_{k_r},\vartheta_{k_{\vartheta}-N_{\vartheta}},z_{k_z}) \,\,\,\, \text{if} \,\,\,\, k_{\vartheta} > N_{\vartheta} \,, \\
\tilde{\phi}^{\nu}_{i}(r_{k_r},\vartheta_{k_{\vartheta}},z_{k_z}) &= 0 \,\,\,\, \text{if} \,\,\,\, k_r \notin \{1,2,\ldots, N_r\} \,\, \text{or} \,\, k_z \notin \{1,2,\ldots, N_z\}\,.
\end{align}
Analogously, the Poisson equation in Eq.~\ref{electrostatics_cylindrical_fd} can be written in discrete form as:
\begin{align}
\nonumber
- \frac{1}{4 \pi} \bigg( D^{2}_r\,\widetilde{V}_{\text{es}}(r_{k_r},\vartheta_{k_{\vartheta}},z_{k_z})  &+ \frac{1}{r_{k_r}} D_r\,\widetilde{V}_{\text{es}}(r_{k_r},\vartheta_{k_{\vartheta}},z_{k_z}) + \frac{1}{r_{k_r}^2}D^{2}_{\vartheta}\,\widetilde{V}_{\text{es}}(r_{k_r},\vartheta_{k_{\vartheta}},z_{k_z})\\ &+ D^{2}_z\,\widetilde{V}_{\text{es}}(r_{k_r},\vartheta_{k_{\vartheta}},z_{k_z}) \bigg) = \tilde{\rho}(r_{k_r},\vartheta_{k_{\vartheta}},z_{k_z})+\tilde{b}(r_{k_r},\vartheta_{k_{\vartheta}},z_{k_z}) \,,
\end{align}
with the discrete representation of the boundary conditions in Eqs.~\ref{Electrostatic_BC_in_cylindrical_coordinates_1} and \ref{Electrostatic_BC_in_cylindrical_coordinates_2} being:\footnote{For $k_r < 1$ or $k_r > N_r$, the point $r_{k_r}$ is used to denote the extension of the mesh in the radial direction, beyond the points $R_{\text{in}}$ and $R_{\text{out}}$, respectively. Thus, $k_r = 0$ denotes points with radial coordinate $R_{\text{in}}-\Delta r$, while $k_r = N_r + 1$ denotes points with radial coordinate $R_{\text{out}}+\Delta r$. Similar considerations hold in the $z$-direction.}
\begin{align}
\widetilde{V}_{\text{es}}(r_{k_r},\vartheta_{k_{\vartheta}},z_{k_z}) &= \widetilde{V}_{\text{es}}(r_{k_r},\vartheta_{k_{\vartheta}+N_{\vartheta}},z_{k_z}) \,\,\,\, \text{if} \,\,\,\, k_{\vartheta} < 1 \,, \\
\widetilde{V}_{\text{es}}(r_{k_r},\vartheta_{k_{\vartheta}},z_{k_z}) &= \widetilde{V}_{\text{es}}(r_{k_r},\vartheta_{k_{\vartheta}-N_{\vartheta}},z_{k_z}) \,\,\,\, \text{if} \,\,\,\, k_{\vartheta} > N_{\vartheta} \,, \\
\widetilde{V}_{\text{es}}(r_{k_r},\vartheta_{k_{\vartheta}},z_{k_z}) &= \mathfrak{v}(r_{k_r},\vartheta_{k_{\vartheta}},z_{k_z}) \,\,\,\, \text{if} \,\,\,\, k_r \notin \{1,2,\ldots, N_r\} \,\, \text{or} \,\, k_z \notin \{1,2,\ldots, N_z\} \,.
\end{align}
The values of $\mathfrak{v}(r,\vartheta,z)$ are computed by means of direct numerical quadrature or using a multipole expansion in cylindrical coordinates \citep{cohl1999compact, wiki_cylindrical_multipole}. 

The evaluation of the electronic ground-state free energy via Eq.~\ref{Harris_Foulkes_Energy_per_cell} and the atomic forces via Eq.~\ref{force_formula_FD} requires a recipe for evaluating integrals over the fundamental domain. To this end, we employ the quadrature rule:
\begin{align}
\nonumber
\int_{\calD_{\calG}}\!f(\bfx)\,\mathrm{d\bfx} &= \int_{r=R_{\text{in}}}^{R_{\text{out}}}\!\int_{\vartheta=0}^{\frac{2\pi}{\mathfrak{N}}}\!\int_{z=0}^{H}f(r,\vartheta,z)\,r\,\mathrm{dr\,d\vartheta\,dz} \\ &\approx \sum_{k_r=1}^{N_r}\sum_{k_{\vartheta}=1}^{N_{\vartheta}}\sum_{k_z=1}^{N_z}f(r_{k_r},\vartheta_{k_{\vartheta}},z_{k_z})\,r_{k_r} \Delta r\, \Delta \vartheta \Delta z\,.
\label{integration_discrete}
\end{align}
The numerical evaluation of the gradient operator for computation of the atomic forces merits further consideration. Since the derivation of Eq.~\ref{force_formula_1} (and consequently, Eq.~\ref{force_formula_FD}) implicitly assumes that the gradient operator is expressed in Cartesian coordinates (for e.g., see \citep{Phanish_SPARC_1}), we need to express the Cartesian gradient in cylindrical coordinates so that Eq.~\ref{force_formula_FD} can be evaluated by means of the finite-difference operators $D_r$, $D_{\vartheta}$ and $D_{z}$. For example, the computation of the $\bfe_1$ component of the force requires application of the operator $\pd{}{x}$, or equivalently,  $\cos(\vartheta) \pd{}{r} - \sin(\vartheta) \pd{}{\vartheta} \approx \cos(\vartheta) D_r - \sin(\vartheta) D_{\vartheta}$. Such an approach \citep{my_analog_of_planewaves} results in the atomic forces being evaluated in Cartesian coordinates directly, which is more convenient for atomic relaxation and molecular dynamics calculations. 

We note that the Laplacian and Hamiltonian matrices resulting from the above discretization are non-Hermitian, even though the infinite-dimensional operators from which they arise are Hermitian (i.e., self-adjoint). Having said this, each matrix does approach a Hermitian matrix as the discretization is refined and/or the finite difference order $n_{o}$ is increased. This issue is well known in the literature (for example, in the context of adaptive coordinates \citep{gygi1995real, Octopus_1}) and has been shown to not interfere with the quality of the solution obtained in practical electronic structure calculations. In particular, the eigenvalues of the Hamiltonian matrix (Eq.~\ref{discretized_KS_Hamiltonian}), which are required for computation of the band energy (Eq.~\ref{Harris_Foulkes_Energy_per_cell}), turn out to be real valued (or have vanishingly small imaginary parts that can be ignored) as required.

An alternative discretization scheme to the one proposed above is the spectral method introduced in \citep{My_PhD_Thesis, banerjee2015spectral} for cluster systems. This scheme --- identical to the plane-wave method in many respects --- is capable of leveraging arbitrary point group symmetries. It is therefore capable of solving the equations of Cyclic DFT on the fundamental domain \citep{My_PhD_Thesis}. However, the basis functions are global in nature (like plane-waves) and therefore the approach is not well suited for simulating large bent structures in which the atoms may be  located far from the origin.\footnote{This issue can be handled by making suitable modifications in the radial basis functions.} In this respect, the localized nature of the proposed finite-difference approach (each node interacts with only a small set of neighboring points) enables the degrees of freedom (i.e., grid points) to be judiciously expended close to regions of interest while avoiding empty regions (i.e., grid points can be placed in regions close to a bent structure), thus improving the overall efficiency of the discretization. With regards to accuracy, it has been shown that the use of high-order finite-difference stencils within the pseudocharge formulation allows the systematic and accurate computation of total energies and forces with minimal interference from numerical issues such as the egg-box effect \citep{Phanish_SPARC_1,ghosh2016sparc2}. 

It is worth mentioning that an alternative real-space discretization to the one adopted here is that based on finite-elements. In particular, the finite-element method allows various non-conventional domains, boundary conditions, and geometries to be easily handled. Moreover, high-order finite-elements have been shown to be an efficient choice for performing DFT calculations \citep{Pask_FEM_review_1,Gavini_higher_order, motamarri2014subquadratic}. However, the relatively simple and easy-to-implement high-order finite-difference discretization employed in this work is accurate and highly computationally efficient\footnote{This can be mainly attributed to the particularly compact representation of the Laplacian compared to other real-space alternatives for achieving the accuracies desired in electronic structure calculations.}, which enables reliable mechanistic simulations (Section \ref{subsec:silicene_results}) to be performed. Indeed, as suggested by the anonymous reviewer, investigating the efficacy of alternate real-space discretization strategies (such as finite-elements and wavelets \citep{BigDFT}) in the present context is a worthy subject of future work.

We mention in passing that even though the finite-difference discretization has been considered by several workers for performing electronic structure calculations \citep{Chelikowsky_Saad_1, Octopus_1, Phanish_SPARC_1}, to the best our knowledge, the present work is the first to do so within the cylindrical coordinate system. This choice of coordinate system presents its own challenges for the finite-difference method. For example, even in the Cartersian coordinate system where the mesh is uniformly spaced, calculation of accurate atomic forces is challenging \cite{ono1999timesaving,bobbitt2015high}. This and a number of such issues have been overcome in the formulation and implementation of Cyclic DFT so as to enable the accurate and efficient evaluation of energies and forces, as verified in Section \ref{Subsec:Verification}.


\section{Results and discussion} \label{Sec:Results}
\subsection{Implementation of Cyclic DFT} \label{subsec:implementation_details}
We implement the strategies and algorithms developed in the previous section using the MATLAB software package. In all the simulations, we utilize the pseudopotentials introduced in \citep{Carter_bulk_pseudo_1, Carter_bulk_pseudo_2}, a smearing of $\sigma = 0.0862$ eV (i.e., an electronic temperature of $1000$ K), and the local density approximation (LDA) \citep{KohnSham_DFT} with the Perdew-Wang parametrization \citep{perdew1992accurate} for the correlation functional. In addition, we choose $R_{\text{in}}$, $R_{\text{out}}$, and $H$ such that all atoms are at least $10-12$ Bohr away from the boundary $\partial\calD_{\calG}^{0}$ (i.e., the boundary on which Dirichlet boundary conditions are applied), so as to allow sufficient decay of the orbitals and the electron density.\footnote{In \citep{Phanish_SPARC_1}, it has been demonstrated that a vacuum of $10$ Bohr is sufficient to achieve accuracy of $10^{-5}$ Ha/atom in the energy and $10^{-5}$ Ha/Bohr in the forces for even polar systems like CO and H$_2$O.} Also, we employ $12^{th}$ order accurate finite-differences (i.e., $n_o = 6$) for all our calculations. The local nature of the finite-difference scheme results in the Laplacian and Hamiltonian matrices being sparse, and these are stored as such. In what follows, rather than use the three discretization parameters $\Delta r$, $\Delta \vartheta$ and $\Delta z$ (or equivalently $N_{r}$, $N_{\vartheta}$ and $N_{z}$), we utilize a single parameter to characterize the discretization: $h = \displaystyle \max \left\{ \Delta r, \frac{(R_{\text{in}}+R_{\text{out}})}{2}\Delta\vartheta, \Delta z \right\}$.

We generate the guess electron density at the start of the SCF iteration by summing the electron densities corresponding to the isolated-atom Kohn-Sham solutions. Random numbers are used to generate the initial guess for the orbitals, which are then orthonormalized for the purposes of stability. We use the Chebyshev polynomial filtered subspace iteration (CheFSI) method \citep{Serial_Chebyshev, Parallel_Chebyshev, zhou2014chebyshev} (with filter orders of $80-120$) for computing approximations to the Kohn-Sham orbitals in each SCF iteration.\footnote{Due to the nature of the cylindrical coordinate system, all the finite-difference mesh points are not uniformly spaced. Consequently, the discretized operators in this case are not as well conditioned as the operators arising from uniform grids in the Cartesian coordinate system. This makes it necessary to employ relatively high Chebyshev polynomial filter orders here.} As an alternative, we retain the option of using the Generalized Preconditioned Locally Harmonic Residual (GPLHR)\footnote{This method can be thought of as an analog of the LOBPCG algorithm \cite{LOBPCG_1} for the case of non-Hermitian matrices. Due to the ability of this method to use preconditioners (based on incomplete LU factorization, for example) it becomes computationally advantageous to use this method whenever the Hamiltonian is poorly conditioned. This can happen, for example, when studying a severely bent nanostructure since $R_{\text{in}}$ is likely to be close to $0$ in this case. This will lead to excessive clustering of finite difference nodes near $R_{\text{in}}$, thus resulting in the poor conditioning. The use of CheFSI in this case would involve very high filter orders, resulting in the repeated computation of the product of the Hamiltonian matrix with a block of vectors which would make the calculation expensive. In contrast, the use of GPLHR, particularly in the early SCF iterations allows the overall computational cost to be kept manageable. We would like to thank Eugene Vecharynski (Lawrence Berkeley National Lab) for his help with the use of the GPLHR method in Cyclic DFT.} algorithm \citep{vecharynski2015generalized}. We solve the Poisson equation using the Generalized Minimal Residual (GMRES) method \citep{saad1986gmres} with an incomplete-LU factorization based preconditioner \citep{saad2003iterative}, while retaining the option of using the AAJ \citep{pratapa2016anderson} and rPJ \citep{pratapa2015restarted} linear solvers.\footnote{These alternative solvers tend to produce better performance while dealing with poorly-conditioned problems \citep{pratapa2016anderson,pratapa2015restarted,suryanarayana2016alternating}.} We use a relative convergence tolerance of $10^{-5}-10^{-6}$ on the effective potential for convergence of the SCF method, and accelerate it using the Periodic Pulay mixing scheme \citep{banerjee2016periodic}.

We perform all computations using a single node of the Mesabi cluster of the Minnesota Supercomputing Institute. Each node of Mesabi has 24 Intel Haswell E5-2680v3 processors operating at $2.50$ GHz and sharing $64$-GB of RAM.


\subsection{Verification studies for the accuracy and efficiency of Cyclic DFT} \label{Subsec:Verification}
\begin{figure}[ht]
\centering
\includegraphics[width=0.45\linewidth]{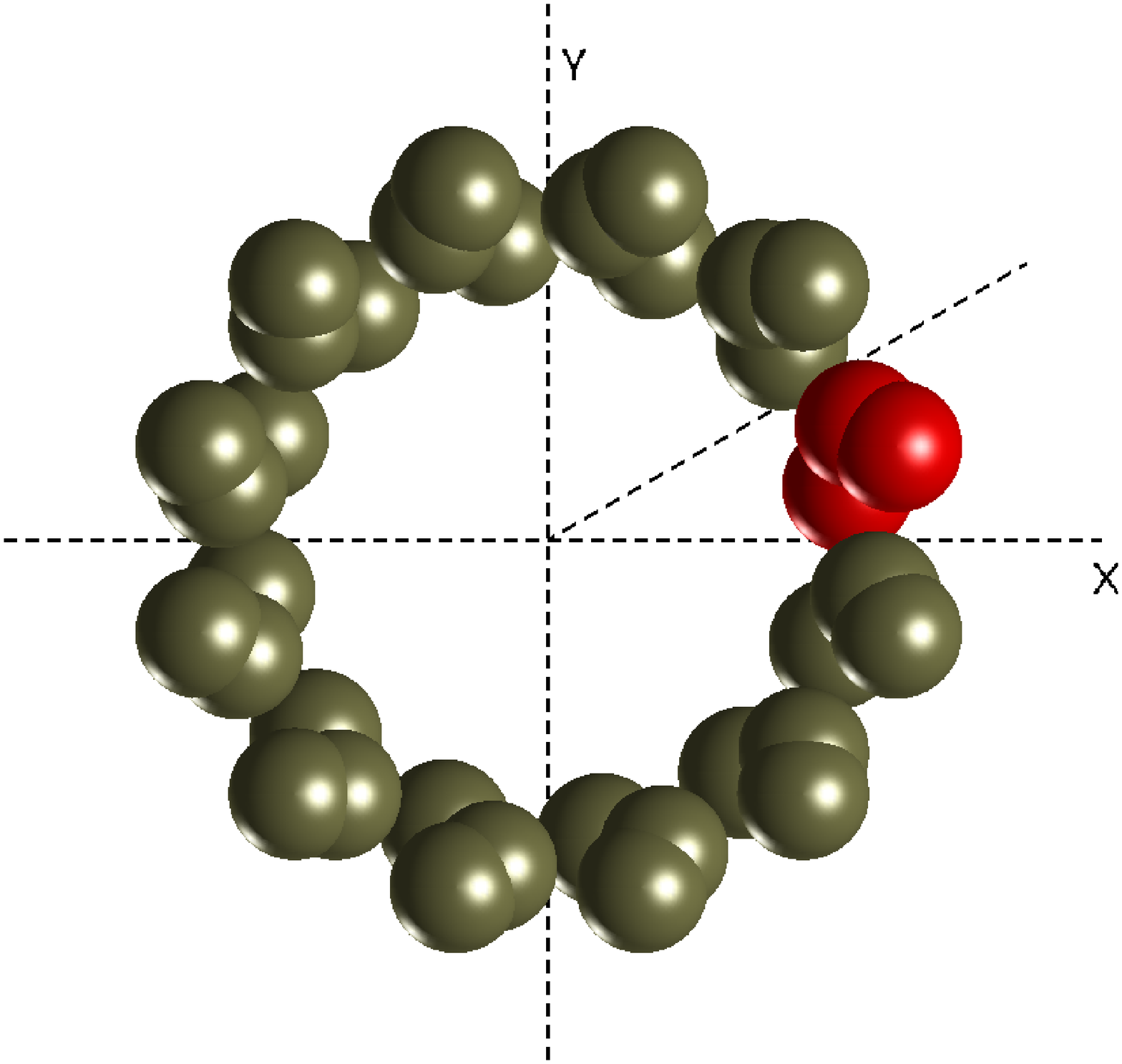}
\caption{Cyclic aluminum nanostructure with $M_{\calP}=3$ and group order $\mathfrak{N}=12$. The atoms within the fundamental domain are colored red.}
\label{fig:aluminum_ring}
\end{figure}

We now verify the accuracy and efficiency of Cyclic DFT. As the representative example, we choose a cyclic aluminum nanostructure with $M_{\calP}=3$ atoms in the fundamental domain and symmetry group order of $\mathfrak{N}=12$, as shown in Fig.~\ref{fig:aluminum_ring}. The atoms are positioned randomly within the fundamental domain such that the radial coordinates are between $11-13$ Bohr and $z$-coordinates are between $10-13$ Bohr.

First, we confirm the convergence of the computed energy and atomic forces to plane-wave results determined using ABINIT \citep{Gonze_ABINIT_1} --- a well-established and optimized code for performing DFT calculations.  In ABINIT, we use a plane-wave energy cutoff of $20$ Ha and a cubic supercell with edge of $40$ Bohr, which results in energy and forces that are converged to within $2.5 \times 10^{-7}$ Ha/atom and $10^{-5}$ Ha/Bohr, respectively. In Cyclic DFT, we choose the dimensions of the annular cylinder $\calC$ to be $H=23$ Bohr, $R_{\text{in}} = 1$ Bohr, and $R_{\text{out}}=23$ Bohr. We present in Table \ref{Table:Convergence} the error in computed energy and atomic forces as the spatial discretization is refined. It is clear that there is systematic convergence in energy and forces to the reference plane-wave result, with accuracies of even $10^{-4}$ Ha/atom and $10^{-4}$ Ha/Bohr being readily achieved. This verifies the accuracy of Cyclic DFT in determining the electronic ground-state energy and atomic forces of cyclic structures. 

\begin{table}[ht]
\centering
\begin{tabular}{|c|c|c|}
\hline 
$h$ & Error in energy & Error in atomic force \\
(Bohr) & ($\mu$Ha/atom) & ($\mu$Ha/Bohr) \\
\hline
$1.00$ & $6980$ & $130537$ \\ 
\hline
$0.71$ & $740$ & $47000$ \\  
\hline
$0.48$ & $27$ & $4300$ \\  
\hline
$0.39$ & $12$ & $658$ \\ 
\hline
$0.30$ & $4$ & $93$ \\
\hline 
\end{tabular} 
\caption{Convergence of the energy and atomic forces computed by Cyclic DFT to the reference plane-wave result. The system under consideration is a cyclic aluminum nanostructure with $M_{\calP}=3$ and group order $\mathfrak{N}=12$. The error in the energy denotes the absolute value of the difference and the error in the atomic force denotes the maximum magnitude of the difference among all the atoms. }
\label{Table:Convergence}
\end{table}

Next, we investigate the efficiency of the symmetry cell reduction in Cyclic DFT. To do so, for the aforementioned cyclic aluminum nanostructure, we consider unit cells where the sectors of the annular cylinder have angles that are $12$, $6$, $4$, $3$, $2$, and $1$ times the angle of the fundamental domain (i.e., $\pi/6$).\footnote{The $12$-fold cyclic symmetry of the system allows us to choose $1$, $2$, $3$, $4$, $6$, and $12$-fold cyclic symmetry subgroups.} This translates to exploiting the cyclic symmetry reduction from groups of order $\mathfrak{N}=1$, $2$, $3$, $4$, $6$, and $12$, respectively. For these cases, we present in Table \ref{Table:BreakdownTimeCyclic} the computational time taken by the various components of Cyclic DFT for $h=0.43$ Bohr. We observe that there is significant reduction in the computational time as the value of $\mathfrak{N}$ is increased.\footnote{During the SCF iterations using CheFSI, the most computationally intensive step tends to be computation of the action of the discretized Hamiltonian matrix on a block of vectors. It is easy to see that this step would stand to gain a speedup by a factor of $\mathfrak{N}$ when a cyclic group of order $\mathfrak{N}$ is used. This is because the Hamiltonian is represented on a smaller physical domain resulting in a fewer number of grid points (i.e., a factor of $\mathfrak{N}$) while the total number of electronic states involved (i.e., the number of states for every value of $\nu$, times the number of values of $\nu$) effectively remains the same. This trend is quite apparent in column $4$ of Table \ref{Table:BreakdownTimeCyclic}. However, for large enough systems, subspace diagonalization becomes the dominant cost within CheFSI, whereby the reduction in cost due to symmetry is expected to scale quadratically with respect to group order. This is also the case for other eigensolvers such as GPLHR, all of which asymptotically scale cubically with system size. Since subspace diagonalization can consume a significant portion of the computational time during large scale electronic structure computations \citep{My_DG_Cheby_paper}, we anticipate that the quadratic speed up obtained by the cyclic group reduction can result in a significant computational saving in large scale calculations. \label{Footnote:QuadraticReduction}} The worse than linear reduction of some of the components within one SCF iteration and the better than linear reduction of the atomic forces can be attributed to the use of inbuilt MATLAB optimizations, as well as the scope for further improvement in the implementation. We note that the energy and atomic forces obtained for the various group orders chosen above are identical to each other within $10^{-8}$ Ha/atom and $10^{-8}$ Ha/Bohr, respectively, further verifying the accuracy of Cyclic DFT. 

\begin{table}[ht]
\centering
\begin{tabular}{|c|c|c|c|c|c|c|}
\hline
$\mathfrak{N}$ & Poisson & Electron density & CheFSI & Energy & Total time & Atomic forces  \\
\hline 
$1$ & $26.4$ & $1.05$ & $353$ & $0.057$ & $371$ & $700$ \\ 
\hline 
$2$ & $12.3$ & $0.51$ & $194$ & $0.013$ & $204$  & $316$ \\ 
\hline 
$3$ & $8.7$ & $0.42$ & $118$ & $0.020$ & $129$  & $216$ \\ 
\hline 
$4$ & $6.3$ & $0.25$ & $92$ & $0.007$ & $102$  & $125$ \\ 
\hline 
$6$ & $4.0$ & $0.17$ & $57$ & $0.005$ & $66$  & $58$ \\ 
\hline 
$12$ & $2.6$ & $0.08$ & $33$ & $0.006$ & $42$  & $29$ \\ 
\hline 
\end{tabular} 
\caption{Computational time in seconds for various components in Cyclic DFT as a function of the group order $\mathfrak{N}$ chosen for symmetry reduction. The system under consideration is a aluminum nanostructure with $12$-fold cyclic symmetry. The Poisson time is for one solution with an initial guess that is identically zero. The electron density and CheFSI times are for a representative iteration within the SCF method.}
\label{Table:BreakdownTimeCyclic}
\end{table}

As demonstrated in Table \ref{Table:SCF:Cyclic}, the symmetry cell reduction also reduces the number of iterations required to achieve convergence within the SCF method. A possible explanation for this phenomenon is as follows. In the CheFSI method applied to the full problem (or a problem with reduced group order), the eigenvectors (and eigenvalues) of the linearized Hamiltonian arising in every SCF iteration are calculated only approximately. Therefore, the eigenvectors(of the full problem, or a problem with reduced group order) do not exactly satisfy cyclic-Bloch boundary conditions on the fundamental domain. Consequently, the electron density and the effective potential do not precisely satisfy cyclic boundary conditions on the fundamental domain. This introduces low frequency error components into the SCF fixed-point iteration, which negatively impacts its convergence. Indeed, at the electronic ground-state, the eigenfunctions, electron density, and effective potential all satisfy the desired boundary conditions on the fundamental domain, as verified by the nearly identical results obtained for the various choices of group order. In contrast, directly enforcing the cyclic-Bloch boundary conditions on the fundamental domain using the appropriate group order bypasses the above described issue and results in better SCF convergence.
\begin{table}[ht]
\centering
\begin{tabular}{|c|c|c|c|c|c|c|}
\hline
$\mathfrak{N}$ & $1$ & $2$ & $3$ & $4$ & $6$ & $12$ \\
\hline
SCF iterations & $127$ & $85$ & $44$ & $36$ & $30$ & $24$ \\
\hline
\end{tabular} 
\caption{Number of SCF iterations as a function of the group order $\mathfrak{N}$. The system under consideration is a aluminum nanostructure with $12$-fold cyclic symmetry. }
\label{Table:SCF:Cyclic}
\end{table}


\subsection{Application of Cyclic DFT to the bending of a silicene nanoribbon}
\label{subsec:silicene_results}
Silicene\footnote{The silicon analogue of graphene.} is a two-dimensional allotrope of silicon consisting of atoms arranged in a buckled honeycomb lattice \cite{vogt2012silicene}. Silicene has fascinating electronic properties\footnote{Just like graphene, silicene has a linear dispersion relation and is semi-metallic.} that include high charge mobility \cite{shao2013first} and the quantum spin Hall effect \cite{liu2011quantum}. Though there have been some electronic structure studies to understand the effect of strains on the mechanical and electronic properties of silicene \cite{qin2012first,peng2013mechanical}, they have been restricted to uniform in-plane deformations.\footnote{This is because the commonly used plane-wave approaches are restricted to periodic boundary conditions, i.e., translational symmetry.} Using silicene as a representative example, we now show how Cyclic DFT can be used for the first principles study of uniform bending in nanostructures. To the best of our knowledge, this is the first self-consistent ab-initio simulation of such a nature. 

\begin{figure}[ht]
\begin{subfigure}{\linewidth}
\centering
\includegraphics[scale=0.35]{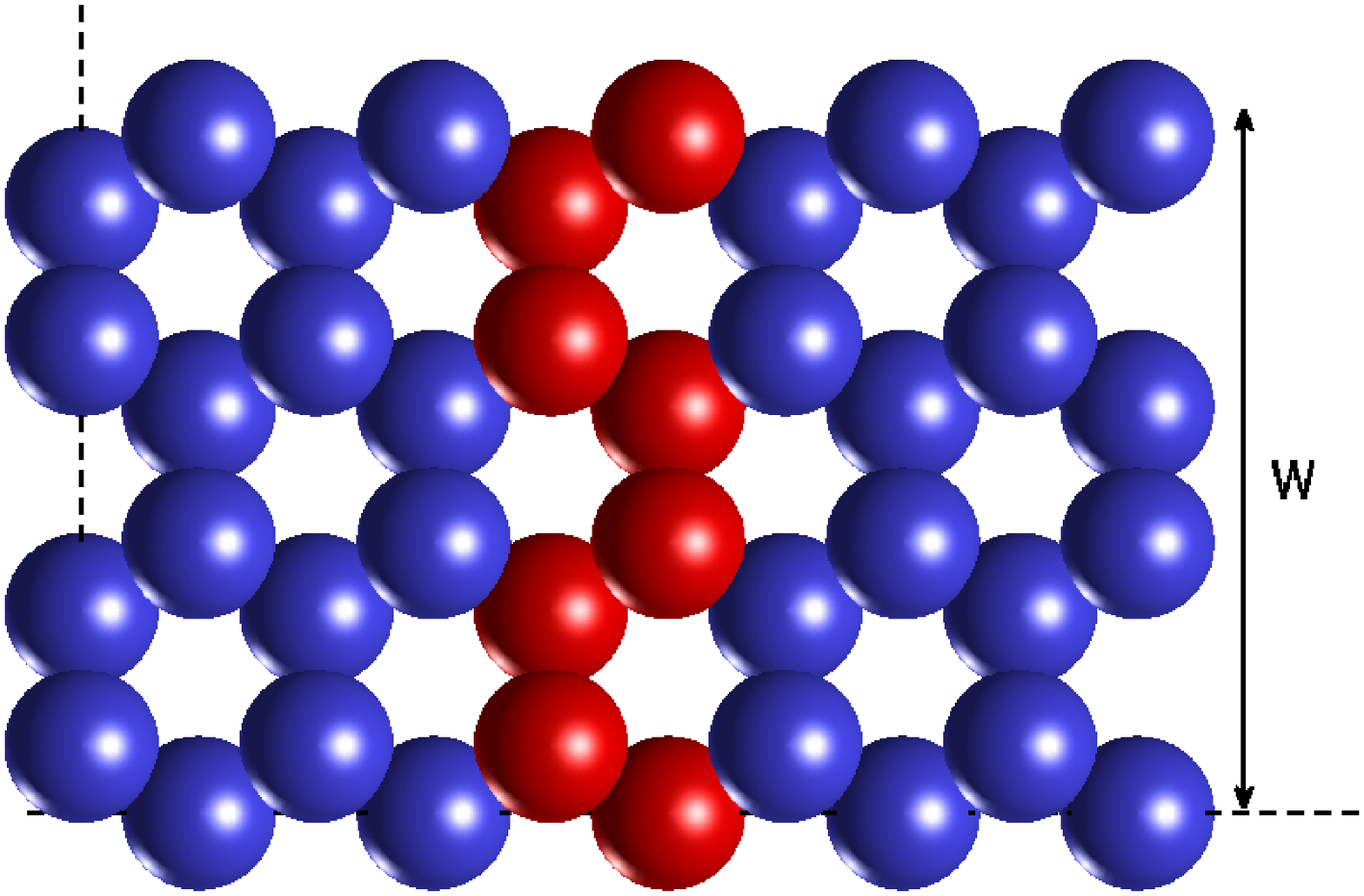}
\caption{Silicene nanoribbon}
\label{Fig:SiliceneSheet}
\end{subfigure}\\
\begin{subfigure}{.45\linewidth}
\centering
\includegraphics[scale=0.28]{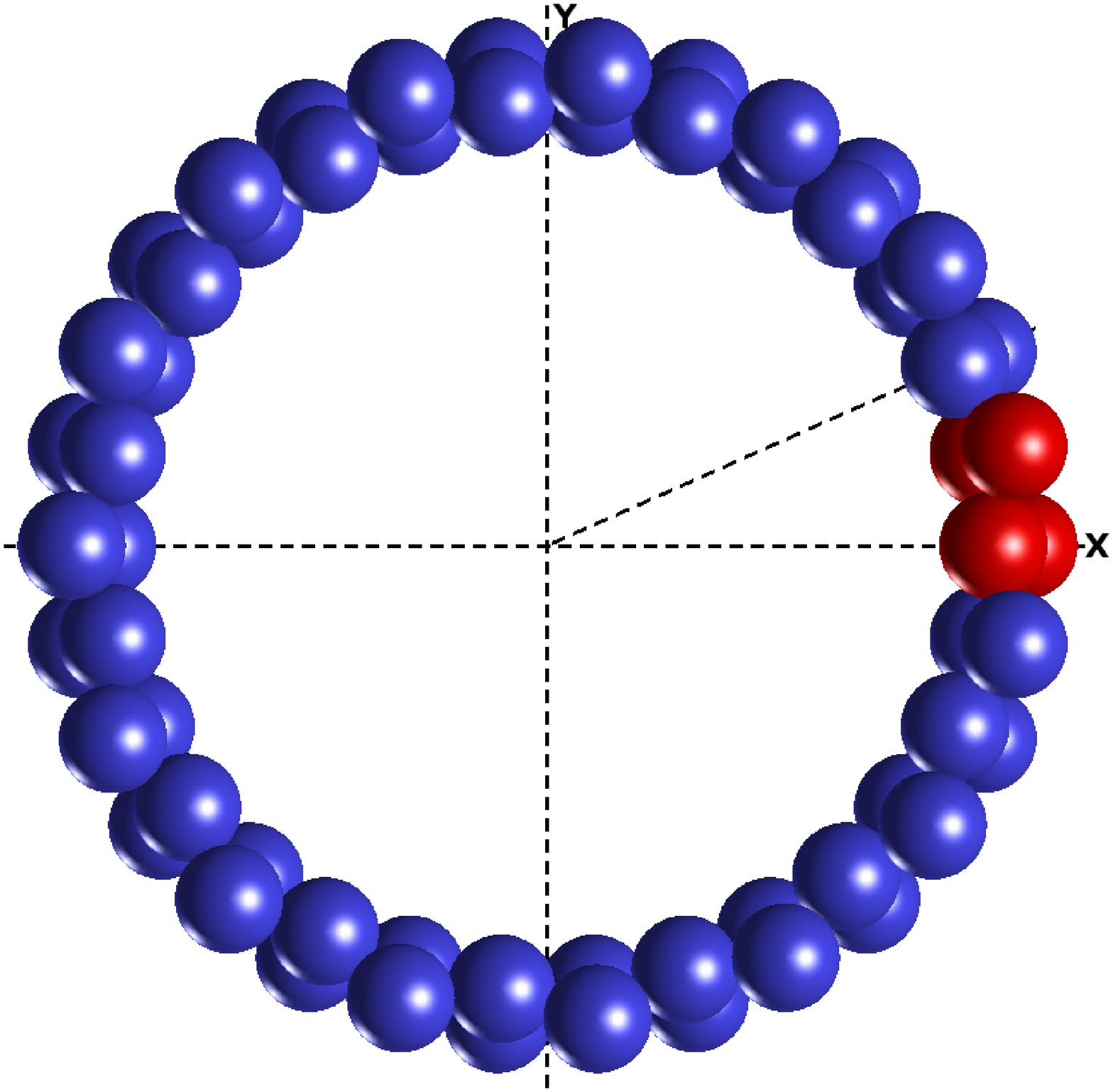}
\caption{Top view of the cyclic structure generated using the silicene nanoribbon.}
\label{Fig:SiliceneBent:Top}
\end{subfigure}%
$\quad$
\begin{subfigure}{.45\linewidth}
\centering
\includegraphics[scale=0.28]{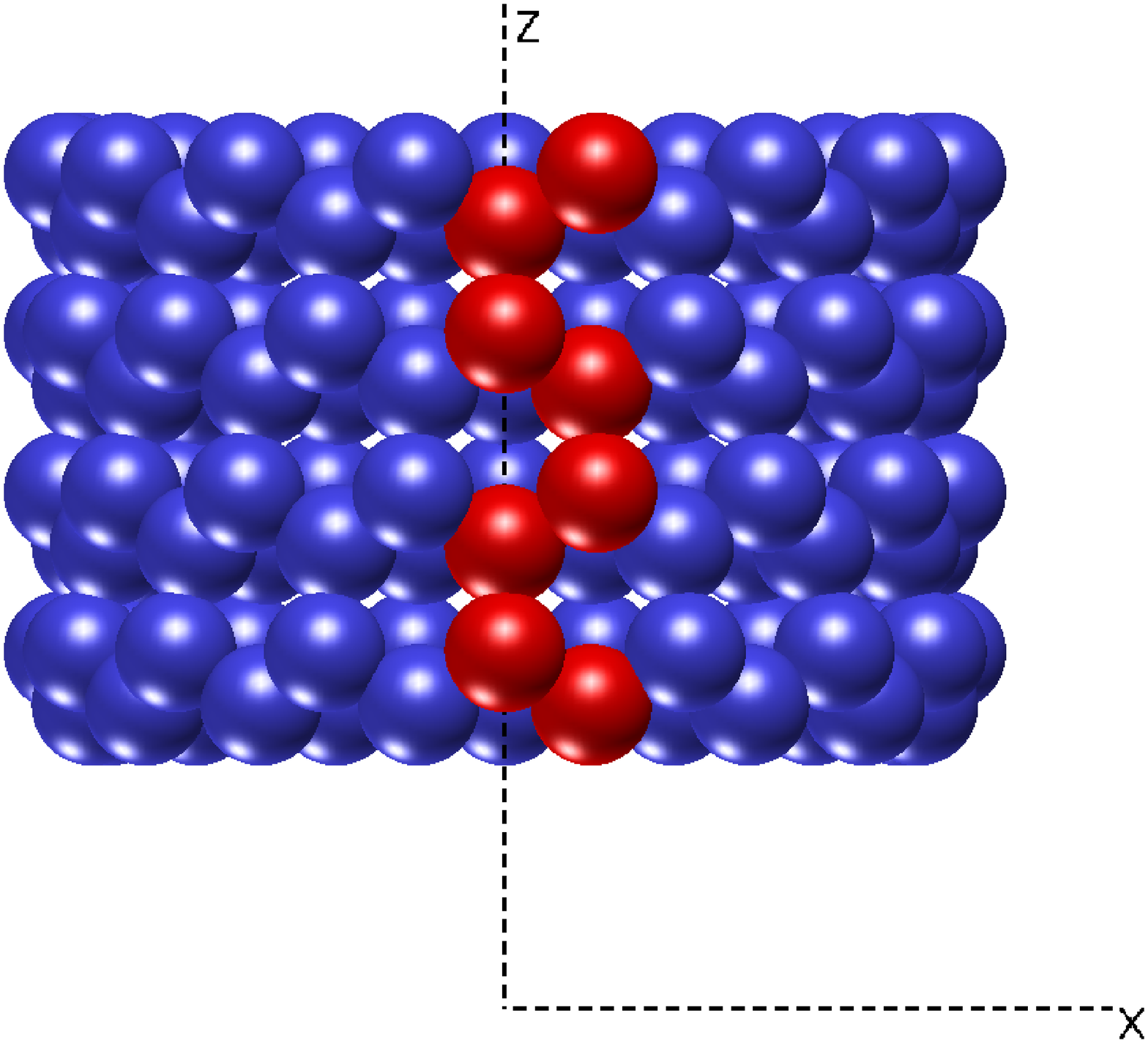}
\caption{Side view of the cyclic structure generated using the silicene nanoribbon.}
\label{Fig:SiliceneBent:Side}
\end{subfigure}
\caption{Schematic for the silicene nanoribbon bending simulation using Cyclic DFT. The radius of the cyclic structure corresponds to the bending radius of curvature $\kappa^{-1}$. The atoms within the fundamental domain are colored red.}
\label{Fig:SiliceneBending}
\end{figure}

The uniform bending simulation of a silicene nanoribbon in Cyclic DFT --- shown schematically in Fig.~\ref{Fig:SiliceneBending} --- proceeds as follows. First, we determine the equilibrium configuration of an infinite silicene sheet, using which we generate a nanoribbon of the desired width $W$. Next, depending on the desired bending curvature $\kappa$, we generate a cyclic structure with radius $\kappa^{-1}$ by `rolling' the silicene nanoribbon. For this cyclic system, we determine the fundamental domain, the atoms which belong to it, and the resulting group order. Finally, we compute the cyclic structure's electronic ground-state and energy using the formulation described in Section~\ref{Sec:Formulation}. We note that the use of a cyclic structure is expected to provide an accurate representation of pure bending,\footnote{For small enough curvatures.} since it replicates the desired bending curvature locally, and furthermore, the electronic interactions are short ranged (i.e., matter is near-sighted \cite{prodan2005nearsightedness}). 

\begin{figure}[ht]
\begin{subfigure}{0.45\linewidth}
\centering
\includegraphics[scale=0.48]{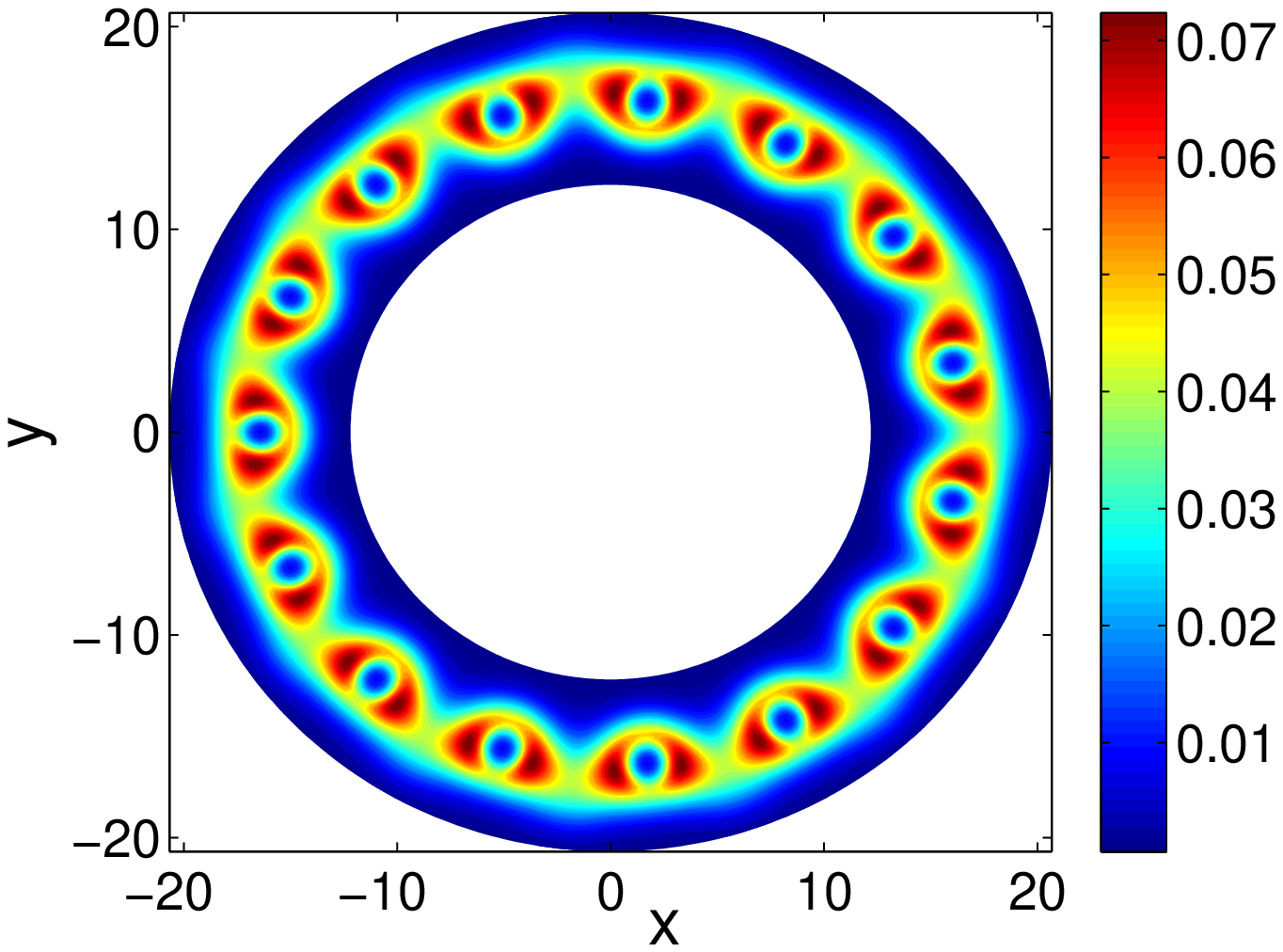}
\caption{Electron density contours on $z \approx 25$ Bohr plane.}
\label{Fig:ContourPlot}
\end{subfigure}%
$\quad\quad$
\begin{subfigure}{0.48\linewidth}
\centering
\includegraphics[scale=0.48]{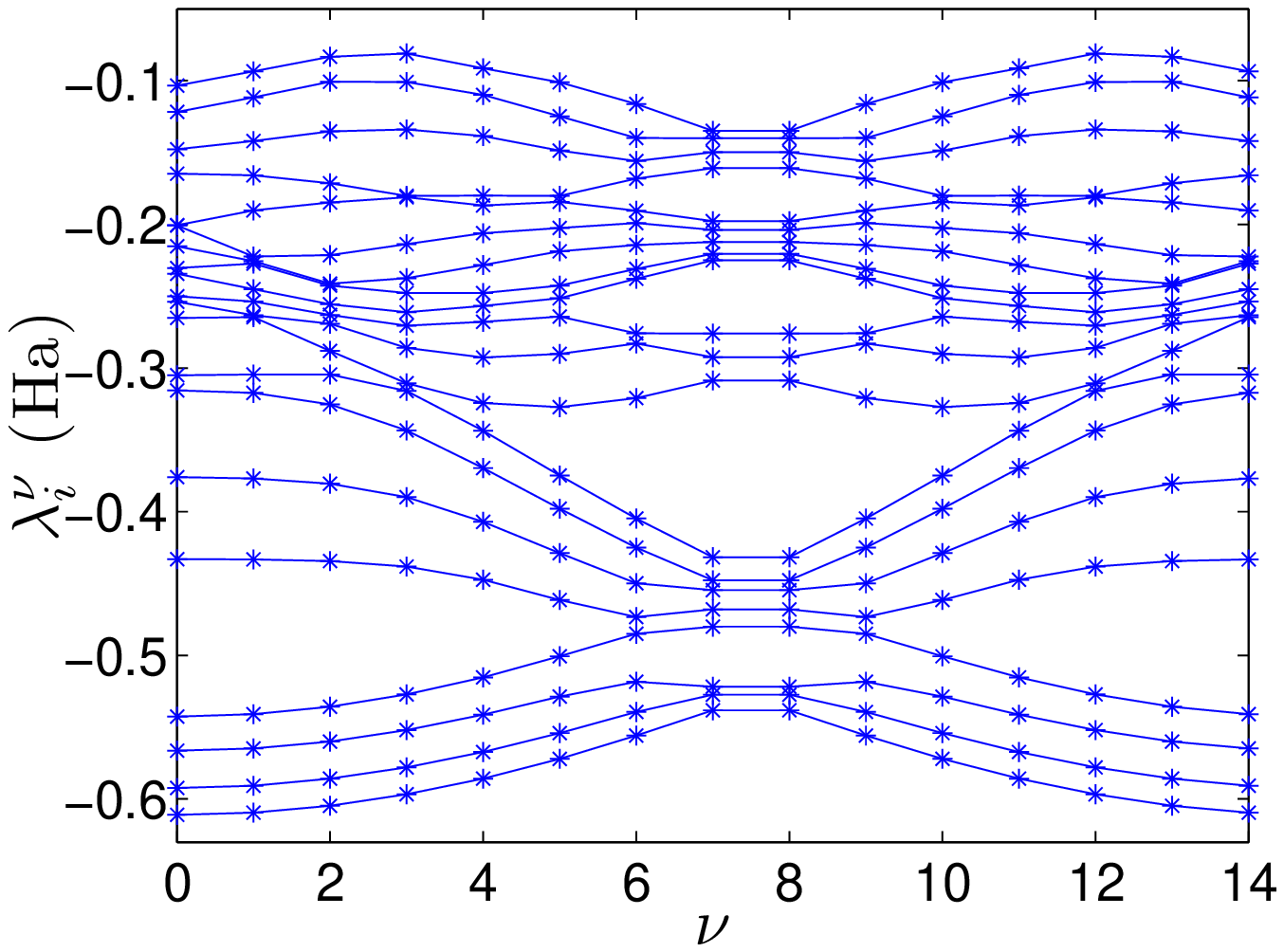}
\caption{Cyclic band structure plot.}
\label{Fig:BandPlot}
\end{subfigure}
\caption{Results for bending of a silicene nanoribbon of width $W=20.79$ Bohr and radius of curvature $\kappa^{-1} = 17.49$ Bohr. The cyclic structure has $M_{\calP}=8$ atoms in the fundamental domain and group order $\mathfrak{N}=15$.}
\label{Fig:BendingSimulationResults}
\end{figure}

We first use the above procedure to determine the electronic ground-state of a silicene nanoribbon of width $W=20.79$ Bohr with bending radius of curvature $\kappa^{-1}=17.49$ Bohr. The corresponding cyclic structure has $M_{\calP}=8$ atoms in the fundamental domain and group order $\mathfrak{N}=15$. We present the results obtained by Cyclic DFT in Fig.~\ref{Fig:BendingSimulationResults}. Specifically, in Fig.~\ref{Fig:ContourPlot}, we present the contours of the electron density on the $z \approx 25$ Bohr plane. In Fig.~\ref{Fig:BandPlot}, we present the cyclic band structure plot\footnote{Analogue of the traditional band structure plot for systems with translational symmetry.} for this system, i.e., the eigenvalues $\lambda_i^{\nu}$ for $i=1, 2, \ldots, 20$ and $\nu=0, 1, \ldots \mathfrak{N}-1$. From this figure\footnote{Disregarding the $\nu = 0$ point, the symmetry of the curves about a vertical line passing through the point $\nu = 7.5$ can be explained in terms of the theoretical framework presented in \citep{My_Elliott_Symmetry_paper}.}, we can see that there is a negligible bandgap for the bent system. In addition, the system appears to be electronically stable. An analogous plot of the phonon eigenvalues can be used to study the structural stability of such systems, and can therefore be used to predict the onset and modes of instabilities\footnote{See \citep{aghaei2012symmetry} for examples of phonon-band structure diagrams in nano systems with helical symmetries, computed using empirical inter-atomic potentials.}. This is a worthy topic of research and is currently being pursued by the authors. 

\begin{figure}[ht]
\centering
\includegraphics[width=0.6\textwidth]{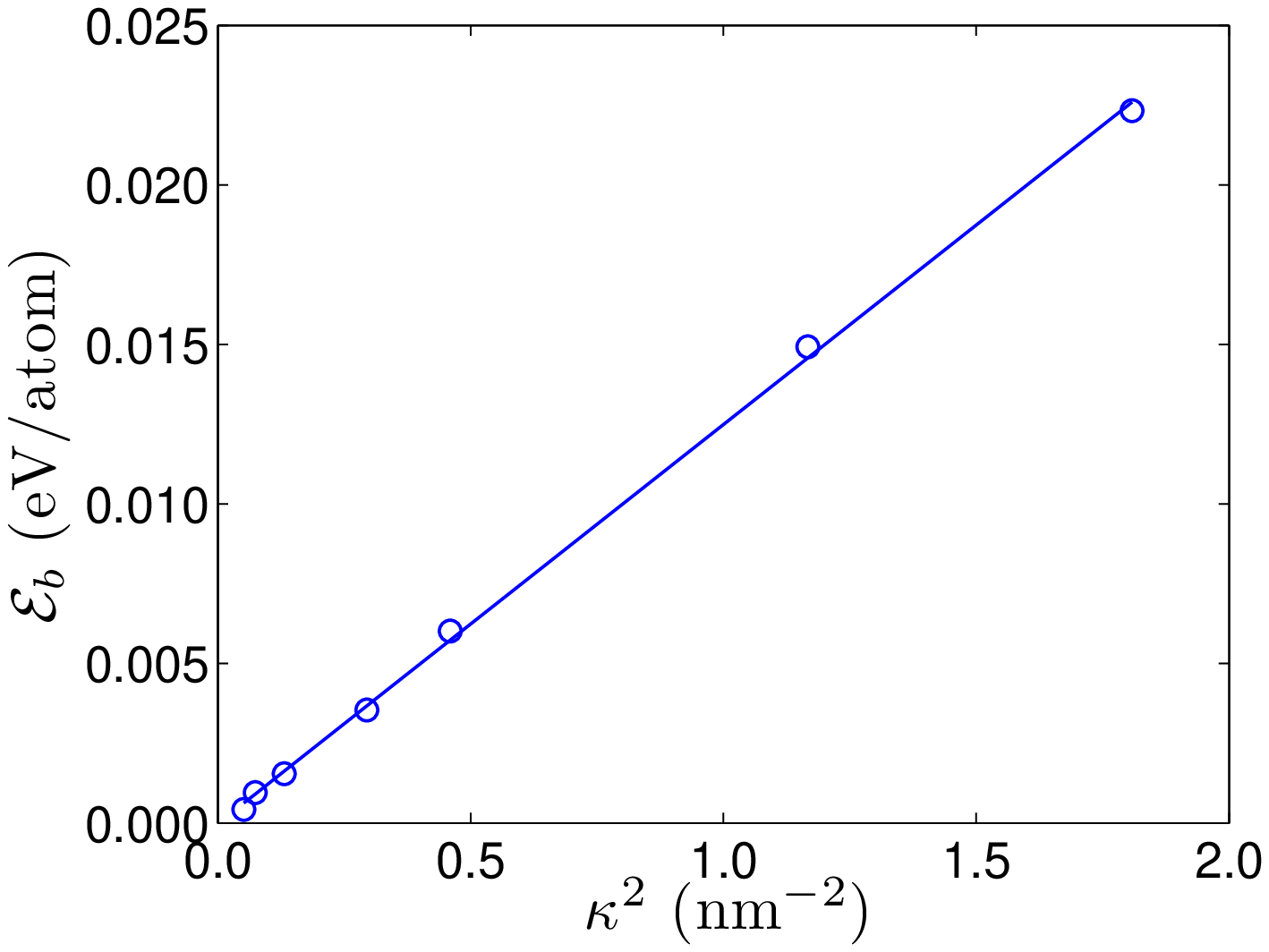}
\caption{Bending energy as a function of the curvature for a silicene nanoribbon of width $W=20.79$ Bohr.}
\label{fig:bendingEnergyCurvature}
\end{figure}

Next, we use Cyclic DFT to study the variation of the silicene nanoribbon's bending energy $\mathcal{E}_b$ with its curvature $\kappa$. We define the strain energy due to bending at a given $\kappa$ to be the difference between the free energy of two configurations --- flat silicene nanoribbon\footnote{The free energy is computed using ABINIT.} (i.e. $\kappa^{-1} \rightarrow 0$) and the cyclic silicene nanostructure with radius of curvature $\kappa^{-1}$. For this study, we choose a silicene nanoribbon of width $W=20.79$ Bohr and bending radii of $\kappa^{-1}=14.05, 17.49, 27.87, 34.80, 52.14, 69.50, \, \text{and} \, 83.39$ Bohr. The corresponding cyclic structures have $M_{\calP}=8$ atoms in the fundamental domain and cyclic group orders of $\mathfrak{N}=12, 15, 30, 45, 60, \, \text{and} \, 72$, respectively. In order to be able to accurately calculate the bending energy, we choose all the parameters within Cyclic DFT so as to achieve an overall accuracy of $10^{-5}$ Ha/atom in the energy. Anticipating Euler-Bernoulli type bending behavior, we plot the bending energy $\mathcal{E}_b$ so calculated as a function of $\kappa^{-2}$ in Fig.~\ref{fig:bendingEnergyCurvature}. We observe that $\mathcal{E}_b$ is proportional to $\kappa^{-2}$ --- consistent with Euler-Bernoulli bending\footnote{For a majority of the data points in  Fig.~\ref{fig:bendingEnergyCurvature}, the straight line fit replicates the simulation data to accuracies of over $98 \%$.} --- with the predicted bending modulus being $32.6$ eV/atom $\text{\AA}^2$. It is worth noting that at the electronic level, we find that there is no noticeable bandgap opening in the silicene nanoribbon for the aforementioned bending curvatures.

On normalizing the calculated bending stiffness with the surface area \citep{nikiforov2014tight, jiang2013elastic}, we obtain a value of $D = 6.16$ eV, which lies between the values of $1.45$ eV and $9.61$ eV that have  been obtained in the literature for graphene \citep{nikiforov2014tight, kudin2001c} and molybdenum disulphide \citep{jiang2013elastic}, respectively. To get an intuitive understanding of this observation, we recall that according to continuum theories of bending, there is usually a strong influence of the effective cross section on the bending stiffness of a material specimen. Due to the presence of out of plane atoms in a single layer relaxed silicene structure, it is likely that this material has an effective thickness\footnote{If one assumes that continuum shell theory is applicable in the present setting, we have the following relation for the effective thickness \citep{shenderova2002carbon}:
\begin{align}
t = \sqrt{\frac{12 D (1-\nu^2)}{Y}} \,,
\end{align}
where $Y$ is the in-plane Young's modulus, and $\nu$ is the Poisson's ratio. On substituting the values of $D$, $\nu$, and $Y$ for silicene into the above expression, we obtain $t \sim 0.4$ nm, which is in good agreement with the values that have been typically employed in literature \citep{pei2014effects,peng2013mechanical}. We note that the effective thickess is sometimes also referred to as the intrinsic finite thickness \citep{zhang2015elastic}.} that is intermediate when compared to the planar single atomic layer graphene structure and the three atomic layer molybdenum disulphide structure, thus leading to a similar trend in the bending stiffness in these three materials. On similar lines, we comment that due to the overall similarities in structure between the buckled silicene geometry studied here and the puckered phosphorene structures studied in the literature \citep{zhang2015elastic, yang2015temperature}, the bending stiffness of these two materials turns out to be quite similar (i.e., our value of $6.16$ eV for silicene vs. the values of $4.88 - 7.99$ eV obtained for phosphorene in the literature \citep{zhang2015elastic}).

We should mention in passing that our simulations do not involve actual relaxation of the atoms in the bent geometry. This is representative of practical scenarios where the movement of the atoms is constrained, e.g. when substrates are utilized to impart the desired strain state \citep{kerszberg2015ab,ding2010stretchable}. Additionally, relaxation effects appear to be unimportant in systems subjected to low curvatures \citep{Dumitrica_Bending_Graphene, nikiforov2014tight}. Edge effects and/or out of plane atomic displacements are likely to influence the mechanical behavior when atomic relaxations are allowed in systems subjected to large curvatures --- particularly so, when the arrangement of atoms within the fundamental domain is more complicated than the relatively simple nanoribbon geometry considered here. We hope to investigate these effects in future work.

Finally, it is worth pointing out that the silicene nanoribbon bending simulations as described above would be extremely challenging to perform --- if not practically impossible --- with existing conventional first principles techniques (especially when large values of the radius of curvature are involved), even if high performance computing resources are used. In contrast, our serial MATLAB implementation of Cyclic DFT enables these studies to be carried out conveniently within a few hours of simulation wall time, thus demonstrating the utility of the approach.

\section{Conclusions and future directions} \label{Conclusions}
In this work, we have presented the theoretical foundations as well as the numerical formulation and implementation of Cyclic DFT --- a self-consistent first principles simulation method for nanostructures with cyclic symmetries. The Cyclic DFT methodology allows us to exploit the cyclic symmetry in a systematic and efficient manner. Additionally, it enables us to probe the behavior of various nanosystems under uniform bending deformations and therefore to directly obtain the associated materials properties from first principles. We have demonstrated these capabilities of Cyclic DFT through bending simulations of silicene nanoribbons.

With the foundational work on Cyclic DFT complete, we now briefly touch on future extensions and applications:
\begin{itemize}
\item \textbf{\small{Extending the range of applicability of Cyclic DFT:}} We are currently implementing accurate norm conserving pseudopotentials \citep{hamann1979norm, troullier1991efficient} and more elaborate exchange-correlation functionals \cite{perdew1996generalized,perdew1996rationale} into Cyclic DFT. We are also developing a well-optimized, large-scale parallel implementation in C/C++. Among other attractive parallelization features of Cyclic DFT, we are making use of the attribute that the $\mathfrak{N}$ eigenvalue problems associated with the different values of $\nu \in \Gamma_{\mathfrak{N}}$ can be solved independent of one another. This lends itself to an embarrassingly parallel implementation of Cyclic DFT. Together, these developments are going to allow us to study a wide variety of materials systems using Cyclic DFT and to reach large system sizes by making effective use of high performance computing resources.
\item  \textbf{\small{Mechanistic simulation studies of low-dimensional nanomaterials under bending deformations:}} Single layers of low dimensional nanomaterials such as graphene, silicene, germanene, phosporene and hexagonal boron nitride (both ribbons and sheets), as well as their multi-layered counterparts have risen to scientific prominence in recent years due to their unique material properties \citep{xu2013graphene,butler2013progress}. Cyclic DFT provides a natural means of studying the effect of uniform bending deformations on such materials and understanding their mechanical behavior through ab-initio atomic relaxation and molecular dynamics simulations \citep{Hutter_abinitio_MD}. 

An extension of Cyclic DFT to study non-uniform bending in low-dimensional nanomaterials can be effectively accomplished by appealing to coarse-gaining ideas developed previously for the study of crystal defects using DFT \citep{suryanarayana2013coarse, ponga2016sublinear}. The general idea is to treat non-uniform bending as a defect which breaks the cyclic symmetry of the system. Therefore, coarse-graining may be achieved by considering the non-uniformly bent structure to be under uniform bending locally. Such an approach has already been developed for atomistic systems \citep{hakobyan2012objective} using quasicontinuum type \citep{tadmor1996quasicontinuum} ideas. We anticipate that it may be also carried out at the level of electronic structure calculations.

\item \textbf{\small{Investigation of multi-physics coupling in nanosystems:}} Since Cyclic DFT is a true first principles simulation methodology, it allows for the possibility of investigating the effect of bending deformations on electronic, magnetic, transport and optical materials properties in nanosystems of interest. The study of electronic properties of two-dimensional nanomaterials systems under deformations has received much scientific attention in recent years \cite{naumov2011gap,kerszberg2015ab,johari2012tuning}. In line with this, the investigation of (for example) whether an electronic band gap can be introduced in existing two dimensional materials or nanotubes by subjecting them to bending deformations is a topic worthy of further research. Importantly, there is also the tantalizing possibility that some novel nanomaterial might develop a significant magnetization or polarization in the cyclic unit cell when subjected to a bending deformation. Such materials, if discovered, are likely to have a profound impact on the design of future sensors or energy conversion technologies. We anticipate that such studies can be conveniently accomplished using Cyclic DFT.

On these lines, we also find it worthwhile to mention the possible use of Cyclic DFT in studying the nanoscale flexoelectric effect \citep{deng2014flexoelectricity, ahmadpoor2015flexoelectricity, kalinin2008electronic, nguyen2013nanoscale, dumitricua2002curvature}. Conventional first principles  calculations of flexoelectric coefficients \citep{hong2013first, ponomareva2012finite, hong2010flexoelectricity} have usually relied on Plane-wave DFT. The computation of the polarization in the periodic unit cell of an infinite crystal involves both theoretical and computational complications \citep{king1993theory, resta2007theory, spaldin2012beginner, resta1994macroscopic}. Additionally, since simulation of the flexoelectric effect, by definition, requires inhomogeneous strains to be imposed on the system under study, the setup of the simulation is further complicated \citep{xu2013direct, chandratre2012coaxing}. In contrast, estimation of the flexoelectric coefficient (for example, of a nanoribbon)  using Cyclic DFT naturally addresses these two issues since the system being simulated is of finite extent (which is likely to help in bypassing the difficulties of computing polarization) and inhomogenous strains can be introduced via uniform bending of the nanoribbon. 
\end{itemize}
\begin{center}
---
\end{center}

\section*{Acknowledgement}
Support for this work, in part, was provided through Scientific Discovery through Advanced Computing (SciDAC) program funded by U.S. Department of Energy, Office of Science, Advanced Scientific Computing Research and Basic Energy Sciences. PS acknowledges the support of the National Science Foundation under Grant Number 1553212. This work was partially carried out while ASB was at the University of Minnesota, Minneapolis. ASB acknowledges support from the following grants while at Minnesota: AFOSR FA9550-15-1-0207, NSF-PIRE OISE-0967140, ONR N00014-14-1-0714 and the MURI project FA9550-12-1-0458 (administered by AFOSR). The authors would like to acknowledge informative discussions with Richard James (Univ. of Minnesota), Ryan Elliott (Univ. of Minnesota), Kaushik Bhattacharya (Caltech), Lin Lin (Univ. of California, Berkeley) and Chao Yang (Lawrence Berkeley National Lab). The authors express their gratitude to the anonymous reviewers and the Editor for their comments and suggestions on the manuscript. The authors would also like to thank the Minnesota Supercomputing Institute for making the computing resources used in this work available. 
\bibliographystyle{elsarticle-num}

\end{document}

%% file: preamble.tex
\newcommand{\rz}{\mathbb{R}}

\newcommand{\cz}{\mathbb{C}}

\newcommand{\bfe}{{\bf e}}
\newcommand{\bff}{{\bf f}}

\newcommand{\bfk}{{\bf k}}

\newcommand{\bfx}{{\bf x}}
\newcommand{\bfy}{{\bf y}}

\newcommand{\bfz}{{\bf z}}

\newcommand{\beq}{\begin{equation}}
\newcommand{\eeq}{\end{equation}}
\newcommand{\beqs}{\begin{eqnarray}}
\newcommand{\eeqs}{\end{eqnarray}}
\newcommand{\beql}{\begin{equation} \label}

\newcommand{\half}{\frac{1}{2}}

\newcommand{\calC}{{\cal C}}
\newcommand{\calD}{{\cal D}}

\newcommand{\calG}{{\cal G}}

\newcommand{\calI}{{\cal I}}

\newcommand{\calM}{{\cal M}}

\newcommand{\calP}{{\cal P}}

\newcommand{\calS}{{\cal S}}

\newcommand{\hamil}{\mathfrak{H}}

\newtheorem{theorem}{Theorem}[section]
\newtheorem{lemma}[theorem]{Lemma}
\newtheorem{proposition}[theorem]{Proposition}
\newtheorem{corollary}[theorem]{Corollary}

\theoremstyle{definition}

\theoremstyle{definition}

\theoremstyle{remark}

\newenvironment{myproof}[1][Proof:]{\begin{trivlist}
\item[\hskip \labelsep {\bfseries #1}]}{\qed\end{trivlist}}

\newcommand{\abs}[1]{\lvert#1\rvert}

\newcommand{\innprod}[3]{\langle#1,#2\rangle_{#3}}

\newcommand{\infm}[2]{\displaystyle\inf_{#1}{#2}}

\newcommand{\pd}[2]{\frac{\partial#1}{\partial#2}}
\newcommand{\hpd}[3]{\frac{\partial^{#3}#1}{\partial#2^{#3}}}

\let\oldFootnote\footnote
\newcommand\nextToken\relax

\renewcommand\footnote[1]{%
    \oldFootnote{#1}\futurelet\nextToken\isFootnote}

\newcommand\isFootnote{%
    \ifx\footnote\nextToken\textsuperscript{,}\fi}

%% file: Manuscript.bbl
\begin{thebibliography}{100}
\expandafter\ifx\csname url\endcsname\relax
  \def\url#1{\texttt{#1}}\fi
\expandafter\ifx\csname urlprefix\endcsname\relax\def\urlprefix{URL }\fi
\expandafter\ifx\csname href\endcsname\relax
  \def\href#1#2{#2} \def\path#1{#1}\fi

\bibitem{HK_DFT}
P.~C. Hohenberg, W.~Kohn, Inhomogenous electron gas, Physical Review 136~(3B)
  (1964) 864--871.

\bibitem{KohnSham_DFT}
W.~Kohn, L.~J. Sham, Self-consistent equations including exchange and
  correlation effects, Physical Review 140~(4A) (1965) 1133--1138.

\bibitem{Kresse_abinitio_iterative}
G.~Kresse, J.~Furthmuller, Efficient iterative schemes for ab initio
  total-energy calculations using a plane-wave basis set, Physical Review B 54
  (1996) 11169--11186.

\bibitem{CASTEP_1}
M.~D. Segall, P.~J.~D. Lindan, M.~J. Probert, C.~J. Pickard, P.~J. Hasnip,
  S.~J. Clark, M.~C. Payne, First-principles simulation: ideas, illustrations
  and the {CASTEP} code, Journal of Physics: Condensed Matter 14~(11) (2002)
  2717.

\bibitem{Quantum_Espresso_1}
P.~Giannozzi, S.~Baroni, N.~Bonini, M.~Calandra, R.~Car, C.~Cavazzoni,
  D.~Ceresoli, G.~L. Chiarotti, M.~Cococcioni, I.~Dabo, A.~D. Corso,
  S.~de~Gironcoli, S.~Fabris, G.~Fratesi, R.~Gebauer, U.~Gerstmann,
  C.~Gougoussis, A.~Kokalj, M.~Lazzeri, L.~Martin-Samos, N.~Marzari, F.~Mauri,
  R.~Mazzarello, S.~Paolini, A.~Pasquarello, L.~Paulatto, C.~Sbraccia,
  S.~Scandolo, G.~Sclauzero, A.~P. Seitsonen, A.~Smogunov, P.~Umari, R.~M.
  Wentzcovitch, {QUANTUM ESPRESSO}: a modular and open-source software project
  for quantum simulations of materials, Journal of Physics: Condensed Matter
  21~(39).

\bibitem{Gonze_ABINIT_1}
X.~Gonze, J.-M. Beuken, R.~Caracas, F.~Detraux, M.~Fuchs, G.-M. Rignanese,
  L.~Sindic, M.~Verstraete, G.~Zerah, F.~Jollet, M.~Torrent, A.~Roy, M.~Mikami,
  P.~Ghosez, J.-Y. Raty, D.~Allan, First-principles computation of material
  properties: the {ABINIT} software project, Computational Materials Science
  25~(3) (2002) 478 -- 492.

\bibitem{Martin_ES}
R.~M. Martin, Electronic Structure: Basic Theory and Practical Methods, 1st
  Edition, Cambridge University Press, 2004.

\bibitem{Octopus_1}
A.~Castro, H.~Appel, M.~Oliveira, C.~Rozzi, X.~Andrade, F.~Lorenzen,
  M.~Marques, E.~Gross, A.~Rubio, Octopus: A tool for the application of
  time-dependent density functional theory, Physica Status Solidi (B) Basic
  Research 243~(11) (2006) 2465--2488.

\bibitem{PARSEC}
L.~Kronik, A.~Makmal, M.~L. Tiago, M.~M.~G. Alemany, M.~Jain, X.~Huang,
  Y.~Saad, J.~R. Chelikowsky, Parsec – the pseudopotential algorithm for
  real-space electronic structure calculations: recent advances and novel
  applications to nano-structures, Physica Status Solidi (b) 243~(5) (2006)
  1063--1079.

\bibitem{Phanish_SPARC_1}
S.~Ghosh, P.~Suryanarayana, {SPARC}: Accurate and efficient finite-difference
  formulation and parallel implementation of density functional theory:
  {I}solated clusters, arXiv preprint arXiv:1603.04334.

\bibitem{ghosh2016sparc2}
S.~Ghosh, P.~Suryanarayana, {SPARC}: Accurate and efficient finite-difference
  formulation and parallel implementation of density functional theory. {Part
  II}: {P}eriodic systems, arXiv preprint arXiv:1603.04339.

\bibitem{Gavini_Kohn_Sham}
P.~Suryanarayana, V.~Gavini, T.~Blesgen, K.~Bhattacharya, M.~Ortiz,
  Non-periodic finite-element formulation of {K}ohn-{S}ham density functional
  theory, Journal of the Mechanics and Physics of Solids 58~(2) (2010)
  256--280.

\bibitem{Pask_FEM_review_1}
J.~Pask, P.~Sterne, Finite element methods in ab initio electronic structure
  calculations, Modelling and Simulation in Materials Science and Engineering
  13~(3) (2005) R71.

\bibitem{Pask_FEM_review_2}
J.~Pask, B.~Klein, P.~Sterne, C.~Fong, Finite-element methods in
  electronic-structure theory, Computer Physics Communications 135~(1) (2001)
  1--34.

\bibitem{Gavini_higher_order}
P.~Motamarri, M.~Nowak, K.~Leiter, J.~Knap, V.~Gavini, Higher-order adaptive
  finite-element methods for {K}ohn-{S}ham density functional theory, Journal
  of Computational Physics 253 (2013) 308--343.

\bibitem{James_OS}
R.~D. James, Objective structures, Journal of the Mechanics and Physics of
  Solids 54~(11) (2006) 2354--2390.

\bibitem{DEJ_ObjForm}
K.~Dayal, R.~S. Elliott, R.~D. James, Objective formulas, (In preparation)
  (2015).

\bibitem{Dumitrica_James_OMD}
T.~Dumitrica, R.~D. James, Objective molecular dynamics, Journal of the
  Mechanics and Physics of Solids 55~(10) (2007) 2206 -- 2236.

\bibitem{dayal2010nonequilibrium}
K.~Dayal, R.~D. James, Nonequilibrium molecular dynamics for bulk materials and
  nanostructures, Journal of the Mechanics and Physics of Solids 58~(2) (2010)
  145--163.

\bibitem{aghaei2012symmetry}
A.~Aghaei, K.~Dayal, R.~S. Elliott, Symmetry-adapted phonon analysis of
  nanostructures, Journal of the Mechanics and Physics of Solids 61 (2012)
  557--578.

\bibitem{My_PhD_Thesis}
A.~S. Banerjee, Density functional methods for {O}bjective {S}tructures: Theory
  and simulation schemes, Ph.D. thesis, University of Minnesota, Minneapolis
  (2013).

\bibitem{banerjee2015spectral}
A.~S. Banerjee, R.~S. Elliott, R.~D. James, A spectral scheme for kohn--sham
  density functional theory of clusters, Journal of Computational Physics 287
  (2015) 226--253.

\bibitem{wei2012bending}
Y.~Wei, B.~Wang, J.~Wu, R.~Yang, M.~L. Dunn, Bending rigidity and gaussian
  bending stiffness of single-layered graphene, Nano letters 13~(1) (2012)
  26--30.

\bibitem{naumov2011gap}
I.~Naumov, A.~Bratkovsky, Gap opening in graphene by simple periodic
  inhomogeneous strain, Physical Review B 84~(24) (2011) 245444.

\bibitem{Dumitrica_Bending_Graphene}
D.~Zhang, E.~Akatyeva, T.~Dumitric\ifmmode~\u{a}\else \u{a}\fi{}, Bending
  ultrathin graphene at the margins of continuum mechanics, Phys. Rev. Lett.
  106 (2011) 255503.

\bibitem{ma2015thermal}
J.~Ma, Y.~Ni, S.~Volz, T.~Dumitric{\u{a}}, Thermal transport in single-walled
  carbon nanotubes under pure bending, Physical Review Applied 3~(2) (2015)
  024014.

\bibitem{Pekka_Efficient_Approach}
P.~Koskinen, O.~O. Kit, Efficient approach for simulating distorted materials,
  Physical review letters 105~(10) (2010) 106401.

\bibitem{Pekka_CNT_Bending}
P.~Koskinen, Electronic and optical properties of carbon nanotubes under pure
  bending, Phys. Rev. B 82 (2010) 193409.

\bibitem{Pekka_GNR_Bending}
P.~Koskinen, Graphene nanoribbons subject to gentle bends, Physical Review B
  85~(20) (2012) 205429.

\bibitem{Pekka_Revised_Periodic}
O.~O. Kit, L.~Pastewka, P.~Koskinen, Revised periodic boundary conditions:
  Fundamentals, electrostatics, and the tight-binding approximation, Physical
  Review B 84~(15) (2011) 155431.

\bibitem{zhang2008stability}
D.~Zhang, M.~Hua, T.~Dumitric{\u{a}}, Stability of polycrystalline and wurtzite
  si nanowires via symmetry-adapted tight-binding objective molecular dynamics,
  The Journal of chemical physics 128~(8) (2008) 084104.

\bibitem{zhang2009electromechanical}
D.~Zhang, R.~James, T.~Dumitric{\u{a}}, Electromechanical characterization of
  carbon nanotubes in torsion via symmetry adapted tight-binding objective
  molecular dynamics, Physical Review B 80~(11) (2009) 115418.

\bibitem{kit2012twisting}
O.~Kit, T.~Tallinen, L.~Mahadevan, J.~Timonen, P.~Koskinen, Twisting graphene
  nanoribbons into carbon nanotubes, Physical Review B 85~(8) (2012) 085428.

\bibitem{Pekka_Twisted_GNR}
P.~Koskinen, Electromechanics of twisted graphene nanoribbons, Applied Physics
  Letters 99~(1) (2011) 013105.

\bibitem{ismail2000ab}
S.~Ismail-Beigi, T.~Arias, Ab initio study of screw dislocations in {M}o and
  {T}a: a new picture of plasticity in bcc transition metals, Physical Review
  Letters 84~(7) (2000) 1499.

\bibitem{hauch1999dynamic}
J.~A. Hauch, D.~Holland, M.~Marder, H.~L. Swinney, Dynamic fracture in single
  crystal silicon, Physical Review Letters 82~(19) (1999) 3823.

\bibitem{cocco2010gap}
G.~Cocco, E.~Cadelano, L.~Colombo, Gap opening in graphene by shear strain,
  Physical Review B 81~(24) (2010) 241412.

\bibitem{koskinen2009density}
P.~Koskinen, V.~M{\"a}kinen, Density-functional tight-binding for beginners,
  Computational Materials Science 47~(1) (2009) 237--253.

\bibitem{senechal1996quasicrystals}
M.~Senechal, Quasicrystals and geometry, CUP Archive, 1996.

\bibitem{wiki_cyclic_compound}
Wikipedia, \href{https://en.wikipedia.org/wiki/Cyclic_compound}{Cyclic compound
  --- {W}ikipedia{,} the free encyclopedia}, [Online; accessed 12-May-2016]
  (2016).
\newline\urlprefix\url{https://en.wikipedia.org/wiki/Cyclic_compound}

\bibitem{hargittai2009symmetry}
M.~Hargittai, I.~Hargittai, Symmetry through the Eyes of a Chemist, Springer
  Science \& Business Media, 2009.

\bibitem{willock2009}
D.~J. Willock, Molecular Symmetry, John Wiley \& Sons, 2009.

\bibitem{go1973ring}
N.~Go, H.~A. Scheraga, Ring closure in chain molecules with {Cn}, {I}, and
  {S{2n}} symmetry, Macromolecules 6~(2) (1973) 273--281.

\bibitem{altmann1994point}
S.~Altmann, P.~Herzig, Point-group theory tables, Oxford, 1994.

\bibitem{My_MS_Thesis}
A.~S. Banerjee, Harmonic analysis on isometry groups of {O}bjective
  {S}tructures and its applications to {O}bjective {D}ensity {F}unctional
  {T}heory, Master's thesis, University of Minnesota, Minneapolis (2011).

\bibitem{slater1974self}
J.~C. Slater, The self-consistent field for molecules and solids, Vol.~4,
  McGraw-Hill New York, 1974.

\bibitem{LCAO_famous}
J.~C. Slater, G.~F. Koster, Simplified {LCAO} method for the periodic potential
  problem, Phys. Rev. 94 (1954) 1498--1524.

\bibitem{Roothan_SCF_Symmetry}
C.~C.~J. Roothaan, Self-consistent field theory for open shells of electronic
  systems, Reviews of Modern Physics 32 (1960) 179--185.

\bibitem{LCAO_1}
C.~M. Zicovich-Wilson, R.~Dovesi, On the use of symmetry-adapted crystalline
  orbitals in scf-lcao periodic calculations. i. the construction of the
  symmetrized orbitals, International Journal of Quantum Chemistry 67~(5)
  (1998) 299--309.

\bibitem{LCAO_2}
C.~M. Zicovich-Wilson, R.~Dovesi, On the use of symmetry-adapted crystalline
  orbitals in scf-lcao periodic calculations. ii. implementation of the
  self-consistent-field scheme and examples, International Journal of Quantum
  Chemistry 67~(5) (1998) 311--320.

\bibitem{atkins2011molecular}
P.~W. Atkins, R.~S. Friedman, Molecular quantum mechanics, Oxford university
  press, 2011.

\bibitem{Reed_Simon4}
M.~Reed, B.~Simon, Analysis of Operators, Vol.~IV of Methods of Modern
  Mathematical Physics, Academic Press, 1978.

\bibitem{LeBris_Defranceschi_ReviewBook}
M.~Defranceschi, C.~{Le Bris} (Eds.), Mathematical Models and Methods for Ab
  Initio Quantum Chemistry, Vol.~74 of Lecture Notes in Chemistry, Springer,
  2000.

\bibitem{zhou2014chebyshev}
Y.~Zhou, J.~R. Chelikowsky, Y.~Saad, Chebyshev-filtered subspace iteration
  method free of sparse diagonalization for solving the kohn--sham equation,
  Journal of Computational Physics 274 (2014) 770--782.

\bibitem{Mermin_Finite_Temp}
N.~D. Mermin, Thermal properties of the inhomogeneous electron gas, Physical
  Review 137 (1965) A1441--A1443.

\bibitem{Kohanoff}
J.~Kohanoff, Electronic structure calculations for solids and molecules: Theory
  and computational methods, 1st Edition, Cambridge University Press, 2006.

\bibitem{suryanarayana2014augmented}
P.~Suryanarayana, D.~Phanish, Augmented lagrangian formulation of orbital-free
  density functional theory, Journal of Computational Physics 275 (2014)
  524--538.

\bibitem{Pask2005}
J.~E. Pask, P.~A. Sterne, Real-space formulation of the electrostatic potential
  and total energy of solids, Phys. Rev. B 71 (2005) 113101.

\bibitem{wavefunc_decay1}
M.~Hoffmann-Ostenhof, T.~Hoffmann-Ostenhof, R.~Ahlrichs, J.~Morgan, On the
  exponential fall off of wavefunctions and electron densities, in:
  K.~Osterwalder (Ed.), Mathematical Problems in Theoretical Physics, Vol. 116
  of Lecture Notes in Physics, Springer Berlin / Heidelberg, 1980, pp. 62--67.

\bibitem{wavefunc_decay2}
R.~Ahlrichs, M.~Hoffmann-Ostenhof, T.~Hoffmann-Ostenhof, J.~D. Morgan, Bounds
  on the decay of electron densities with screening, Physical Review A 23~(5)
  (1981) 2106--2117.

\bibitem{LeBris_ReviewBook}
C.~{Le Bris} (Ed.), Computational Chemistry, Vol.~X of Handbook of Numerical
  Analysis, North-Holland, 2003.

\bibitem{suryanarayana2013optimized}
P.~Suryanarayana, Optimized purification for density matrix calculation,
  Chemical Physics Letters 555 (2013) 291--295.

\bibitem{suryanarayana2013spectral}
P.~Suryanarayana, On spectral quadrature for linear-scaling density functional
  theory, Chemical Physics Letters 584 (2013) 182--187.

\bibitem{pratapa2015spectral}
P.~P. Pratapa, P.~Suryanarayana, J.~E. Pask, Spectral quadrature method for
  accurate o (n) electronic structure calculations of metals and insulators,
  Computer Physics Communications.

\bibitem{harris1985simplified}
J.~Harris, Simplified method for calculating the energy of weakly interacting
  fragments, Physical Review B 31~(4) (1985) 1770.

\bibitem{foulkes1989tight}
W.~M.~C. Foulkes, R.~Haydock, Tight-binding models and density-functional
  theory, Physical review B 39~(17) (1989) 12520.

\bibitem{Finnis_book}
M.~Finnis, Interatomic Forces in Condensed Matter, 1st Edition, Oxford
  University Press, 2003.

\bibitem{Parr_Yang}
R.~G. Parr, W.~Yang, Density-Functional Theory of Atoms and Molecules, Vol.~16
  of International Series of Monographs on Chemistry, Oxford University Press,
  USA, 1994.

\bibitem{Defranceschi_LeBris}
M.~Defranceschi, C.~{Le Bris}, Computing a molecule: A mathematical viewpoint,
  Journal of Mathematical Chemistry 21~(1) (1997) 1--30.

\bibitem{rhodes2010crystallography}
G.~Rhodes, Crystallography made crystal clear: a guide for users of
  macromolecular models, Academic press, 2010.

\bibitem{rohrer2001structure}
G.~S. Rohrer, Structure and bonding in crystalline materials, Cambridge
  University Press, 2001.

\bibitem{giustino2014materials}
F.~Giustino, Materials modelling using density functional theory: properties
  and predictions, Oxford University Press (UK), 2014.

\bibitem{peierls1955quantum}
R.~E. Peierls, Quantum theory of solids, Oxford University Press, 1955.

\bibitem{gruner2000density}
G.~Gruner, Density waves in solids, Vol.~89, Westview Press, 2000.

\bibitem{kennedy2004proof}
T.~Kennedy, E.~H. Lieb, Proof of the peierls instability in one dimension, in:
  Condensed Matter Physics and Exactly Soluble Models, Springer, 2004, pp.
  85--88.

\bibitem{Evans_PDE}
L.~C. Evans, Partial Differential Equations, Vol.~19 of Graduate Studies in
  Mathematics, American Mathematical Society, 1998.

\bibitem{Renardy_Rogers}
M.~Renardy, R.~C. Rogers, An Introduction to Partial Differential Equations,
  2nd Edition, Vol.~13 of Texts in Applied Mathematics, Springer, 2004.

\bibitem{Kato}
T.~Kato, Perturbation Theory for Linear Operators, Classics in Mathematics,
  Springer, 1995.

\bibitem{grisvard2011elliptic}
P.~Grisvard, Elliptic problems in nonsmooth domains, Vol.~69, SIAM, 2011.

\bibitem{Folland_Harmonic}
G.~B. Folland, A Course in Abstract Harmonic Analysis, 1st Edition, Studies in
  Advanced Mathematics, Taylor {\&} Francis, 1994.

\bibitem{Barut_Reps}
A.~O. Barut, R.~Raczka, Theory of Group Representations and Applications,
  second revised Edition, World Scientific Publishing Company, 1986.

\bibitem{Bossavit_Old}
A.~Bossavit, Symmetry, groups, and boundary value problems. a progressive
  introduction to noncommutative harmonic analysis of partial differential
  equations in domains with geometrical symmetry, Computer Methods in Applied
  Mechanics and Engineering 56~(2) (1986) 167--215.

\bibitem{Bossavit_New}
A.~Bossavit, Boundary value problems with symmetry and their approximation by
  finite elements, SIAM Journal on Applied Mathematics 53~(5) (1993)
  1352--1380.

\bibitem{My_Elliott_Symmetry_paper}
A.~S. Banerjee, R.~S. Elliott., A systematic framework for the study of a
  certain class of frequently occurring non-generic degeneracies, (in
  preparation) (2016).

\bibitem{Ashcroft_Mermin}
N.~W. Ashcroft, N.~D. Mermin, Solid State Physics, 1st Edition, Brooks Cole,
  1976.

\bibitem{Odeh_Keller}
F.~Odeh, J.~B. Keller, Partial differential equations with periodic
  coefficients and bloch waves in crystals, Journal of Mathematical Physics 5
  (1964) 1499--1504.

\bibitem{Sattlegger_Thesis}
D.~Sattlegger, A generalization of bloch’s theorem to objective atomic
  structures, (In preparation) (2007).

\bibitem{Pekka_1}
P.~Koskinen, O.~O. Kit, Efficient approach for simulating distorted materials,
  Phys. Rev. Lett. 105 (2010) 106401.

\bibitem{Pekka_2}
O.~O. Kit, L.~Pastewka, P.~Koskinen, Revised periodic boundary conditions:
  Fundamentals, electrostatics, and the tight-binding approximation, Phys. Rev.
  B 84 (2011) 155431.

\bibitem{mohseni2000numerical}
K.~Mohseni, T.~Colonius, Numerical treatment of polar coordinate singularities,
  Journal of Computational Physics 157~(2) (2000) 787--795.

\bibitem{lai2001note}
M.-C. Lai, A note on finite difference discretizations for poisson equation on
  a disk, Numerical Methods for Partial Differential Equations 17~(3) (2001)
  199--203.

\bibitem{mazziotti1999spectral}
D.~A. Mazziotti, Spectral difference methods for solving differential
  equations, Chemical physics letters 299~(5) (1999) 473--480.

\bibitem{cohl1999compact}
H.~S. Cohl, J.~E. Tohline, A compact cylindrical green’s function expansion
  for the solution of potential problems, The astrophysical journal 527~(1)
  (1999) 86.

\bibitem{wiki_cylindrical_multipole}
Wikipedia,
  \href{https://en.wikipedia.org/wiki/Cylindrical_multipole_moments}{Cylindrical
  multipole moments --- {W}ikipedia{,} the free encyclopedia}, [Online;
  accessed 12-May-2016] (2009).
\newline\urlprefix\url{https://en.wikipedia.org/wiki/Cylindrical_multipole_moments}

\bibitem{my_analog_of_planewaves}
A.~S. Banerjee, R.~S. Elliott, R.~D. James, An analog of the plane-wave method
  for isolated systems., (In preparation) (2016).

\bibitem{gygi1995real}
F.~Gygi, G.~Galli, Real-space adaptive-coordinate electronic-structure
  calculations, Physical Review B 52~(4) (1995) R2229.

\bibitem{motamarri2014subquadratic}
P.~Motamarri, V.~Gavini, Subquadratic-scaling subspace projection method for
  large-scale kohn-sham density functional theory calculations using spectral
  finite-element discretization, Physical Review B 90~(11) (2014) 115127.

\bibitem{BigDFT}
L.~Genovese, A.~Neelov, S.~Goedecker, T.~Deutsch, S.~A. Ghasemi, A.~Willand,
  D.~Caliste, O.~Zilberberg, M.~Rayson, A.~Bergman, et~al., Daubechies wavelets
  as a basis set for density functional pseudopotential calculations, The
  Journal of Chemical Physics 129 (2008) 014109.

\bibitem{Chelikowsky_Saad_1}
J.~R. Chelikowsky, N.~Troullier, K.~Wu, Y.~Saad, {Higher order} {finite
  difference} pseudopotential method: An application to diatomic molecules,
  Phys. Rev. B 50 (1994) 11355--11364.

\bibitem{ono1999timesaving}
T.~Ono, K.~Hirose, Timesaving double-grid method for real-space
  electronic-structure calculations, Physical review letters 82~(25) (1999)
  5016.

\bibitem{bobbitt2015high}
N.~S. Bobbitt, G.~Schofield, C.~Lena, J.~R. Chelikowsky, High order forces and
  nonlocal operators in a {K}ohn--{S}ham {H}amiltonian, Physical Chemistry
  Chemical Physics 17~(47) (2015) 31542--31549.

\bibitem{Carter_bulk_pseudo_1}
C.~Huang, E.~A. Carter, Transferable local pseudopotentials for magnesium,
  aluminum and silicon, Physical Chemistry Chemical Physics 10~(47) (2008)
  7109--7120.

\bibitem{Carter_bulk_pseudo_2}
B.~Zhou, Y.~A. Wang, E.~A. Carter, Transferable local pseudopotentials derived
  via inversion of the kohn-sham equations in a bulk environment, Physical
  Review B 69~(12) (2004) 125109.

\bibitem{perdew1992accurate}
J.~P. Perdew, Y.~Wang, Accurate and simple analytic representation of the
  electron-gas correlation energy, Physical Review B 45~(23) (1992) 13244.

\bibitem{Serial_Chebyshev}
Y.~Zhou, Y.~Saad, M.~L. Tiago, J.~R. Chelikowsky, Self-consistent-field
  calculations using {Chebyshev}-filtered subspace iteration, Journal of
  Computational Physics 219 (2006) 172--184.

\bibitem{Parallel_Chebyshev}
Y.~Zhou, Y.~Saad, M.~L. Tiago, J.~R. Chelikowsky, Parallel
  self-consistent-field calculations via {Chebyshev}-filtered subspace
  acceleration, Phys. Rev. E 74 (2006) 066704.

\bibitem{LOBPCG_1}
A.~Knyazev, Toward the optimal preconditioned eigensolver: Locally optimal
  block preconditioned conjugate gradient method, SIAM Journal on Scientific
  Computing 23~(2) (2001) 517--541.

\bibitem{vecharynski2015generalized}
E.~Vecharynski, C.~Yang, F.~Xue, Generalized preconditioned locally harmonic
  residual method for non-hermitian eigenproblems, SIAM Journal on Scientific
  Computing 38 (2015) A500--–A527.

\bibitem{saad1986gmres}
Y.~Saad, M.~H. Schultz, Gmres: A generalized minimal residual algorithm for
  solving nonsymmetric linear systems, SIAM Journal on scientific and
  statistical computing 7~(3) (1986) 856--869.

\bibitem{saad2003iterative}
Y.~Saad, Iterative methods for sparse linear systems, 2nd Edition, {SIAM},
  2003.

\bibitem{pratapa2016anderson}
P.~P. Pratapa, P.~Suryanarayana, J.~E. Pask, Anderson acceleration of the
  jacobi iterative method: An efficient alternative to {K}rylov methods for
  large, sparse linear systems, Journal of Computational Physics 306 (2016)
  43--54.

\bibitem{pratapa2015restarted}
P.~P. Pratapa, P.~Suryanarayana, Restarted {P}ulay mixing for efficient and
  robust acceleration of fixed-point iterations, Chemical Physics Letters 635
  (2015) 69--74.

\bibitem{suryanarayana2016alternating}
P.~Suryanarayana, P.~P. Pratapa, J.~E. Pask, Alternating anderson-richardson
  method: An efficient alternative to preconditioned {K}rylov methods for large,
  sparse linear systems, arXiv preprint arXiv:1606.08740.

\bibitem{banerjee2016periodic}
A.~S. Banerjee, P.~Suryanarayana, J.~E. Pask, Periodic {P}ulay method for robust
  and efficient convergence acceleration of self-consistent field iterations,
  Chemical Physics Letters 647 (2016) 31--35.

\bibitem{My_DG_Cheby_paper}
A.~S. Banerjee, L.~Lin, W.~Hu, C.~Yang, J.~E. Pask, Chebyshev polynomial
  filtered subspace iteration in the {D}iscontinuous {G}alerkin method for
  large-scale electronic structure calculations, arXiv preprint
  arXiv:1606.03416. (\url{http://arxiv.org/abs/1606.03416})

\bibitem{vogt2012silicene}
P.~Vogt, P.~De~Padova, C.~Quaresima, J.~Avila, E.~Frantzeskakis, M.~C. Asensio,
  A.~Resta, B.~Ealet, G.~Le~Lay, Silicene: compelling experimental evidence for
  graphenelike two-dimensional silicon, Physical review letters 108~(15) (2012)
  155501.

\bibitem{shao2013first}
Z.-G. Shao, X.-S. Ye, L.~Yang, C.-L. Wang, First-principles calculation of
  intrinsic carrier mobility of silicene, Journal of Applied Physics 114~(9)
  (2013) 093712.

\bibitem{liu2011quantum}
C.-C. Liu, W.~Feng, Y.~Yao, Quantum spin {H}all effect in silicene and
  two-dimensional germanium, Physical review letters 107~(7) (2011) 076802.

\bibitem{qin2012first}
R.~Qin, C.-H. Wang, W.~Zhu, Y.~Zhang, First-principles calculations of
  mechanical and electronic properties of silicene under strain, Aip Advances
  2~(2) (2012) 022159.

\bibitem{peng2013mechanical}
Q.~Peng, X.~Wen, S.~De, Mechanical stabilities of silicene, Rsc Advances 3~(33)
  (2013) 13772--13781.

\bibitem{prodan2005nearsightedness}
E.~Prodan, W.~Kohn, Nearsightedness of electronic matter, Proceedings of the
  National Academy of Sciences of the United States of America 102~(33) (2005)
  11635--11638.

\bibitem{nikiforov2014tight}
I.~Nikiforov, E.~Dontsova, R.~James, T.~Dumitric{\u{a}}, Tight-binding theory
  of graphene bending, Physical Review B 89~(15) (2014) 155437.

\bibitem{jiang2013elastic}
J.-W. Jiang, Z.~Qi, H.~S. Park, T.~Rabczuk, Elastic bending modulus of
  single-layer molybdenum disulfide ($mos2$): finite thickness effect,
  Nanotechnology 24~(43) (2013) 435705.

\bibitem{kudin2001c}
K.~N. Kudin, G.~E. Scuseria, B.~I. Yakobson, {C}2{F}, {BN}, and {C} nanoshell
  elasticity from ab initio computations, Physical Review B 64~(23) (2001)
  235406.

\bibitem{shenderova2002carbon}
O.~Shenderova, V.~Zhirnov, D.~Brenner, Carbon nanostructures, Critical Reviews
  in Solid State and Material Sciences 27~(3-4) (2002) 227--356.

\bibitem{pei2014effects}
Q.-X. Pei, Z.-D. Sha, Y.-Y. Zhang, Y.-W. Zhang, Effects of temperature and
  strain rate on the mechanical properties of silicene, Journal of Applied
  Physics 115~(2) (2014) 023519.

\bibitem{zhang2015elastic}
H.-Y. Zhang, J.-W. Jiang, Elastic bending modulus for single-layer black
  phosphorus, Journal of Physics D: Applied Physics 48~(45) (2015) 455305.

\bibitem{yang2015temperature}
Z.~Yang, J.~Zhao, N.~Wei, Temperature-dependent mechanical properties of
  monolayer black phosphorus by molecular dynamics simulations, Applied Physics
  Letters 107~(2) (2015) 023107.

\bibitem{kerszberg2015ab}
N.~Kerszberg, P.~Suryanarayana, Ab initio strain engineering of graphene:
  opening bandgaps up to 1 ev, RSC Advances 5~(54) (2015) 43810--43814.

\bibitem{ding2010stretchable}
F.~Ding, H.~Ji, Y.~Chen, A.~Herklotz, K.~Dörr, Y.~Mei, A.~Rastelli, O.~G.
  Schmidt, Stretchable graphene: a close look at fundamental parameters through
  biaxial straining, Nano letters 10~(9) (2010) 3453--3458.

\bibitem{hamann1979norm}
D.~Hamann, M.~Schl{\"u}ter, C.~Chiang, Norm-conserving pseudopotentials,
  Physical Review Letters 43~(20) (1979) 1494.

\bibitem{troullier1991efficient}
N.~Troullier, J.~L. Martins, Efficient pseudopotentials for plane-wave
  calculations, Physical review B 43~(3) (1991) 1993.

\bibitem{perdew1996generalized}
J.~P. Perdew, K.~Burke, M.~Ernzerhof, Generalized gradient approximation made
  simple, Physical review letters 77~(18) (1996) 3865.

\bibitem{perdew1996rationale}
J.~P. Perdew, M.~Ernzerhof, K.~Burke, Rationale for mixing exact exchange with
  density functional approximations, The Journal of Chemical Physics 105~(22)
  (1996) 9982--9985.

\bibitem{xu2013graphene}
M.~Xu, T.~Liang, M.~Shi, H.~Chen, Graphene-like two-dimensional materials,
  Chemical reviews 113~(5) (2013) 3766--3798.

\bibitem{butler2013progress}
S.~Z. Butler, S.~M. Hollen, L.~Cao, Y.~Cui, J.~A. Gupta, H.~R. Guti{\'e}rrez,
  T.~F. Heinz, S.~S. Hong, J.~Huang, A.~F. Ismach, et~al., Progress,
  challenges, and opportunities in two-dimensional materials beyond graphene,
  ACS nano 7~(4) (2013) 2898--2926.

\bibitem{Hutter_abinitio_MD}
D.~Marx, J.~Hutter, Ab initio molecular dynamics: basic theory and advanced
  methods, 1st Edition, Cambridge University Press, 2009.

\bibitem{suryanarayana2013coarse}
P.~Suryanarayana, K.~Bhattacharya, M.~Ortiz, Coarse-graining kohn--sham density
  functional theory, Journal of the Mechanics and Physics of Solids 61~(1)
  (2013) 38--60.

\bibitem{ponga2016sublinear}
M.~Ponga, K.~Bhattacharya, M.~Ortiz, A sublinear-scaling approach to
  density-functional-theory analysis of crystal defects, Journal of the
  Mechanics and Physics of Solids In Press, Accepted Manuscript.

\bibitem{hakobyan2012objective}
Y.~Hakobyan, E.~Tadmor, R.~James, Objective quasicontinuum approach for rod
  problems, Physical Review B 86~(24) (2012) 245435.

\bibitem{tadmor1996quasicontinuum}
E.~B. Tadmor, M.~Ortiz, R.~Phillips, Quasicontinuum analysis of defects in
  solids, Philosophical Magazine A 73~(6) (1996) 1529--1563.

\bibitem{johari2012tuning}
P.~Johari, V.~B. Shenoy, Tuning the electronic properties of semiconducting
  transition metal dichalcogenides by applying mechanical strains, ACS nano
  6~(6) (2012) 5449--5456.

\bibitem{deng2014flexoelectricity}
Q.~Deng, L.~Liu, P.~Sharma, Flexoelectricity in soft materials and biological
  membranes, Journal of the Mechanics and Physics of Solids 62 (2014) 209--227.

\bibitem{ahmadpoor2015flexoelectricity}
F.~Ahmadpoor, P.~Sharma, Flexoelectricity in two-dimensional crystalline and
  biological membranes, Nanoscale 7~(40) (2015) 16555--16570.

\bibitem{kalinin2008electronic}
S.~V. Kalinin, V.~Meunier, Electronic flexoelectricity in low-dimensional
  systems, Physical Review B 77~(3) (2008) 033403.

\bibitem{nguyen2013nanoscale}
T.~D. Nguyen, S.~Mao, Y.-W. Yeh, P.~K. Purohit, M.~C. McAlpine, Nanoscale
  flexoelectricity, Advanced Materials 25~(7) (2013) 946--974.

\bibitem{dumitricua2002curvature}
T.~Dumitric{\u{a}}, C.~M. Landis, B.~I. Yakobson, Curvature-induced
  polarization in carbon nanoshells, Chemical physics letters 360~(1) (2002)
  182--188.

\bibitem{hong2013first}
J.~Hong, D.~Vanderbilt, First-principles theory and calculation of
  flexoelectricity, Physical Review B 88~(17) (2013) 174107.

\bibitem{ponomareva2012finite}
I.~Ponomareva, A.~Tagantsev, L.~Bellaiche, Finite-temperature flexoelectricity
  in ferroelectric thin films from first principles, Physical Review B 85~(10)
  (2012) 104101.

\bibitem{hong2010flexoelectricity}
J.~Hong, G.~Catalan, J.~Scott, E.~Artacho, The flexoelectricity of barium and
  strontium titanates from first principles, Journal of Physics: Condensed
  Matter 22~(11) (2010) 112201.

\bibitem{king1993theory}
R.~King-Smith, D.~Vanderbilt, Theory of polarization of crystalline solids,
  Physical Review B 47~(3) (1993) 1651.

\bibitem{resta2007theory}
R.~Resta, D.~Vanderbilt, Theory of polarization: a modern approach, in: Physics
  of Ferroelectrics, Springer, 2007, pp. 31--68.

\bibitem{spaldin2012beginner}
N.~A. Spaldin, A beginner's guide to the modern theory of polarization, Journal
  of Solid State Chemistry 195 (2012) 2--10.

\bibitem{resta1994macroscopic}
R.~Resta, Macroscopic polarization in crystalline dielectrics: the geometric
  phase approach, Reviews of modern physics 66~(3) (1994) 899.

\bibitem{xu2013direct}
T.~Xu, J.~Wang, T.~Shimada, T.~Kitamura, Direct approach for flexoelectricity
  from first-principles calculations: cases for srtio3 and batio3, Journal of
  Physics: Condensed Matter 25~(41) (2013) 415901.

\bibitem{chandratre2012coaxing}
S.~Chandratre, P.~Sharma, Coaxing graphene to be piezoelectric, Applied Physics
  Letters 100~(2) (2012) 023114.

\end{thebibliography}
